\documentclass[a4paper]{aa}
\usepackage{graphicx}
\usepackage{natbib}
\usepackage{multirow}
\usepackage{longtable,lscape}
\usepackage{txfonts}  %%%
\def\atlas9{{\sc ATLAS9}}

\begin{document}

\title{Spectral optical monitoring of the double peaked emission line AGN Arp 102B: 
II.  Variability of the broad line properties }

\author {L.\v C. Popovi\'c\inst{1,2} \and A.I. Shapovalova\inst{3} \and D.Ili\'c
\inst{2,4} \and A.N.Burenkov\inst{3} \and V.H. Chavushyan\inst{5}  \and W. Kollatschny\inst{6} \and A. Kova\v
cevi\'c\inst{2,4} \and J. R. Vald\'és\inst{5}	\and J. Le\'on-Tavares\inst{5,8} \and N.G. Bochkarev\inst{7} \and
 V. Pati\~no-\'Álvarez\inst{5} \and  J. Torrealba\inst{5}}

\titlerunning{Spectral monitoring of Arp 102B}
\authorrunning{L.\v C. Popovi\'c et al.}
\offprints{A. I. Shapovalova, \\ \email{ashap@sao.ru}\\ }

\institute{
Astronomical Observatory, Volgina 7, 11160 Belgrade 74, Serbia \and
Isaac Newton Institute of Chile, Yugoslavia Branch \and 
Special Astrophysical Observatory of the Russian AS,
Nizhnij Arkhyz, Karachaevo-Cherkesia 369167, Russia \and
Department of Astronomy, Faculty of Mathematics, University
of Belgrade, Studentski trg 16, 11000 Belgrade, Serbia \and Instituto
Nacional de Astrof\'{\i}sica, \'{O}ptica y Electr\'onica, Apartado
Postal 51, CP 72000, Puebla, Pue. M\'exico \and Institut
f\"ur Astrophysik, Friedrich-Hund-Platz 1, G\"ottingen, Germany \and
Sternberg Astronomical Institute, Moscow, Russia \and
Finnish Centre for Astronomy with ESO (FINCA), University of Turku, V\"ais\"al\"antie 20, FI-21500 Piikki\"o, Finland }
\date{Received  / Accepted }

\abstract
% context heading (optional)
{We investigate  a long-term (26 years, from 1987 to 2013) variability in the broad spectral line properties
of the radio galaxy Arp 102B, an active galaxy with
 broad double-peaked emission lines. We use observations presented in Paper I (Shapovalova et al. 2013)
in the period from 1987 to 2011, and a new set of observations performed in 2012--2013.}
%aims heading  (mandatory):
{To explore  the BLR geometry, and clarify some contradictions about
the nature of the BLR  in Arp 102B we explore variations in the H$\alpha$ and H$\beta$ line parameters 
during the monitored period.}
% methods heading (mandatory):
{We fit the broad lines with three broad Gaussian functions finding the positions and intensities of the blue and red peaks
in H$\alpha$ and H$\beta$. Additionally we fit averaged line profiles with the disc model.}
% results heading (mandatory):
{We find that the broad line profiles are double-peaked and have not been changed significantly in shapes, beside
an additional  small peak that,  from time to time can be seen in the blue part of the H$\alpha$ line.
The positions of the blue and red peaks {  have not changed significantly during the monitored period}. 
The H$\beta$ line is broader than H$\alpha$ line in the monitored period.  The disc model is able to 
reproduce the H$\beta$ and H$\alpha$ broad line profiles, however, observed variability in the line parameters 
are not in a good agreement with the emission disc hypothesis.}
% conclusions heading  (optional), leave it empty if necessary
{It seems that the BLR of Arp 102B has a disc-like geometry, but the role of an outflow can also play an important role
in observed variation of the broad line properties.}


\maketitle

\section{Introduction}

Arp 102B, a LINER like object, was the first active galaxy where the broad line
double-peaked profiles have been modeled with emission of an accretion disc \citep{c89,ch89}.
After that, this galaxy has been widely
accepted as a prototype of an AGN with broad lines emitted from the disc
\citep[see][]{ch89,s90,eh94,e97} and has been studied intensively through
different monitoring campaigns \citep[see e.g.][]{mp90, n97,s00,g07,sh13}. 

In general, the monitoring campaigns 
agree that  the broad line region (BLR) in Arp 102B, where the double-peaked broad emission lines are forming,
 seems to have a disc-like geometry, but there are still some open issues and contradictions concerning the disc model
\citep[][]{mp90,n97,s00,g07,sh13}. 
The broad H$\alpha$ line profile variation, observed by \cite{s00}
from 1992 to 1996,  corresponds to gas rotating
in the disc with in-homogeneity in the surface brightness, that is in an agreement with result of \cite{g07}, 
they found that  the 
accretion disc is the most favorable model. However, \cite{g07} concluded that the most of the observed facts 
 fail to explain the variability of the profiles assuming processes in the accretion disc.
Additional disagreement with disc model was reported by \cite{mp90}, they showed  that, at least at some epochs, the
long-wavelength side of the line profile is higher than the short-wavelength side, contrary to what is expected from a
relativistic disc, i.e.  an asymmetric relativistic disc cannot explain the peak ratio observed in some periods.
 A sinusoidal variation of the red-to-blue peak flux ratio is present
 in the H$\alpha$ line profile \citep[][]{n97,sh13}. This variation can be explained as a transient orbiting
hot spot in the accretion disc\footnote{Similar variation  can also be a
consequence of gravitational lensing from a massive body close to
primary black hole \citep[][]{pop01}}. 
 
Additionally, it is confirmed that  high
ionized lines, as Ly$\alpha$ and CIV $\lambda$1550 do not show the
disc-like profile (two peaks) and that the Ly$\alpha$ line is single-peaked \citep[][]{h96}, i.e. that the
Ly$\alpha$/H$\beta$ ratio is less than 0.12 in the displaced peaks. {  The lack of
double-peaked high ionization lines from the disc can indicate that the disc medium 
is too dense for those lines and that they cannot be emitted from the disc.
However, it is interesting that}
in the near-infrared no trace of double-peaked HI lines  \citep[][]{ri06}. Moreover 
both the Bracket and Paschen series are almost absent apart from Pa$\alpha$ and Pa$\beta$.

One of the way to confirm the disc emission in the broad lines is to
observe polarization in broad lines \citep[see e.g.][]{co98,co00,af13},
since the expected position angle of
polarization should be parallel to {  the disc plane}.
This angle is the same as the direction of the radio jet \citep{a96,co98,co00},
that is not expected for the broad lines emitted from an accretion disc.
Also, \cite{c97}  found  discrepancies between the disc
model and the spectropolarimetric data, but they found that if  some
specific conditions  are taken into account, the observations could
be explained with emission of the accretion disc. 

%Finally, note here that \cite{s90} suggested that a bicone model would better fit the Fe
%K$\alpha$ line in the X-ray spectrum, and they gave some arguments
%against disc emission in case of double-peaked broad line profiles, based on the comparison of the
%disc models and the largest available line profile parameters, concluding that in the case
%of Arp 102B the remarkable agreement between the theory and observations could be fortuitous. 

Although the double-peaked broad lines of Arp 102B indicate an accretion disc
emission, there are some deviations in broad
line profile variations that may indicate another geometry.
 In Paper I \citep{sh13}
we present observations and variation in line fluxes and the continuum. In this paper
we discuss the long term spectral line profile variations in order to clarify the nature of the
broad line region  in Arp 102B.
The paper is organized as follows: in \S 2  we describe our observations and used methods of
analysis, in \S 3 we present our results of the line profile variation analysis, 
in \S 4 we discuss our results;
and finally in \S 5 we outline our conclusions.

\section{Observations and methods of analysis}

\subsection{Observations}

Spectra of Arp 102B  were taken during the monitoring period 1987--2010 with 5 different telescopes: 
6 m and 1 m telescopes of the SAO RAS (Russia), two 2.1 m telescopes in Mexico (Guillermo Haro Observatory, 
GHO, and the Observatorio Astronomico Nacional at San Pedro Martir, OAN-SPM), and two telescopes in Spain
(3.5 m and 2.2 m telescopes of Calar Alto observatory). Details of observation and data reduction are given in
Paper I, and here will not be repeated.

\begin{figure}
\centering
\includegraphics[width=9cm]{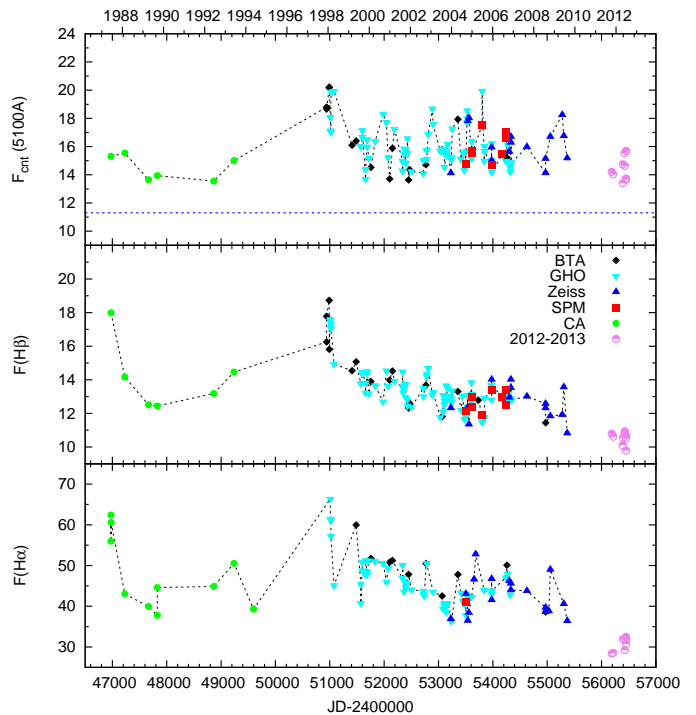}
\caption{Light-curves with the new data in the 2012-2013 period. From top to bottom:  the blue continuum
flux, the H$\beta$ and H$\alpha$ line flux.
Observations with different telescopes are denoted with different
symbols given in the middle plot. The continuum flux is plotted in units
of $10^{-16} \rm erg \ cm^{-2} s^{-1} \AA^{-1}$, and the line flux in
units of $10^{-14} \rm erg \ cm^{-2} s^{-1}$. The dashed line in the blue
and red continuum light-curves mark the contribution of the 
host galaxy starlight-continuum.} \label{lc}
\end{figure}

In addition to the observational data presented in Paper I, we added 10 new spectra 
observed in 2012--2013 period with 2.1 m telescope of GHO. The measurements of 
the continuum at 5100 \AA, H$\alpha$, and H$\beta$ fluxes are given in Table
 \ref{tab1}.  To compare new observations with ones given in Paper I, we plot in Fig. \ref{lc}
 the light curves in the continuum and in the broad lines. As it can be seen from Fig. \ref{lc},  the minimum in the line
 flux is observed in 2012-2013, while the continuum flux is similar as one observed in  the 1989-1992 period.

%%%%%%%%%%%%%%%%%%%%%%%%%%%%%%%%
%%%%%%%%%%%%%%%%%%%%%%%%%%%%%%%%

\begin{table}
\begin{center}
\caption[]{Measurements of  line fluxes of the new set of observations in the 2012--2013 period.
The continuum flux is in units
of $10^{-16} \rm erg \ cm^{-2} s^{-1} \AA^{-1}$, and the line flux in
units of $10^{-14} \rm erg \ cm^{-2} s^{-1}$.}\label{tab1}
\begin{tabular}{lccccc}
\hline \hline
N& UT-date & MJD   & F$_{\rm cnt}$5100\AA  & F(H$\beta$) & F(H$\alpha$) \\
\hline
 1 & 2012Sep17 &  56187.69 & 14.21   & 10.78  & 28.40  \\ 
 2 & 2012Oct14 &  56214.67 & 14.03   & 10.61  & 28.53  \\ 
 3 & 2013Apr07 &  56389.99 & 13.39   & 10.10  & 31.88  \\
 4 & 2013Apr11 &  56393.98 & 14.74   & 10.55  & -      \\  
 5 & 2013May10 &  56422.87 & 15.52   & 10.82  & 32.07  \\ 
 6 & 2013May13 &  56425.82 & 14.63   & 10.94  &        \\ 
 7 & 2013May14 &  56426.81 & -       & -      & 29.22  \\ 
 8 & 2013Jun06 &  56449.86 & 13.65   & 10.52  & 32.49  \\ 
 9 & 2013Jun07 &  56450.87 & 15.69   & 10.69  & 31.76  \\ 
10 & 2013Jun08 &  56451.87 & 13.73   & 9.78   & 30.76  \\ 
\hline
\end{tabular}
\end{center}
\end{table}

%%%%%%%%%%%%%%%%%%%%%%%%%%%%%%%%
\begin{figure}
\centering
\includegraphics[width=9cm]{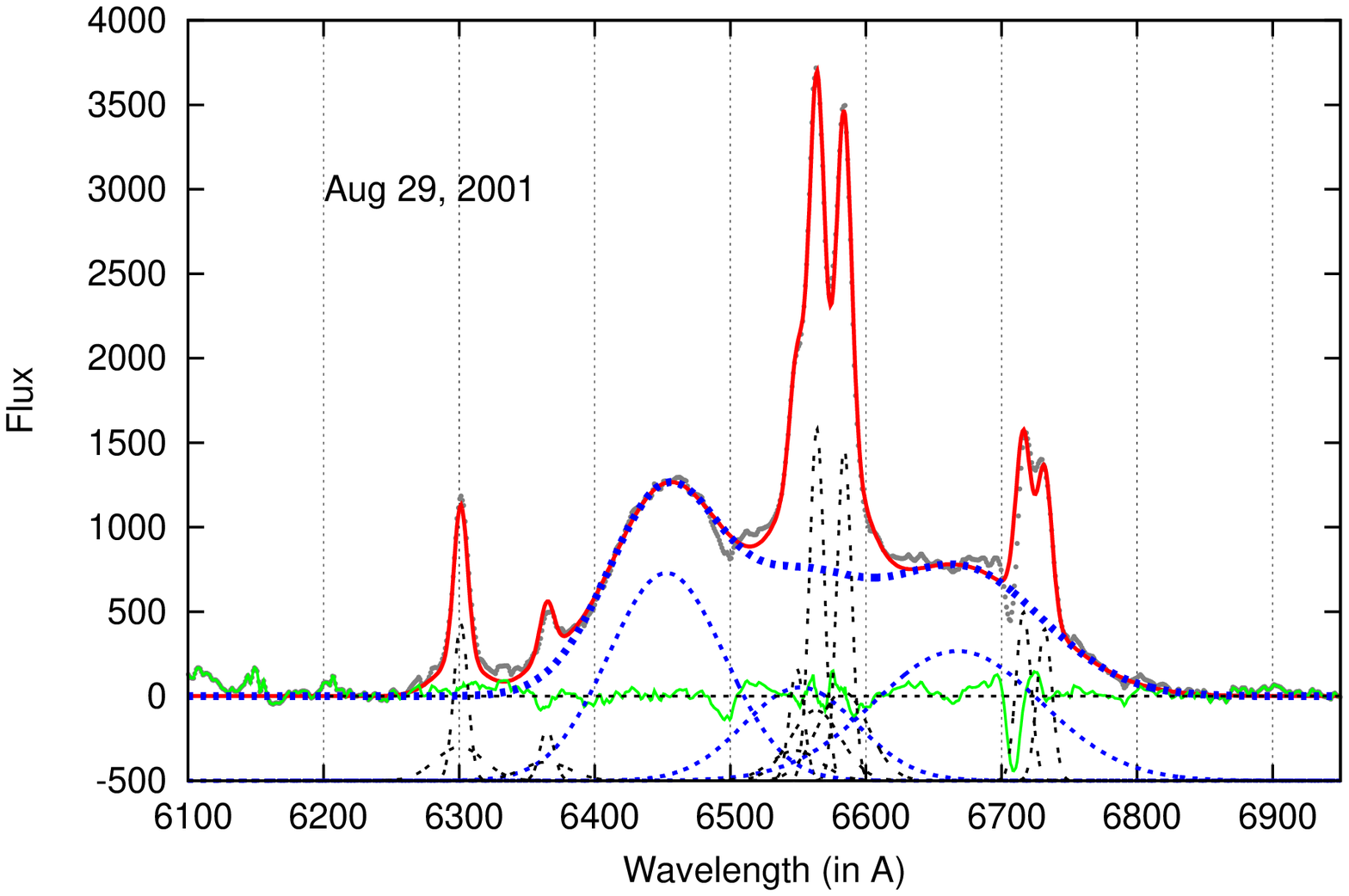}
\includegraphics[width=9cm]{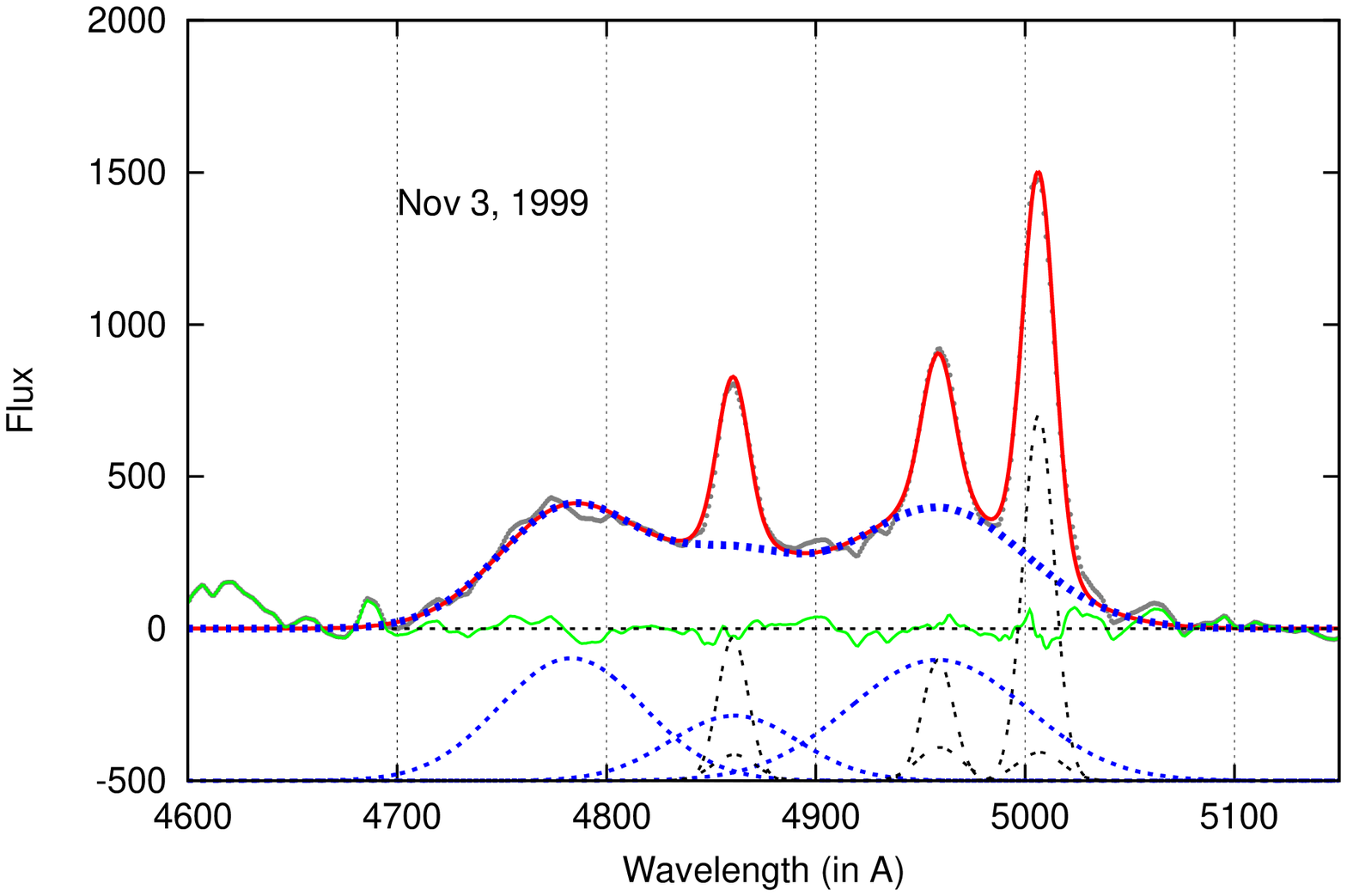}
\caption{Examples of the best Gaussian fit of one epoch H$\alpha$ (upper panel) and 
H$\beta$ line (bottom panel). Below the observed spectrum (dots), model (solid line), and
residual (thin solid line), the Gaussian components are given, where broad components are
denoted with dashed line and narrow components with double-dashed line. The thick 
dashed line reproduce the modeled broad component of the fitted Balmer line.} \label{gauss}
\end{figure}
%%%%%%%%%%%%%%%%%%%%%%%%%%%%%%%%

\subsection{Method of analysis}

\begin{figure*}
\centering
\includegraphics[width=8.cm]{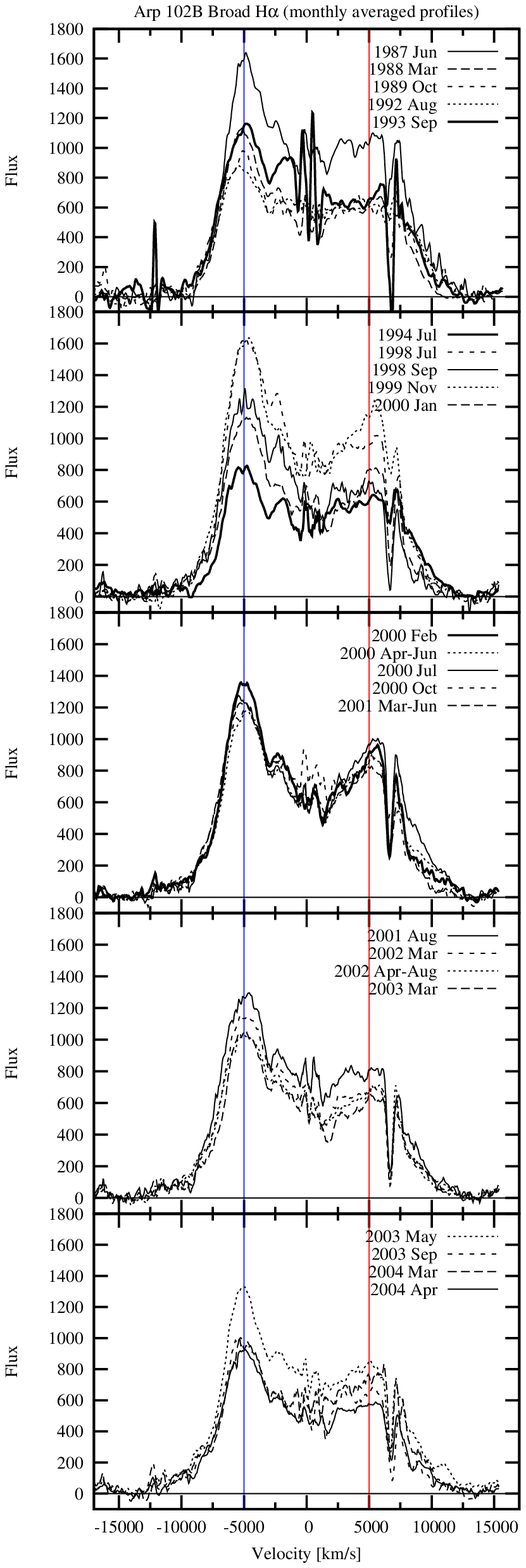}
\includegraphics[width=8.cm]{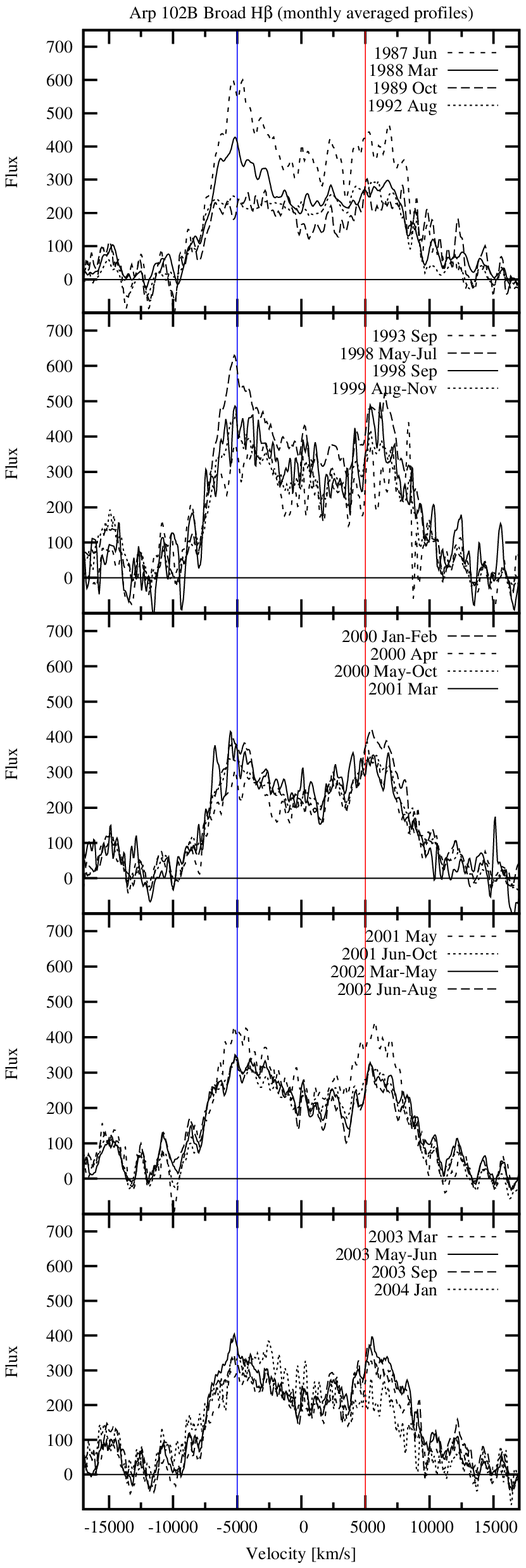}
\caption{Month-averaged profiles of the H$\alpha$ and H$\beta$
        broad emission lines in the period 1987--2004. The abscissa (OX) shows
        radial velocities with respect to the narrow component of the H$\alpha$
        or H$\beta$ line. The ordinate (OY) shows the flux in units of
         $10^{-16} {\rm erg\ cm^{-2}s^{-1}}$\AA$^{-1}$.}\label{month1}
\end{figure*}

\begin{figure*}
\centering
\includegraphics[width=8.cm]{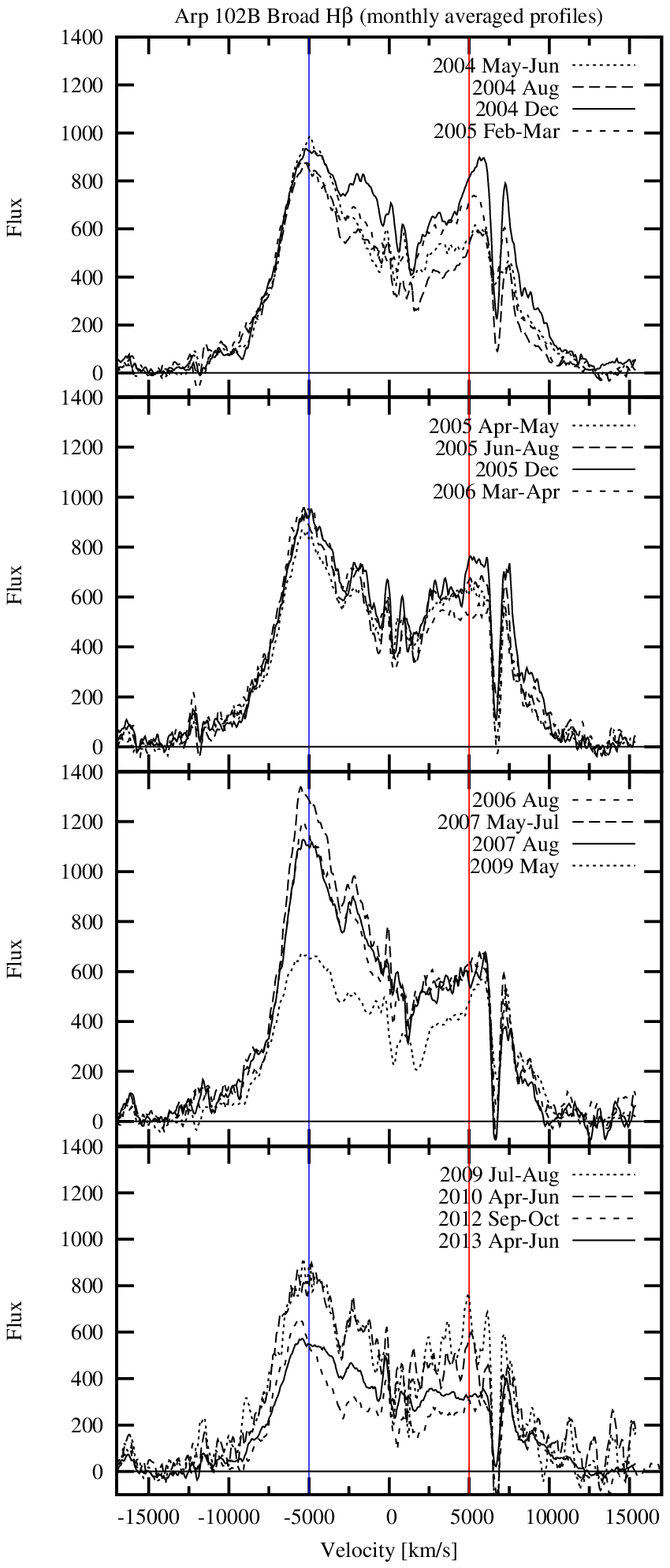}
\includegraphics[width=8.cm]{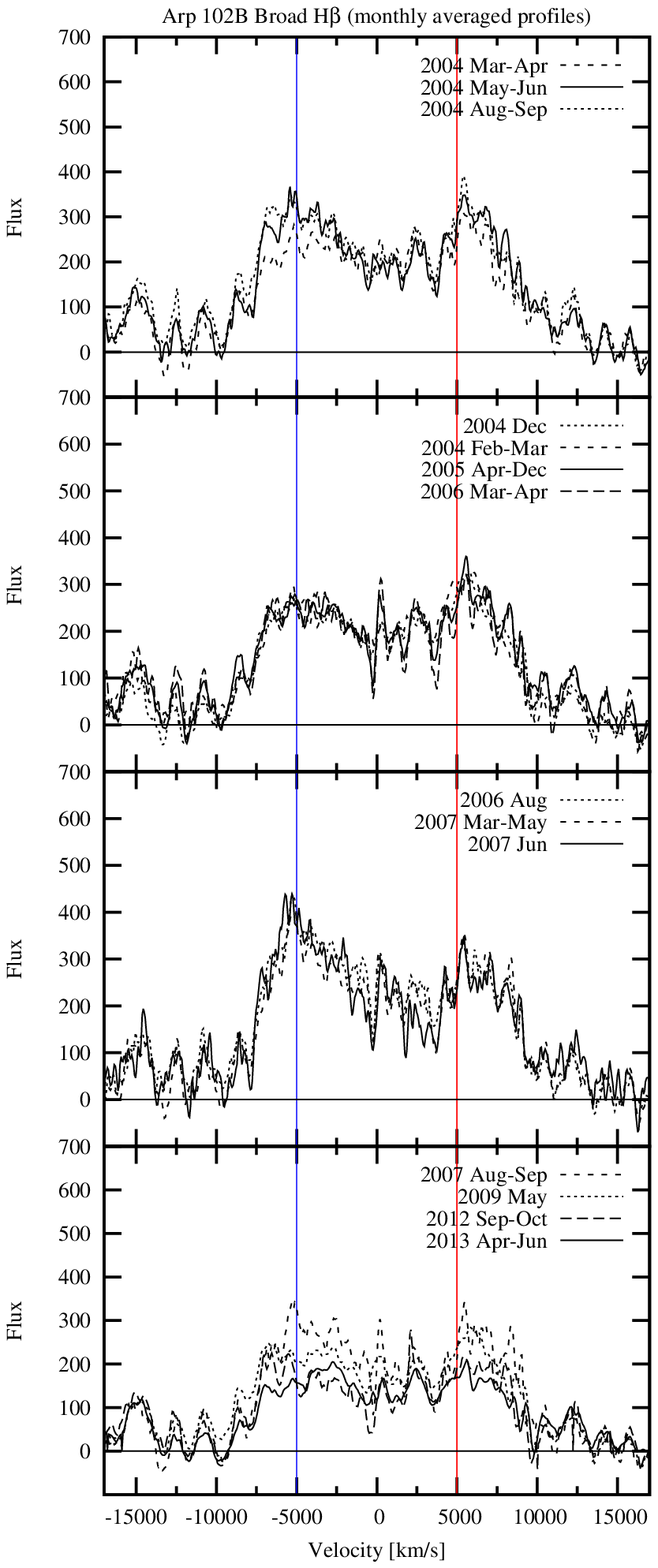}
\caption{The same as in Fig. \ref{month1}, but for period 2004-2013}\label{month2}
\end{figure*}

\begin{figure}
\centering
\includegraphics[width=4cm]{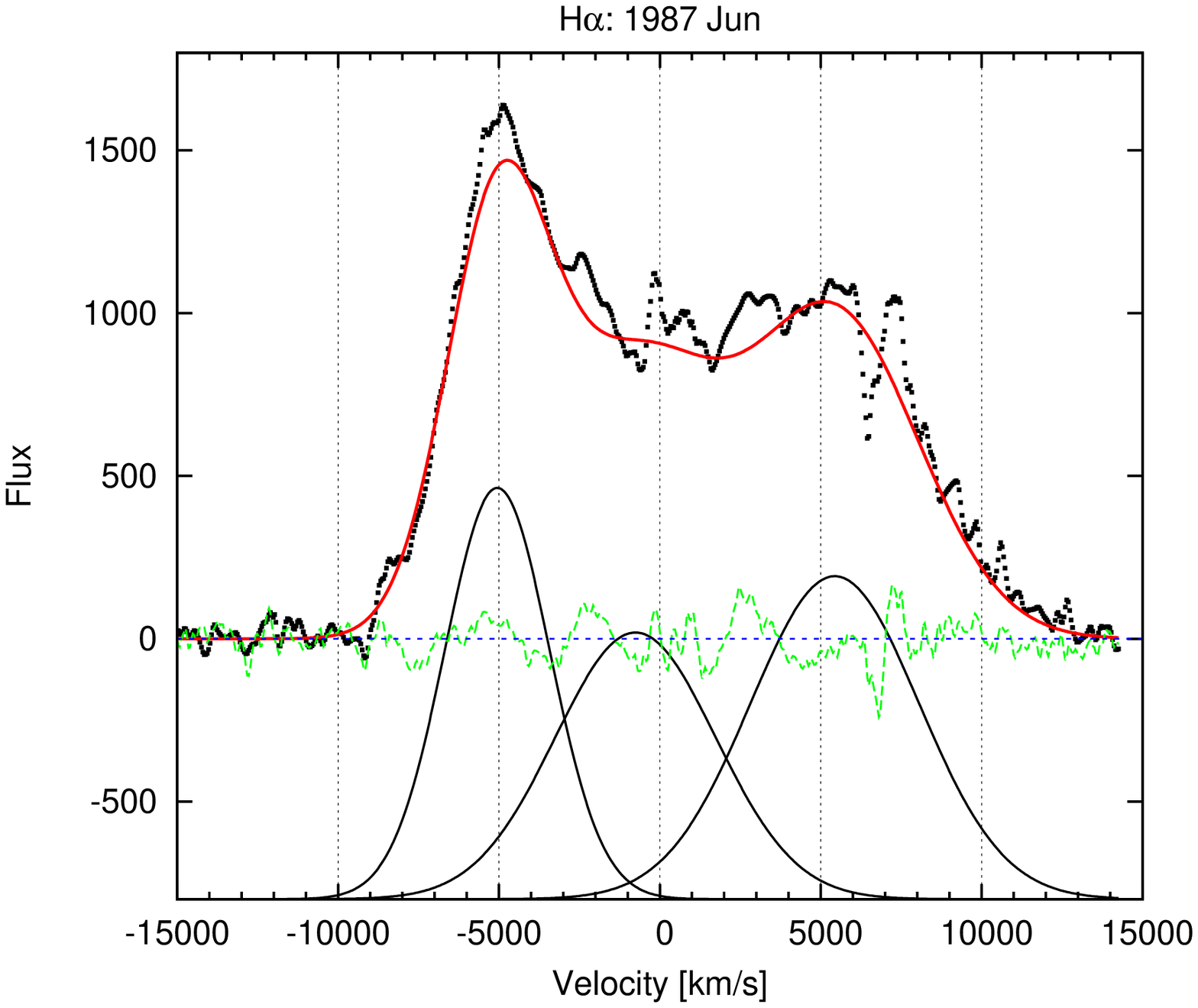}
\includegraphics[width=4cm]{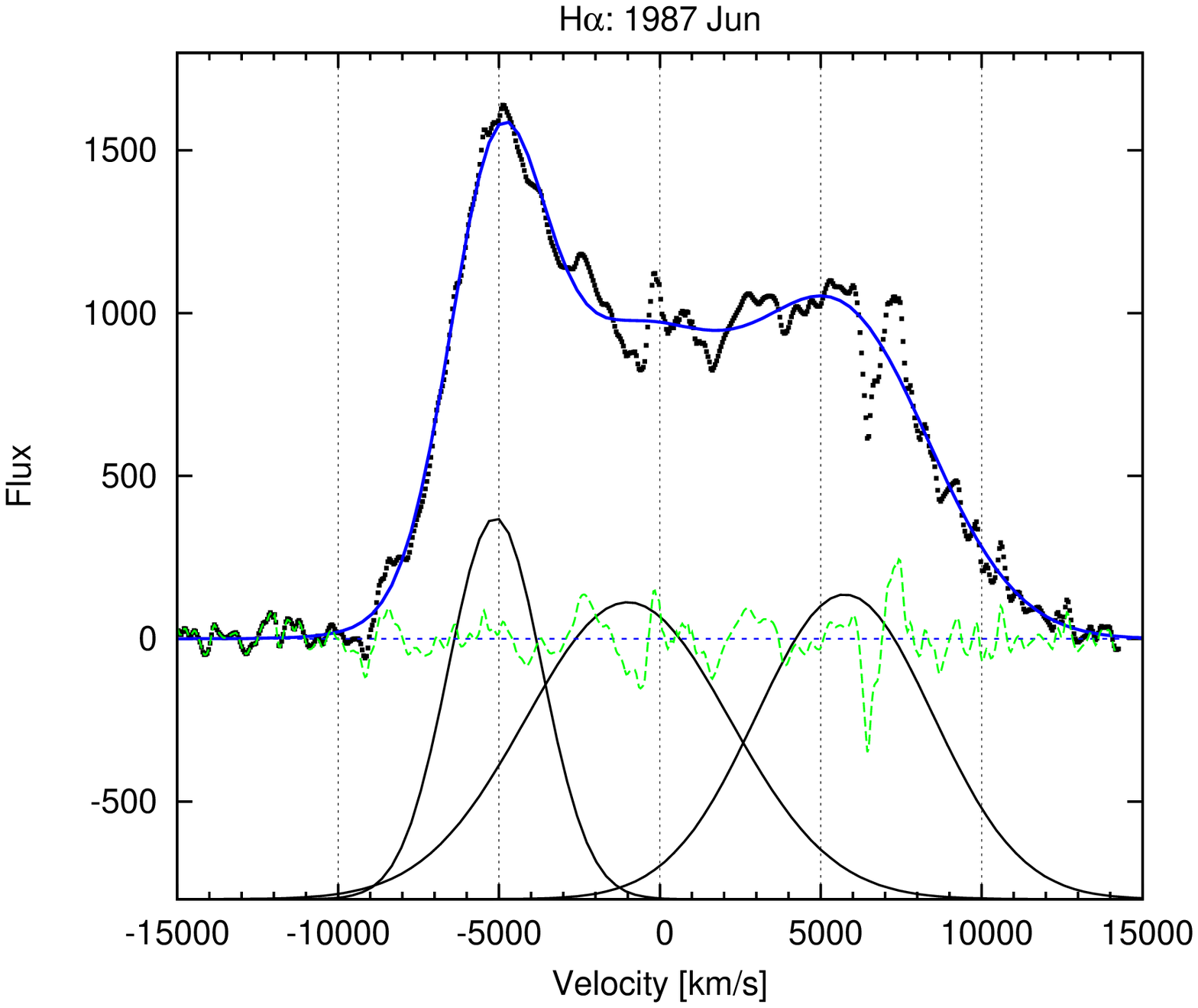}
\includegraphics[width=4cm]{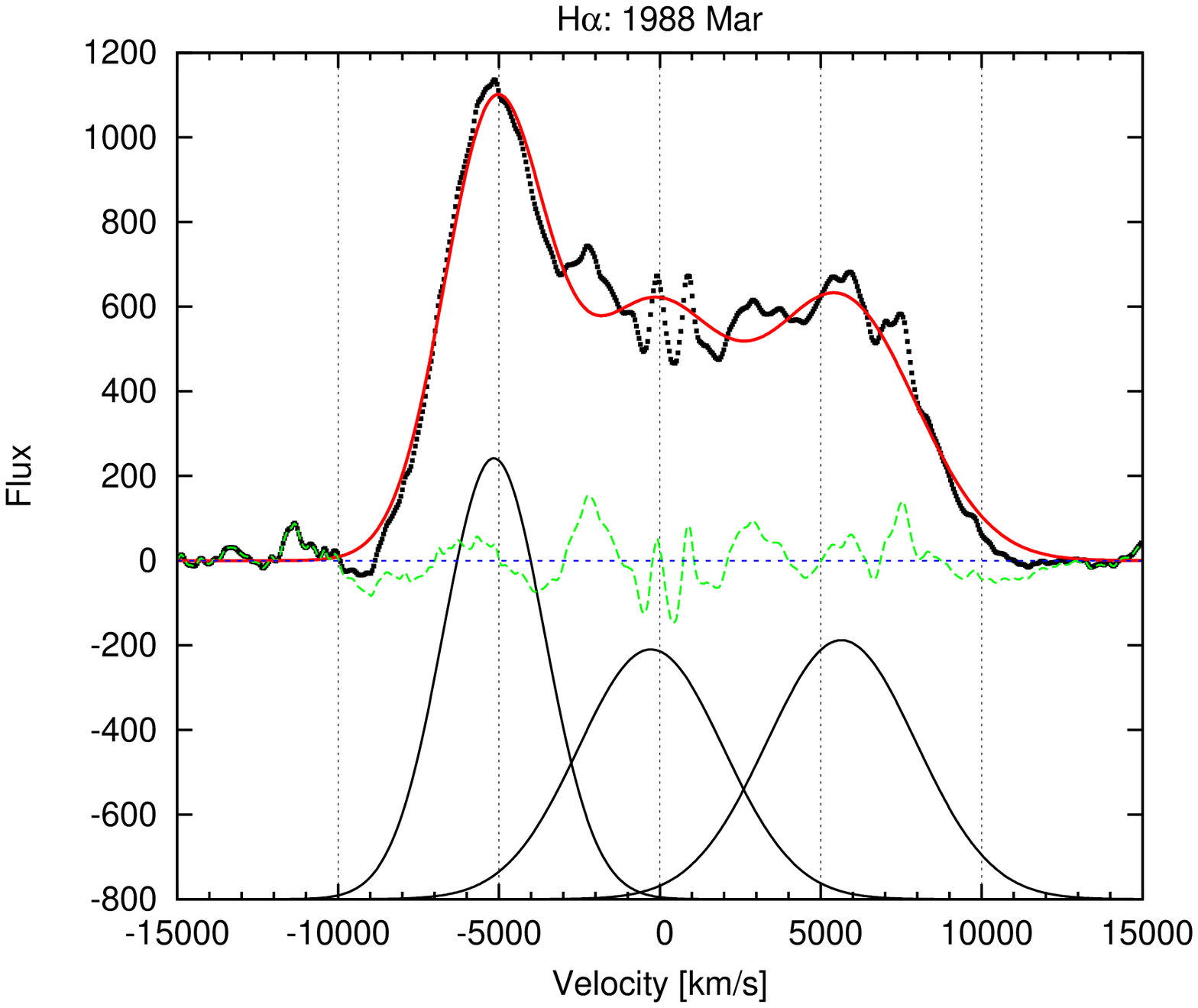}
\includegraphics[width=4cm]{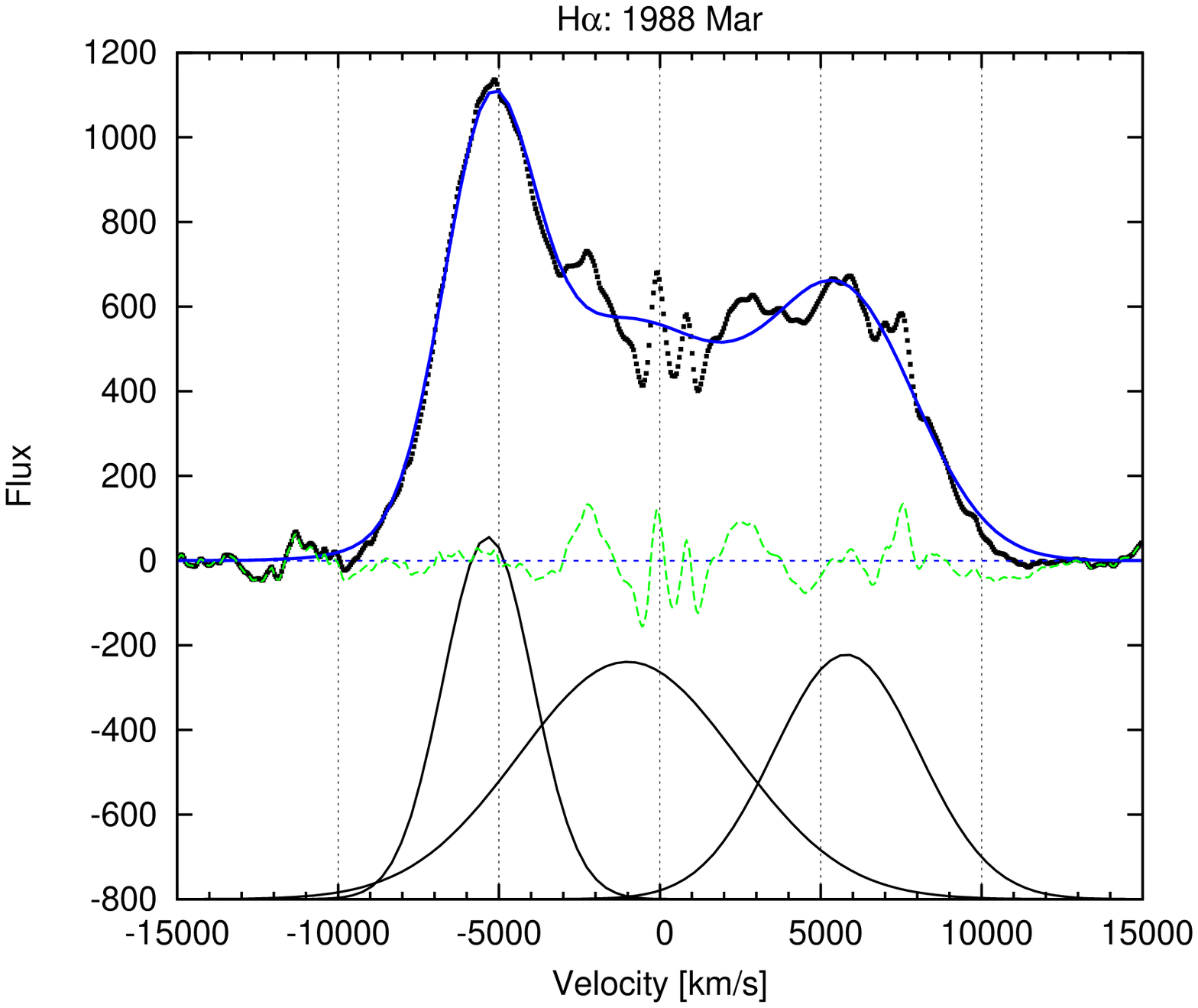}
\includegraphics[width=4cm]{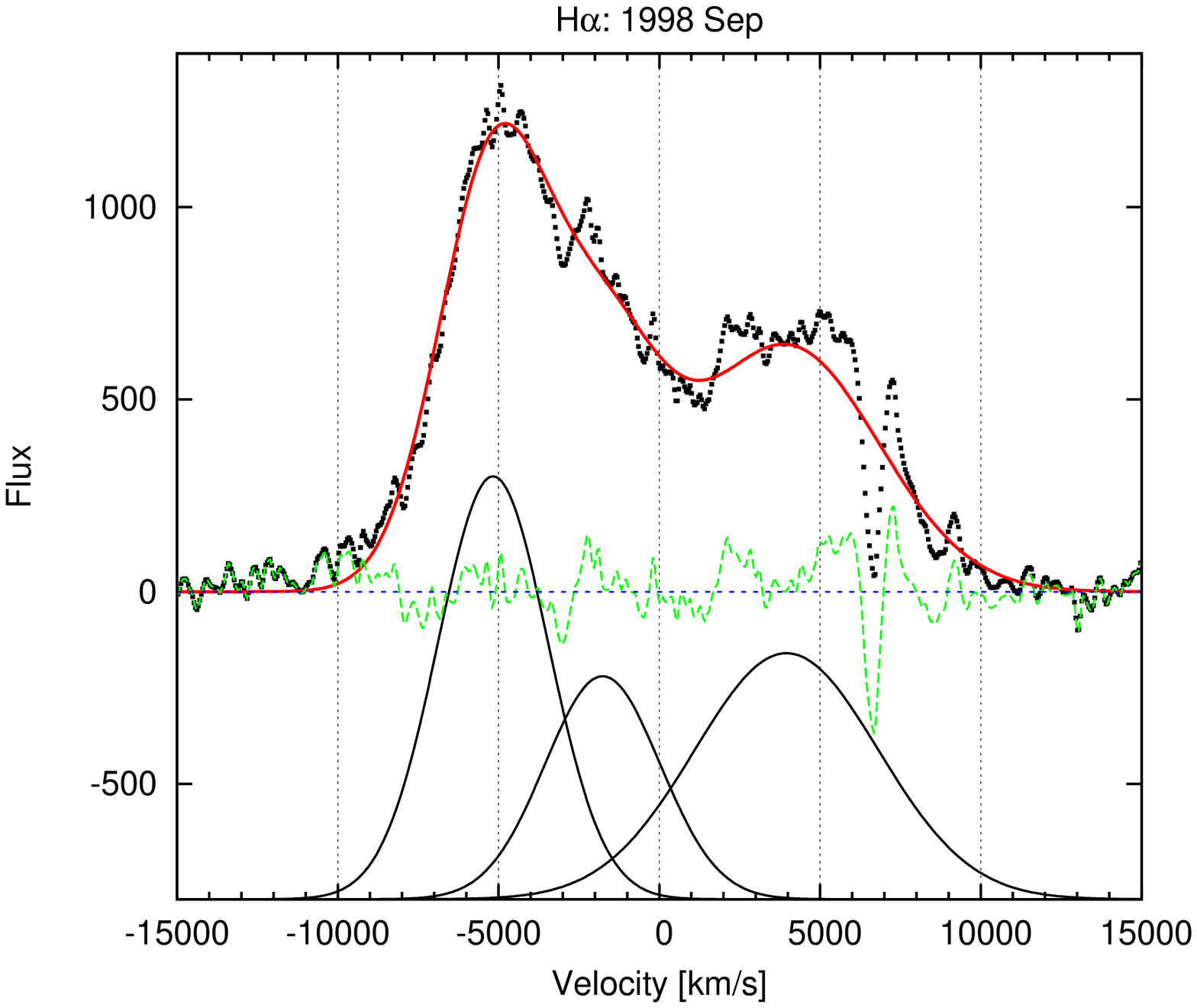}
\includegraphics[width=4cm]{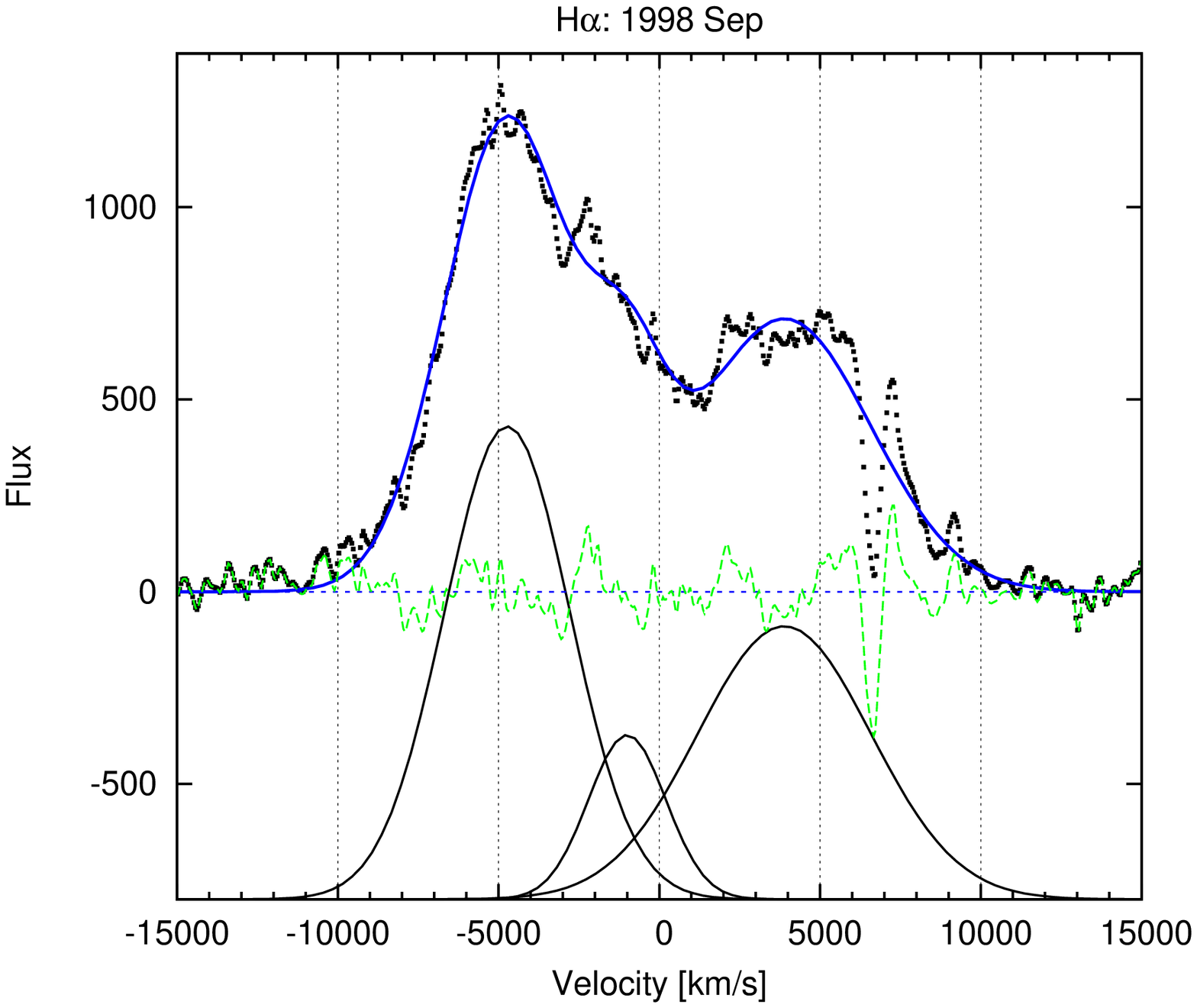}
\caption{Month-averaged H$\alpha$ broad component fitted {  with Gaussians using two different
assumptions: all parameters are free (left panels) and the shift of the central component is 
limited (right panels).}
Below the observed spectrum (dots), model (solid line), and
residual (dashed line), the three broad Gaussian components are given. Each plot is labeled 
with the observed month and year.}\label{fig_ha_broad_gauss}
\end{figure}

To explore the broad line profile variability in details, we performed the following methods of analysis:

a) To subtract the narrow lines and obtain only the broad profiles,
we fitted all H$\alpha$ and H$\beta$ lines with Gaussian functions.
{   There is a problem to remove the narrow lines in Arp 102B,
since they are right on top of the red peak in both broad lines (H$\alpha$ and H$\beta$)
and the measured properties of the red peak can be strongly affected by the 
the narrow line removal procedure. We discuss this in more details in the Appendix, and here we 
shortly describe the subtraction procedure. To estimate the narrow line contribution we fitted
simultaneously the whole H$\alpha$ and H$\beta$ profiles (broad + narrow lines).}
We used three broad Gaussian functions for the broad component of the H$\alpha$ and H$\beta$ lines
(Fig. \ref{gauss}), while for all narrow lines we assumed the same widths and shifts 
\citep[see][]{pop04}. We assumed that the ratio of [OIII]4959 and [OIII]5007 follows the flux ratio
1:3 \citep[][]{dim07}, and the same for [NII] doublet: F[NII]6584/F[NII]6548$\sim$3. In the 
H$\alpha$ wavelength band, the [SII]6717,6731 doublet, and [OI]6300 and [OI]6363 lines
were fitted using single Gaussian. Note here that only one Gaussian cannot properly fit the narrow
line wings, therefore we included an additional, broader, Gaussian (with 
significantly smaller intensity) for fitting the narrow line wings.
As it can be seen in Fig. \ref{gauss}, the three broad components (red and blue
shifted, and a central one) can satisfactory explain the broad line profiles.
{  To check the narrow line subtraction we perform several tests, see Appendix for details.}

b) After subtraction of the narrow lines and continuum, 
we made month-averaged line profiles using similar profiles (same peak positions and total line flux
differs up to $\sim$10\%). These month-averaged line profiles are given for both H$\alpha$ and
H$\beta$ line in Figs. \ref{month1} and \ref{month2}, and we further
continued to measure and analyze properties of these profiles. 

c) To determine 
the peak positions in the H$\alpha$ and H$\beta$ broad line profiles, we performed again the Gaussian
analysis, but now of the month-averaged broad profile. The broad profile was fitted with three Gaussian functions, 
corresponding to the blue, central, and red component (Fig. \ref{fig_ha_broad_gauss}).
{  By the inspection of the parameters obtained from the Gaussian fittings,
we note that the central component is very often changing the position, and 
affects other parameters, especially the intensity of the red peak. Therefore, additionally we 
fitted the month-averaged broad line profiles using the constraint that the central 
component should not be with large shift velocity, i.e. putting the limits to the shift of 
the central component between -1000 kms$^{-1}$  and 600 kms$^{-1}$ (see Figs. \ref{fig_ha_broad_gauss} and
 \ref{fig_hb_broad_gauss}, and the discussion in Appendix). 
The obtained error-bars of the parameters from both fittings are often smaller 
than  the difference between the parameters obtained from these two fitting procedures, 
therefore we accepted for the line parameters the averaged value of the two best-fittings, 
while for their error-bars we took the corresponding standard deviation, i.e. the discrepancy between the averaged
parameters and ones obtained from the fits}.
The {  averaged} parameters {  with their uncertainties} of the Gaussian fittings of the broad 
H$\alpha$ and H$\beta$ components are given in Tables \ref{tab02} and  \ref{tab03}.

%%%%%%%%%%%%%%%%%%%%%%%%%%%%%%%%
%%%%%%%%%%%%%%%%%%%%%%%%%%%%%%%%

%%%%%%%%%%%%%%%%%%%%%%%%%%%%%%%%
%%%%%%%%%%%%%%%%%%%%%%%%%%%%%%%%

d) The broad double-peaked line profiles are usually fitted with some sort of disc model 
\citep[see e.g.][etc.]{eh94, g07, n97, pop11}. Here we use a relativistic 
disc model given by \citet{c89} and \citet{ch89} to model the broad H$\alpha$ profile.
In Fig. \ref{comp} we present disc fitting with our parameters and ones given in \citet{g07}, we will
discuss this in more details in \S 3.2.

e) Finally, we measured the Full Width at Half Maximum (FWHM) and Full Width at Quarter (1/4) Maximum (FWQM). 
These measurements are also given in Tables \ref{tab02} and  \ref{tab03}.

\begin{table*}
\caption[]{Parameters of the Gaussian analysis of  the broad H$\alpha$ line profiles: shifts, intensities (given
in $10^{-16} {\rm erg\ cm^{-2}s^{-1}}$\AA$^{-1}$), width of 
the blue, central, 
and red Gaussian, respectively. Measurements of the FWHM and FWQM of are also given}\label{tab02}
\centering
\resizebox{16.5cm}{!}
{
\begin{tabular}{llccccccccccc}
\hline \hline
Year & Month  & \multicolumn{3}{c}{shift} & \multicolumn{3}{c}{intensity} & \multicolumn{3}{c}{width} & FWHM & FWQM\\ 
     &     &  blue  & central & red   & blue  & central & red  &  blue  & central & red   & &  \\
     &     &  km s$^{-1}$  & km s$^{-1}$    &  km s$^{-1}$ &      &  
           &      & km s$^{-1}$ & km s$^{-1}$ & km s$^{-1}$ & km s$^{-1}$ & km s$^{-1}$ \\
\hline
\hline
1987	&	Jun	&	-5083	$\pm$	45	&	-877	$\pm$	170	&	5595	$\pm$	222	&	1217	$\pm$	65	&	866	$\pm$	66	&	964	$\pm$	40	&	2180	$\pm$	169	&	4005	$\pm$	678	&	3784	$\pm$	115	&	14300	$\pm$	858	&	16700	$\pm$	1002	\\
1988	&	Mar	&	-5250	$\pm$	131	&	-638	$\pm$	512	&	5720	$\pm$	106	&	949	$\pm$	131	&	576	$\pm$	21	&	595	$\pm$	24	&	2095	$\pm$	175	&	3978	$\pm$	1132	&	3224	$\pm$	74	&	14500	$\pm$	870	&	16100	$\pm$	966	\\
1989	&	Oct	&	-5400	$\pm$	4	&	-793	$\pm$	293	&	4983	$\pm$	447	&	893	$\pm$	9	&	424	$\pm$	58	&	611	$\pm$	18	&	2552	$\pm$	21	&	2714	$\pm$	354	&	4800	$\pm$	336	&	15300	$\pm$	918	&	18100	$\pm$	1086	\\
1992	&	Aug	&	-5562	$\pm$	140	&	-475	$\pm$	289	&	6392	$\pm$	319	&	701	$\pm$	127	&	607	$\pm$	18	&	560	$\pm$	63	&	2200	$\pm$	306	&	4385	$\pm$	1216	&	3617	$\pm$	185	&	15800	$\pm$	948	&	18100	$\pm$	1086	\\
1993	&	Sep	&	-5208	$\pm$	54	&	-785	$\pm$	305	&	5263	$\pm$	859	&	1070	$\pm$	28	&	710	$\pm$	72	&	679	$\pm$	30	&	2440	$\pm$	59	&	2770	$\pm$	682	&	4105	$\pm$	853	&	15000	$\pm$	900	&	17200	$\pm$	1032	\\
1994	&	Sep	&	-5241	$\pm$	91	&	-721	$\pm$	395	&	6019	$\pm$	182	&	658	$\pm$	147	&	487	$\pm$	12	&	576	$\pm$	76	&	1937	$\pm$	288	&	4401	$\pm$	1603	&	3641	$\pm$	79	&	14000	$\pm$	840	&	17000	$\pm$	1020	\\
1998	&	Jul	&	-4993	$\pm$	58	&	-1239	$\pm$	338	&	4847	$\pm$	642	&	1317	$\pm$	230	&	817	$\pm$	52	&	876	$\pm$	120	&	2246	$\pm$	44	&	3847	$\pm$	1502	&	3631	$\pm$	526	&	14100	$\pm$	846	&	15500	$\pm$	930	\\
1998	&	Sep	&	-4949	$\pm$	311	&	-1381	$\pm$	539	&	3925	$\pm$	49	&	1165	$\pm$	92	&	504	$\pm$	108	&	675	$\pm$	49	&	2594	$\pm$	243	&	2116	$\pm$	581	&	3925	$\pm$	161	&	14300	$\pm$	858	&	15300	$\pm$	918	\\
1999	&	Nov	&	-4981	$\pm$	194	&	-453	$\pm$	770	&	5281	$\pm$	337	&	1363	$\pm$	336	&	758	$\pm$	152	&	995	$\pm$	78	&	2344	$\pm$	366	&	4163	$\pm$	1783	&	2957	$\pm$	163	&	14600	$\pm$	876	&	15800	$\pm$	948	\\
2000	&	Jan	&	-5040	$\pm$	30	&	-682	$\pm$	450	&	5293	$\pm$	460	&	896	$\pm$	233	&	559	$\pm$	40	&	632	$\pm$	167	&	2163	$\pm$	413	&	4805	$\pm$	2733	&	2430	$\pm$	146	&	13800	$\pm$	828	&	15300	$\pm$	918	\\
2000	&	Feb	&	-5075	$\pm$	109	&	-636	$\pm$	515	&	5214	$\pm$	682	&	1244	$\pm$	79	&	592	$\pm$	40	&	813	$\pm$	51	&	2374	$\pm$	88	&	3084	$\pm$	1273	&	3293	$\pm$	646	&	14100	$\pm$	846	&	15600	$\pm$	936	\\
2000	&	Apr-Jun	&	-5001	$\pm$	188	&	-842	$\pm$	56	&	5059	$\pm$	491	&	1061	$\pm$	99	&	485	$\pm$	95	&	787	$\pm$	35	&	2598	$\pm$	109	&	2820	$\pm$	1193	&	3802	$\pm$	485	&	14300	$\pm$	858	&	16400	$\pm$	984	\\
2000	&	Jul	&	-5095	$\pm$	113	&	-1037	$\pm$	52	&	5193	$\pm$	267	&	1198	$\pm$	46	&	507	$\pm$	60	&	944	$\pm$	39	&	2413	$\pm$	80	&	2437	$\pm$	543	&	4013	$\pm$	350	&	14600	$\pm$	876	&	17400	$\pm$	1044	\\
2000	&	Oct	&	-5122	$\pm$	312	&	-694	$\pm$	433	&	5150	$\pm$	544	&	930	$\pm$	337	&	602	$\pm$	248	&	675	$\pm$	205	&	2406	$\pm$	630	&	4316	$\pm$	2856	&	3070	$\pm$	611	&	14000	$\pm$	840	&	16700	$\pm$	1002	\\
2001	&	Mar-Jun	&	-5237	$\pm$	74	&	-884	$\pm$	165	&	4929	$\pm$	333	&	1147	$\pm$	33	&	644	$\pm$	37	&	776	$\pm$	10	&	2558	$\pm$	43	&	2817	$\pm$	556	&	3299	$\pm$	354	&	14200	$\pm$	852	&	16500	$\pm$	990	\\
2001	&	Aug	&	-5135	$\pm$	119	&	-957	$\pm$	59	&	4459	$\pm$	490	&	1214	$\pm$	47	&	546	$\pm$	117	&	788	$\pm$	21	&	2613	$\pm$	75	&	2514	$\pm$	633	&	4254	$\pm$	445	&	14300	$\pm$	858	&	17000	$\pm$	1020	\\
2002	&	Mar	&	-5058	$\pm$	336	&	-959	$\pm$	121	&	4779	$\pm$	873	&	1041	$\pm$	121	&	511	$\pm$	192	&	661	$\pm$	15	&	2636	$\pm$	341	&	2742	$\pm$	1405	&	3684	$\pm$	784	&	14200	$\pm$	852	&	16100	$\pm$	966	\\
2002	&	Apr-Aug	&	-5027	$\pm$	288	&	-666	$\pm$	172	&	5008	$\pm$	465	&	951	$\pm$	76	&	466	$\pm$	116	&	653	$\pm$	10	&	2906	$\pm$	230	&	2673	$\pm$	955	&	3613	$\pm$	463	&	14400	$\pm$	864	&	16900	$\pm$	1014	\\
2003	&	Mar	&	-5097	$\pm$	238	&	-1160	$\pm$	308	&	5034	$\pm$	370	&	978	$\pm$	55	&	537	$\pm$	87	&	592	$\pm$	5	&	2311	$\pm$	258	&	2496	$\pm$	689	&	3831	$\pm$	420	&	14000	$\pm$	840	&	16100	$\pm$	966	\\
2003	&	May	&	-5129	$\pm$	124	&	-800	$\pm$	283	&	4717	$\pm$	833	&	1210	$\pm$	13	&	564	$\pm$	193	&	786	$\pm$	8	&	2444	$\pm$	243	&	2553	$\pm$	968	&	4425	$\pm$	1043	&	14100	$\pm$	846	&	16800	$\pm$	1008	\\
2003	&	Sep	&	-5130	$\pm$	126	&	-684	$\pm$	446	&	4708	$\pm$	979	&	930	$\pm$	40	&	421	$\pm$	148	&	640	$\pm$	16	&	2534	$\pm$	365	&	932	$\pm$	4016	&	3374	$\pm$	820	&	14300	$\pm$	858	&	16600	$\pm$	996	\\
2004	&	Mar	&	-5185	$\pm$	66	&	-865	$\pm$	192	&	5568	$\pm$	292	&	698	$\pm$	184	&	533	$\pm$	67	&	593	$\pm$	150	&	2027	$\pm$	406	&	5318	$\pm$	2233	&	3494	$\pm$	363	&	14600	$\pm$	876	&	17600	$\pm$	1056	\\
2004	&	Apr	&	-5085	$\pm$	132	&	-816	$\pm$	251	&	4818	$\pm$	824	&	857	$\pm$	26	&	349	$\pm$	123	&	542	$\pm$	40	&	2551	$\pm$	343	&	2851	$\pm$	1284	&	4043	$\pm$	652	&	14400	$\pm$	864	&	16500	$\pm$	990	\\
2004	&	May-Jun	&	-5015	$\pm$	232	&	-774	$\pm$	464	&	5143	$\pm$	315	&	860	$\pm$	100	&	374	$\pm$	80	&	538	$\pm$	33	&	2637	$\pm$	206	&	3181	$\pm$	1081	&	3970	$\pm$	217	&	14100	$\pm$	846	&	16400	$\pm$	984	\\
2004	&	Aug	&	-5070	$\pm$	194	&	-649	$\pm$	76	&	5269	$\pm$	605	&	811	$\pm$	41	&	411	$\pm$	53	&	528	$\pm$	17	&	2798	$\pm$	208	&	2553	$\pm$	1049	&	3132	$\pm$	566	&	14600	$\pm$	876	&	16500	$\pm$	990	\\
2004	&	Dec	&	-4995	$\pm$	93	&	-788	$\pm$	300	&	5321	$\pm$	531	&	860	$\pm$	61	&	519	$\pm$	21	&	775	$\pm$	6	&	2837	$\pm$	12	&	2717	$\pm$	1004	&	3686	$\pm$	634	&	14800	$\pm$	888	&	18000	$\pm$	1080	\\
2005	&	Feb-Mar	&	-5150	$\pm$	124	&	-631	$\pm$	522	&	4893	$\pm$	482	&	802	$\pm$	39	&	375	$\pm$	42	&	654	$\pm$	60	&	3007	$\pm$	76	&	2907	$\pm$	1185	&	3490	$\pm$	501	&	14600	$\pm$	876	&	16900	$\pm$	1014	\\
2005	&	Apr	&	-5151	$\pm$	219	&	-774	$\pm$	246	&	4757	$\pm$	692	&	797	$\pm$	4	&	395	$\pm$	121	&	611	$\pm$	15	&	2552	$\pm$	641	&	2664	$\pm$	1108	&	3482	$\pm$	685	&	14500	$\pm$	870	&	17000	$\pm$	1020	\\
2005	&	Jun-Aug	&	-5237	$\pm$	240	&	-927	$\pm$	104	&	4887	$\pm$	521	&	838	$\pm$	34	&	389	$\pm$	113	&	645	$\pm$	19	&	2707	$\pm$	423	&	2738	$\pm$	1072	&	3485	$\pm$	515	&	14700	$\pm$	882	&	16900	$\pm$	1014	\\
2005	&	Dec	&	-5128	$\pm$	270	&	-1044	$\pm$	62	&	4884	$\pm$	294	&	855	$\pm$	53	&	420	$\pm$	85	&	690	$\pm$	5	&	2804	$\pm$	162	&	2369	$\pm$	762	&	3990	$\pm$	317	&	14800	$\pm$	888	&	17800	$\pm$	1068	\\
2006	&	Mar-Apr	&	-5065	$\pm$	257	&	-800	$\pm$	283	&	4710	$\pm$	759	&	923	$\pm$	41	&	351	$\pm$	210	&	524	$\pm$	13	&	2826	$\pm$	609	&	2078	$\pm$	878	&	4129	$\pm$	751	&	14500	$\pm$	870	&	17000	$\pm$	1020	\\
2006	&	Aug	&	-5039	$\pm$	273	&	-1106	$\pm$	149	&	4655	$\pm$	752	&	1056	$\pm$	90	&	496	$\pm$	156	&	594	$\pm$	17	&	2442	$\pm$	283	&	2585	$\pm$	1007	&	3684	$\pm$	584	&	13700	$\pm$	822	&	15200	$\pm$	912	\\
2007	&	May-Jul	&	-5164	$\pm$	391	&	-1414	$\pm$	585	&	4688	$\pm$	587	&	1133	$\pm$	172	&	652	$\pm$	172	&	605	$\pm$	2	&	2389	$\pm$	402	&	2583	$\pm$	1024	&	3590	$\pm$	397	&	13400	$\pm$	804	&	15300	$\pm$	918	\\
2007	&	Aug	&	-5163	$\pm$	327	&	-1202	$\pm$	286	&	4643	$\pm$	646	&	1019	$\pm$	67	&	624	$\pm$	146	&	596	$\pm$	8	&	2405	$\pm$	437	&	2497	$\pm$	704	&	3256	$\pm$	370	&	13900	$\pm$	834	&	15700	$\pm$	942	\\
2009	&	May	&	-5265	$\pm$	223	&	-981	$\pm$	21	&	5420	$\pm$	401	&	629	$\pm$	55	&	373	$\pm$	33	&	490	$\pm$	15	&	2718	$\pm$	117	&	2632	$\pm$	933	&	3411	$\pm$	589	&	14900	$\pm$	894	&	17700	$\pm$	1062	\\
2009	&	Jul-Aug	&	-5249	$\pm$	528	&	-1326	$\pm$	461	&	4753	$\pm$	511	&	769	$\pm$	65	&	376	$\pm$	170	&	568	$\pm$	11	&	2927	$\pm$	616	&	2167	$\pm$	1179	&	3654	$\pm$	314	&	14900	$\pm$	894	&	16400	$\pm$	984	\\
2010	&	Apr-Jun	&	-5363	$\pm$	211	&	-1190	$\pm$	268	&	3724	$\pm$	1983	&	704	$\pm$	63	&	335	$\pm$	255	&	478	$\pm$	62	&	2602	$\pm$	165	&	2106	$\pm$	1256	&	5717	$\pm$	3209	&	14700	$\pm$	882	&	16700	$\pm$	1002	\\
2012	&	Sep-Oct	&	-5581	$\pm$	135	&	-1173	$\pm$	244	&	5274	$\pm$	477	&	612	$\pm$	4	&	274	$\pm$	40	&	283	$\pm$	24	&	1981	$\pm$	249	&	2384	$\pm$	415	&	3893	$\pm$	630	&	14700	$\pm$	882	&	16500	$\pm$	990	\\
2013	&	Apr-Jun	&	-5056	$\pm$	86	&	-792	$\pm$	295	&	4676	$\pm$	993	&	537	$\pm$	6	&	241	$\pm$	91	&	333	$\pm$	4	&	2823	$\pm$	66	&	2365	$\pm$	675	&	4499	$\pm$	877	&	14800	$\pm$	888	&	16800	$\pm$	1008	\\
\hline																																				mean	&		&	-5148	$\pm$	143	&	-887	$\pm$	240	&	5016	$\pm$	486	&	946	$\pm$	208	&	504	$\pm$	141	&	649	$\pm$	150	&	2502	$\pm$	269	&	2955	$\pm$	880	&	3728	$\pm$	553	&	14431	$\pm$	448	&	16618	$\pm$	790	\\
\hline
\end{tabular}}
\end{table*}

\begin{table*}
\caption[]{The same as in Table \ref{tab02}, but for the broad H$\beta$ line. }\label{tab03}
\centering
\resizebox{16.5cm}{!}
{
\begin{tabular}{llccccccccccc}
\hline \hline
Year & Month  & \multicolumn{3}{c}{shift} & \multicolumn{3}{c}{intensity} & \multicolumn{3}{c}{width} & FWHM & FWQM\\ 
     &     &  blue  & central & red   & blue  & central & red  &  blue  & central & red   & &  \\
     &     &  km s$^{-1}$  & km s$^{-1}$    &  km s$^{-1}$ &      &  
           &     & km s$^{-1}$ & km s$^{-1}$ & km s$^{-1}$ & km s$^{-1}$ & km s$^{-1}$ \\
\hline
\hline  
1987	&	Jun     	&	-4930	$\pm$	326	&	-440	$\pm$	791	&	5747	$\pm$	434	&	478	$\pm$	112	&	257	$\pm$	76	&	377	$\pm$	34	&	2640	$\pm$	378	&	3907	$\pm$	1692	&	3889	$\pm$	360	&	15300	$\pm$	918	&	17500	$\pm$	1050	\\
1988	&	Mar     	&	-5206	$\pm$	110	&	-575	$\pm$	601	&	6442	$\pm$	583	&	320	$\pm$	109	&	221	$\pm$	9	&	235	$\pm$	63	&	2445	$\pm$	282	&	4831	$\pm$	2834	&	3964	$\pm$	154	&	15500	$\pm$	930	&	17100	$\pm$	1026	\\
1989	&	Oct     	&	-4974	$\pm$	47	&	-515	$\pm$	687	&	6035	$\pm$	54	&	231	$\pm$	4	&	111	$\pm$	10	&	225	$\pm$	7	&	3446	$\pm$	202	&	2486	$\pm$	745	&	4913	$\pm$	1017	&	17800	$\pm$	1068	&	18900	$\pm$	1134	\\
1992	&	Aug     	&	-5714	$\pm$	496	&	-847	$\pm$	216	&	5263	$\pm$	189	&	215	$\pm$	35	&	157	$\pm$	27	&	236	$\pm$	59	&	2961	$\pm$	459	&	3752	$\pm$	1299	&	4050	$\pm$	21	&	16100	$\pm$	966	&	18800	$\pm$	1128	\\
1993	&	Sep     	&	-4968	$\pm$	161	&	378	$\pm$	315	&	6499	$\pm$	1118	&	279	$\pm$	119	&	184	$\pm$	57	&	223	$\pm$	89	&	3248	$\pm$	78	&	5688	$\pm$	4228	&	3354	$\pm$	1184	&	17100	$\pm$	1026	&	18900	$\pm$	1134	\\
1998	&	May-Jul 	&	-5217	$\pm$	156	&	-332	$\pm$	945	&	6533	$\pm$	48	&	444	$\pm$	107	&	380	$\pm$	16	&	416	$\pm$	27	&	2184	$\pm$	447	&	4848	$\pm$	1200	&	2911	$\pm$	34	&	15400	$\pm$	924	&	17100	$\pm$	1026	\\
1998	&	Sep     	&	-5053	$\pm$	454	&	-470	$\pm$	738	&	5969	$\pm$	15	&	406	$\pm$	42	&	273	$\pm$	25	&	374	$\pm$	8	&	2932	$\pm$	394	&	2669	$\pm$	803	&	4043	$\pm$	62	&	16200	$\pm$	972	&	18400	$\pm$	1104	\\
1999	&	Aug-Nov 	&	-5052	$\pm$	456	&	-419	$\pm$	809	&	5998	$\pm$	27	&	391	$\pm$	20	&	269	$\pm$	32	&	372	$\pm$	6	&	3009	$\pm$	502	&	2777	$\pm$	650	&	4010	$\pm$	109	&	15600	$\pm$	936	&	17100	$\pm$	1026	\\
2000	&	Jan-Feb 	&	-4955	$\pm$	226	&	-355	$\pm$	913	&	6172	$\pm$	233	&	285	$\pm$	93	&	194	$\pm$	33	&	329	$\pm$	53	&	2554	$\pm$	695	&	5034	$\pm$	3103	&	3250	$\pm$	326	&	15400	$\pm$	924	&	17600	$\pm$	1056	\\
2000	&	Apr     	&	-4928	$\pm$	96	&	-311	$\pm$	972	&	5608	$\pm$	559	&	264	$\pm$	5	&	139	$\pm$	38	&	294	$\pm$	3	&	2827	$\pm$	59	&	3012	$\pm$	518	&	4150	$\pm$	782	&	15800	$\pm$	948	&	17900	$\pm$	1074	\\
2000	&	May-Oct 	&	-5249	$\pm$	8	&	265	$\pm$	474	&	6161	$\pm$	180	&	243	$\pm$	71	&	206	$\pm$	27	&	241	$\pm$	84	&	2547	$\pm$	485	&	6032	$\pm$	3201	&	3206	$\pm$	770	&	16100	$\pm$	966	&	18800	$\pm$	1128	\\
2001	&	Mar     	&	-5152	$\pm$	210	&	-475	$\pm$	743	&	5274	$\pm$	317	&	332	$\pm$	1	&	183	$\pm$	5	&	320	$\pm$	7	&	2963	$\pm$	266	&	2753	$\pm$	303	&	3889	$\pm$	332	&	15700	$\pm$	942	&	17400	$\pm$	1044	\\
2001	&	May     	&	-5018	$\pm$	162	&	-343	$\pm$	928	&	6086	$\pm$	124	&	300	$\pm$	107	&	248	$\pm$	24	&	358	$\pm$	46	&	2583	$\pm$	544	&	4989	$\pm$	2205	&	2742	$\pm$	168	&	14900	$\pm$	894	&	17900	$\pm$	1074	\\
2001	&	Jun-Oct 	&	-5137	$\pm$	394	&	-448	$\pm$	781	&	6167	$\pm$	441	&	239	$\pm$	98	&	192	$\pm$	52	&	243	$\pm$	45	&	2811	$\pm$	743	&	4792	$\pm$	2630	&	3482	$\pm$	523	&	15700	$\pm$	942	&	18200	$\pm$	1092	\\
2002	&	Mar-May 	&	-4802	$\pm$	125	&	-372	$\pm$	796	&	6385	$\pm$	37	&	276	$\pm$	39	&	133	$\pm$	3	&	263	$\pm$	5	&	4032	$\pm$	161	&	3875	$\pm$	829	&	4064	$\pm$	95	&	16100	$\pm$	966	&	18400	$\pm$	1104	\\
2002	&	Jun-Aug 	&	-4465	$\pm$	470	&	-350	$\pm$	919	&	6196	$\pm$	142	&	314	$\pm$	8	&	83	$\pm$	66	&	260	$\pm$	14	&	4506	$\pm$	861	&	1921	$\pm$	1100	&	3700	$\pm$	187	&	16100	$\pm$	966	&	18900	$\pm$	1134	\\
2003	&	Mar     	&	-4987	$\pm$	486	&	-438	$\pm$	795	&	6071	$\pm$	45	&	239	$\pm$	35	&	176	$\pm$	40	&	281	$\pm$	3	&	2904	$\pm$	426	&	3586	$\pm$	340	&	3415	$\pm$	67	&	15500	$\pm$	930	&	17400	$\pm$	1044	\\
2003	&	May-Jun 	&	-5163	$\pm$	253	&	-469	$\pm$	752	&	5980	$\pm$	178	&	335	$\pm$	12	&	174	$\pm$	15	&	325	$\pm$	3	&	2882	$\pm$	235	&	3205	$\pm$	46	&	3926	$\pm$	206	&	15800	$\pm$	948	&	17300	$\pm$	1038	\\
2003	&	Sep     	&	-5101	$\pm$	1	&	-539	$\pm$	653	&	5656	$\pm$	489	&	284	$\pm$	15	&	118	$\pm$	21	&	287	$\pm$	13	&	3233	$\pm$	129	&	3235	$\pm$	1161	&	4846	$\pm$	486	&	16400	$\pm$	984	&	17900	$\pm$	1074	\\
2004	&	Jan     	&	-4184	$\pm$	462	&	304	$\pm$	412	&	5807	$\pm$	145	&	309	$\pm$	19	&	120	$\pm$	75	&	197	$\pm$	27	&	4465	$\pm$	1227	&	4100	$\pm$	1563	&	3582	$\pm$	22	&	15000	$\pm$	900	&	18200	$\pm$	1092	\\
2004	&	Mar-Apr 	&	-4713	$\pm$	418	&	359	$\pm$	342	&	5726	$\pm$	168	&	243	$\pm$	8	&	119	$\pm$	55	&	254	$\pm$	7	&	3925	$\pm$	468	&	2491	$\pm$	713	&	4019	$\pm$	600	&	15800	$\pm$	948	&	18600	$\pm$	1116	\\
2004	&	May-Jun 	&	-4858	$\pm$	110	&	-130	$\pm$	1230	&	6447	$\pm$	158	&	249	$\pm$	99	&	155	$\pm$	39	&	253	$\pm$	60	&	3027	$\pm$	802	&	6152	$\pm$	4465	&	3212	$\pm$	507	&	16300	$\pm$	978	&	18700	$\pm$	1122	\\
2004	&	Aug-Sep 	&	-5293	$\pm$	28	&	-394	$\pm$	851	&	5831	$\pm$	223	&	304	$\pm$	20	&	122	$\pm$	54	&	300	$\pm$	19	&	3762	$\pm$	410	&	3256	$\pm$	366	&	4010	$\pm$	834	&	16400	$\pm$	984	&	18200	$\pm$	1092	\\
2004	&	Dec     	&	-5308	$\pm$	506	&	-346	$\pm$	926	&	6031	$\pm$	177	&	214	$\pm$	66	&	169	$\pm$	24	&	263	$\pm$	24	&	3107	$\pm$	379	&	4154	$\pm$	1635	&	3449	$\pm$	2	&	16600	$\pm$	996	&	17800	$\pm$	1068	\\
2005	&	Feb-Mar 	&	-5032	$\pm$	301	&	446	$\pm$	9	&	5889	$\pm$	134	&	193	$\pm$	96	&	175	$\pm$	33	&	227	$\pm$	75	&	2926	$\pm$	509	&	5799	$\pm$	3966	&	2813	$\pm$	901	&	15300	$\pm$	918	&	18000	$\pm$	1080	\\
2005	&	Apr-Dec 	&	-5084	$\pm$	270	&	-516	$\pm$	685	&	6247	$\pm$	151	&	181	$\pm$	106	&	169	$\pm$	13	&	224	$\pm$	71	&	2936	$\pm$	431	&	6546	$\pm$	4992	&	3427	$\pm$	593	&	17200	$\pm$	1032	&	18100	$\pm$	1086	\\
2006	&	Mar-Apr 	&	-4921	$\pm$	15	&	-172	$\pm$	1169	&	6495	$\pm$	64	&	190	$\pm$	106	&	168	$\pm$	23	&	200	$\pm$	67	&	2956	$\pm$	908	&	6612	$\pm$	5120	&	3071	$\pm$	753	&	16400	$\pm$	984	&	17700	$\pm$	1062	\\
2006	&	Aug     	&	-4630	$\pm$	134	&	505	$\pm$	134	&	6096	$\pm$	71	&	356	$\pm$	12	&	146	$\pm$	7	&	277	$\pm$	0	&	3893	$\pm$	2	&	2665	$\pm$	453	&	3691	$\pm$	150	&	16400	$\pm$	984	&	16700	$\pm$	1002	\\
2007	&	May     	&	-5096	$\pm$	98	&	-501	$\pm$	705	&	6632	$\pm$	412	&	251	$\pm$	98	&	212	$\pm$	26	&	250	$\pm$	59	&	2504	$\pm$	906	&	5492	$\pm$	3532	&	3000	$\pm$	629	&	16100	$\pm$	966	&	17100	$\pm$	1026	\\
2007	&	Jun     	&	-5138	$\pm$	240	&	-505	$\pm$	701	&	6435	$\pm$	211	&	373	$\pm$	14	&	186	$\pm$	12	&	253	$\pm$	2	&	2757	$\pm$	251	&	2841	$\pm$	232	&	4325	$\pm$	240	&	16300	$\pm$	978	&	16800	$\pm$	1008	\\
2007	&	Aug-Sep 	&	-4957	$\pm$	235	&	-470	$\pm$	750	&	6637	$\pm$	463	&	216	$\pm$	73	&	182	$\pm$	31	&	241	$\pm$	27	&	2790	$\pm$	587	&	4959	$\pm$	2125	&	3166	$\pm$	382	&	16300	$\pm$	978	&	16700	$\pm$	1002	\\
2009	&	May     	&	-4797	$\pm$	414	&	417	$\pm$	260	&	5938	$\pm$	40	&	234	$\pm$	6	&	88	$\pm$	47	&	242	$\pm$	4	&	4595	$\pm$	987	&	2283	$\pm$	613	&	3735	$\pm$	80	&	17300	$\pm$	1038	&	18200	$\pm$	1092	\\
2012	&	Sep-Oct 	&	-5593	$\pm$	308	&	264	$\pm$	475	&	6039	$\pm$	437	&	149	$\pm$	57	&	104	$\pm$	37	&	146	$\pm$	52	&	2587	$\pm$	784	&	6780	$\pm$	5142	&	3194	$\pm$	1567	&	16000	$\pm$	960	&	17700	$\pm$	1062	\\
2013	&	Apr-Jun 	&	-4424	$\pm$	683	&	300	$\pm$	425	&	5363	$\pm$	810	&	156	$\pm$	6	&	69	$\pm$	60	&	157	$\pm$	25	&	3441	$\pm$	37	&	1350	$\pm$	1907	&	4531	$\pm$	989	&	16700	$\pm$	1002	&	18000	$\pm$	1080	\\
\hline																																															
mean	&		&	-5003	$\pm$	298	&	-220	$\pm$	374	&	6054	$\pm$	365	&	279	$\pm$	78	&	173	$\pm$	63	&	269	$\pm$	62	&	3129	$\pm$	616	&	4085	$\pm$	1484	&	3677	$\pm$	544	&	16076	$\pm$	654	&	17882	$\pm$	663	\\

\hline
\end{tabular}}
\end{table*}

%%%%%%%%%%%%%%%%%%%%%%%%%%%%%%%%

\begin{figure}
\centering
\includegraphics[width=4cm]{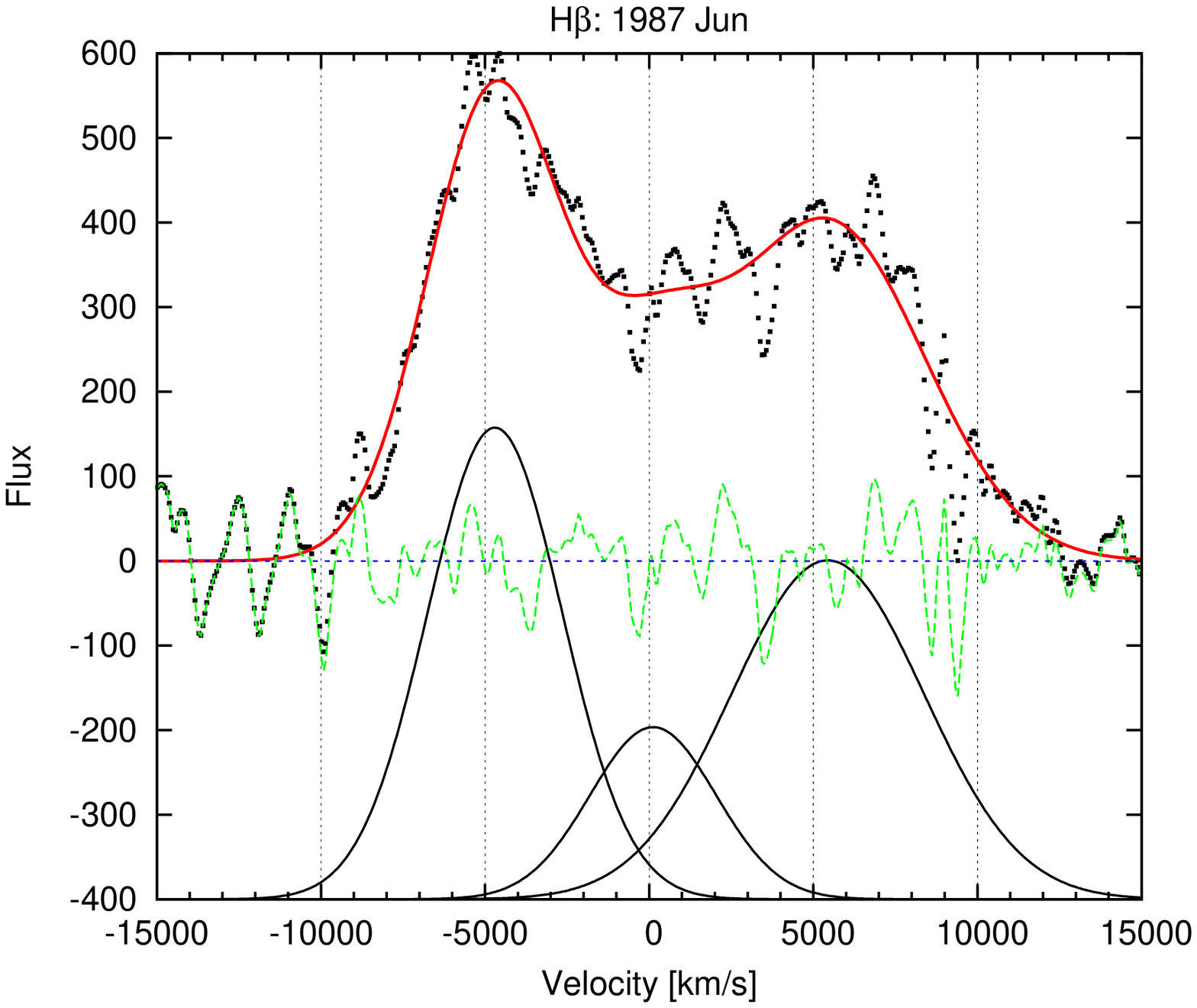}
\includegraphics[width=4cm]{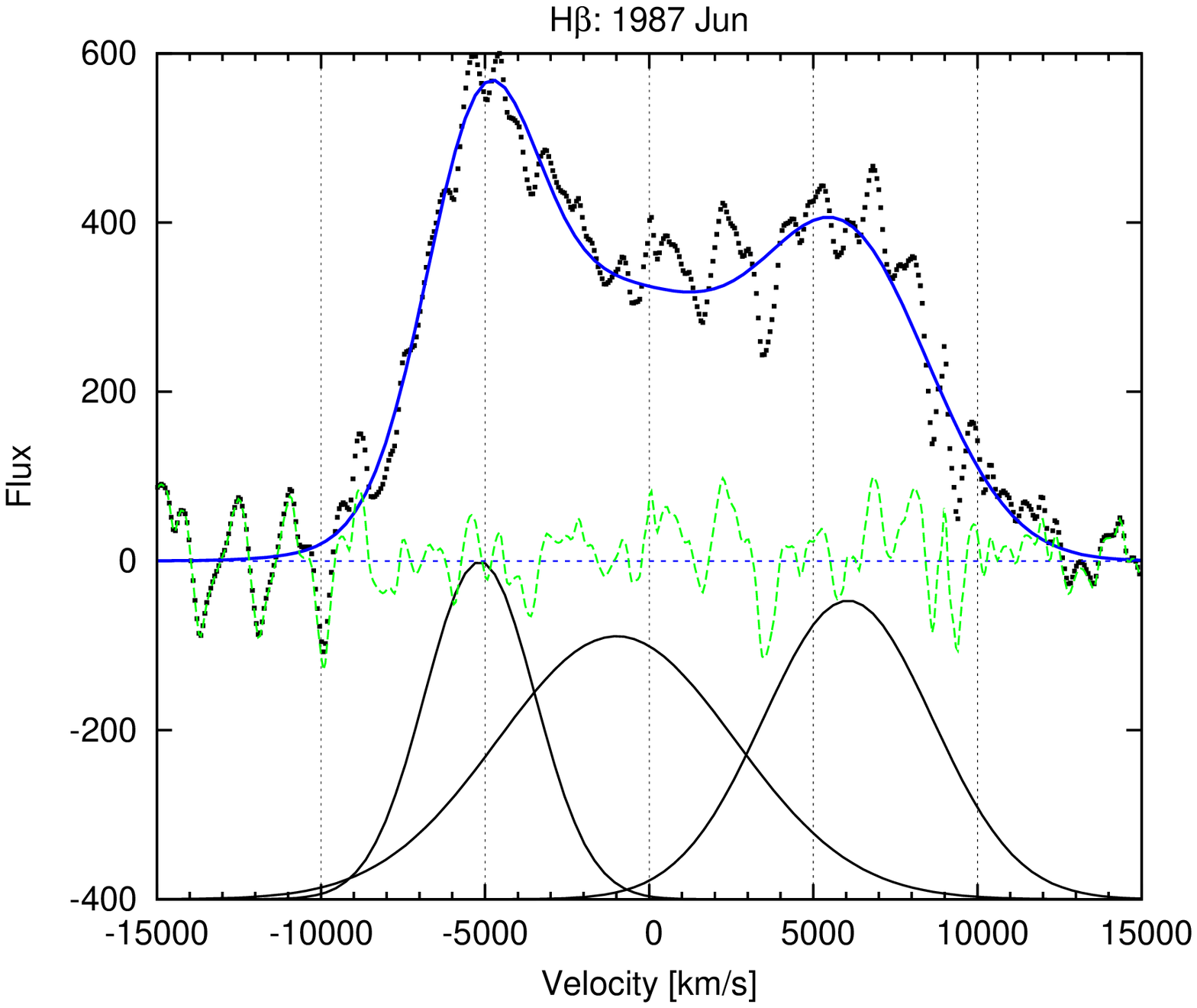}
\includegraphics[width=4cm]{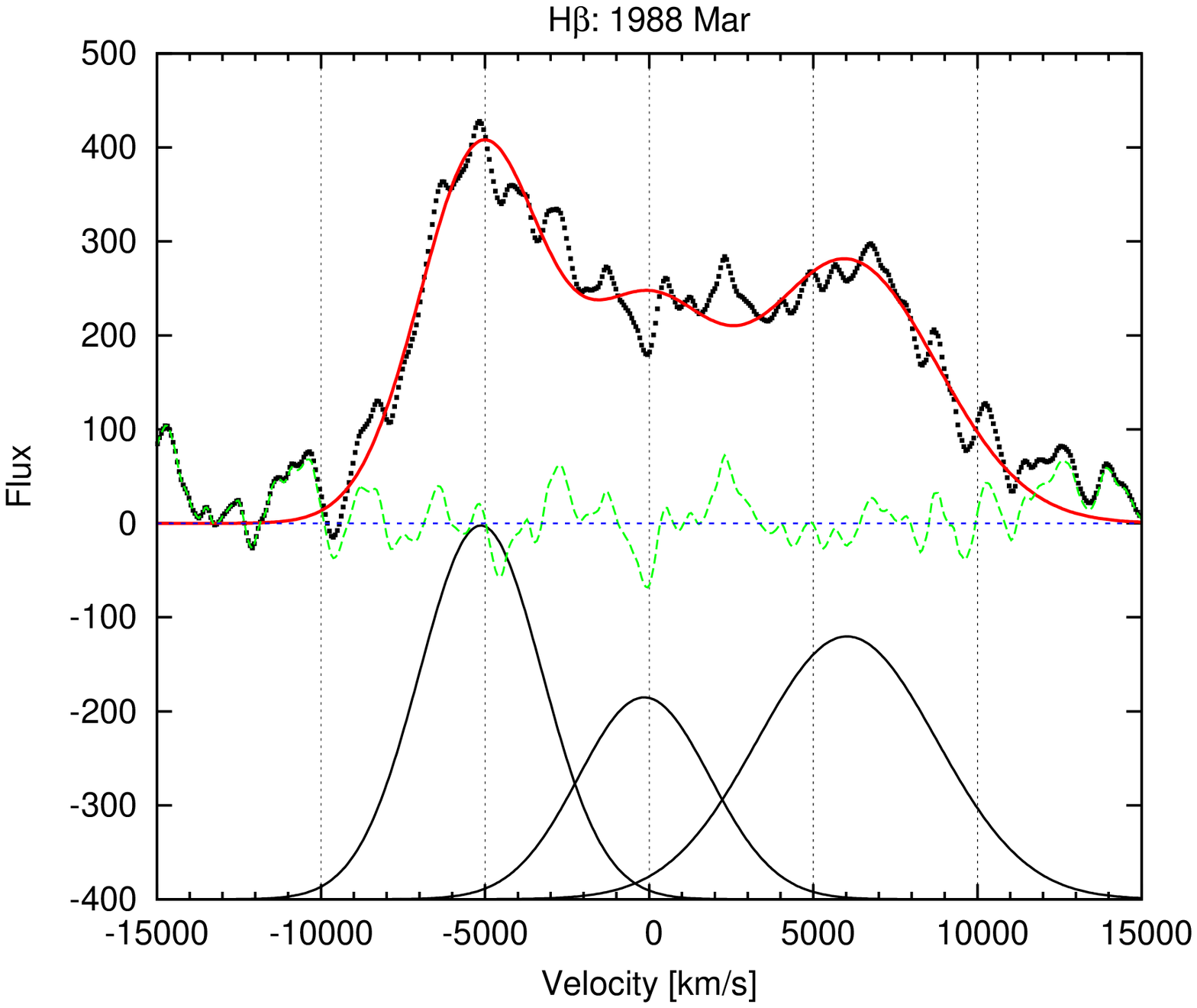}
\includegraphics[width=4cm]{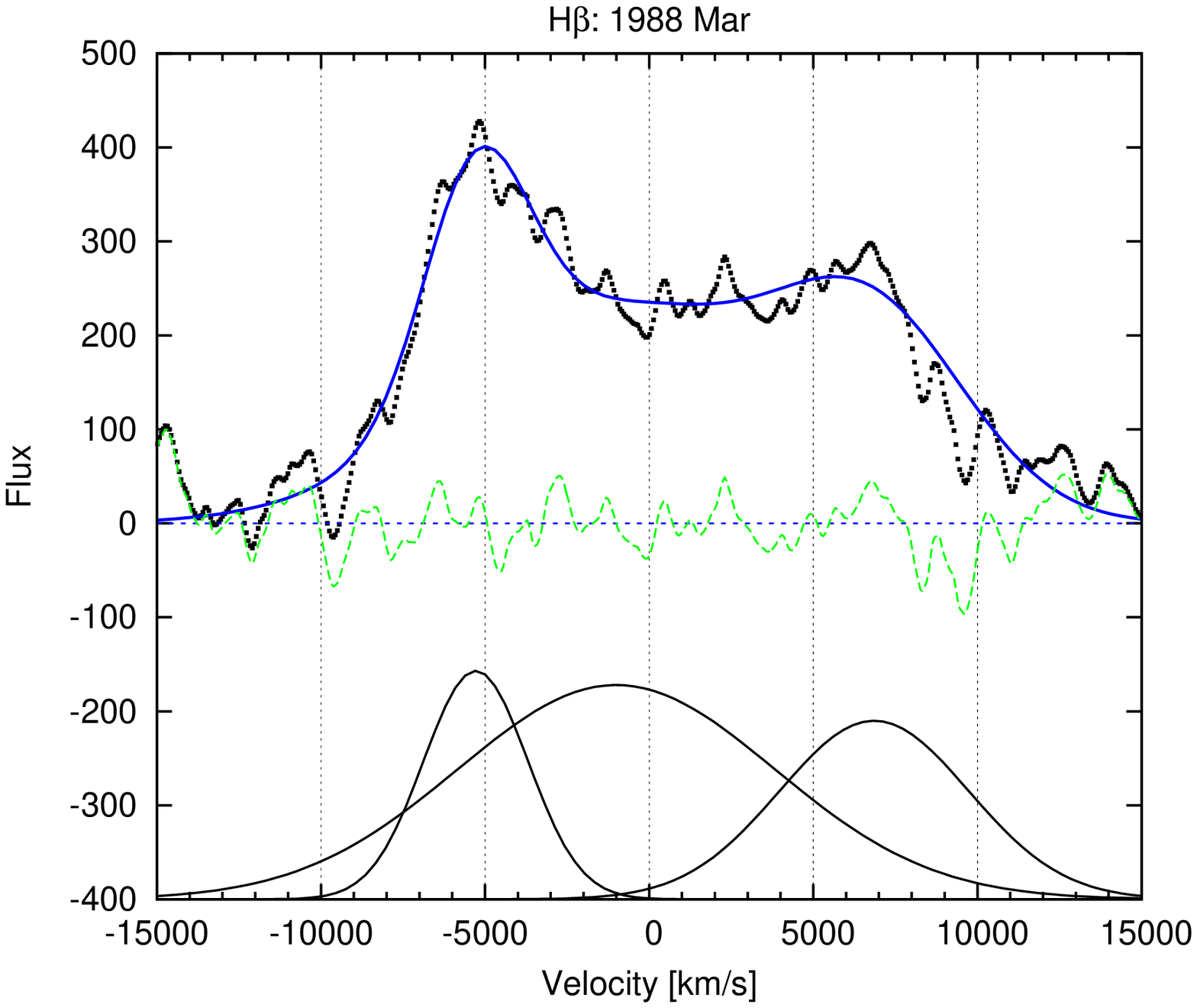}
\includegraphics[width=4cm]{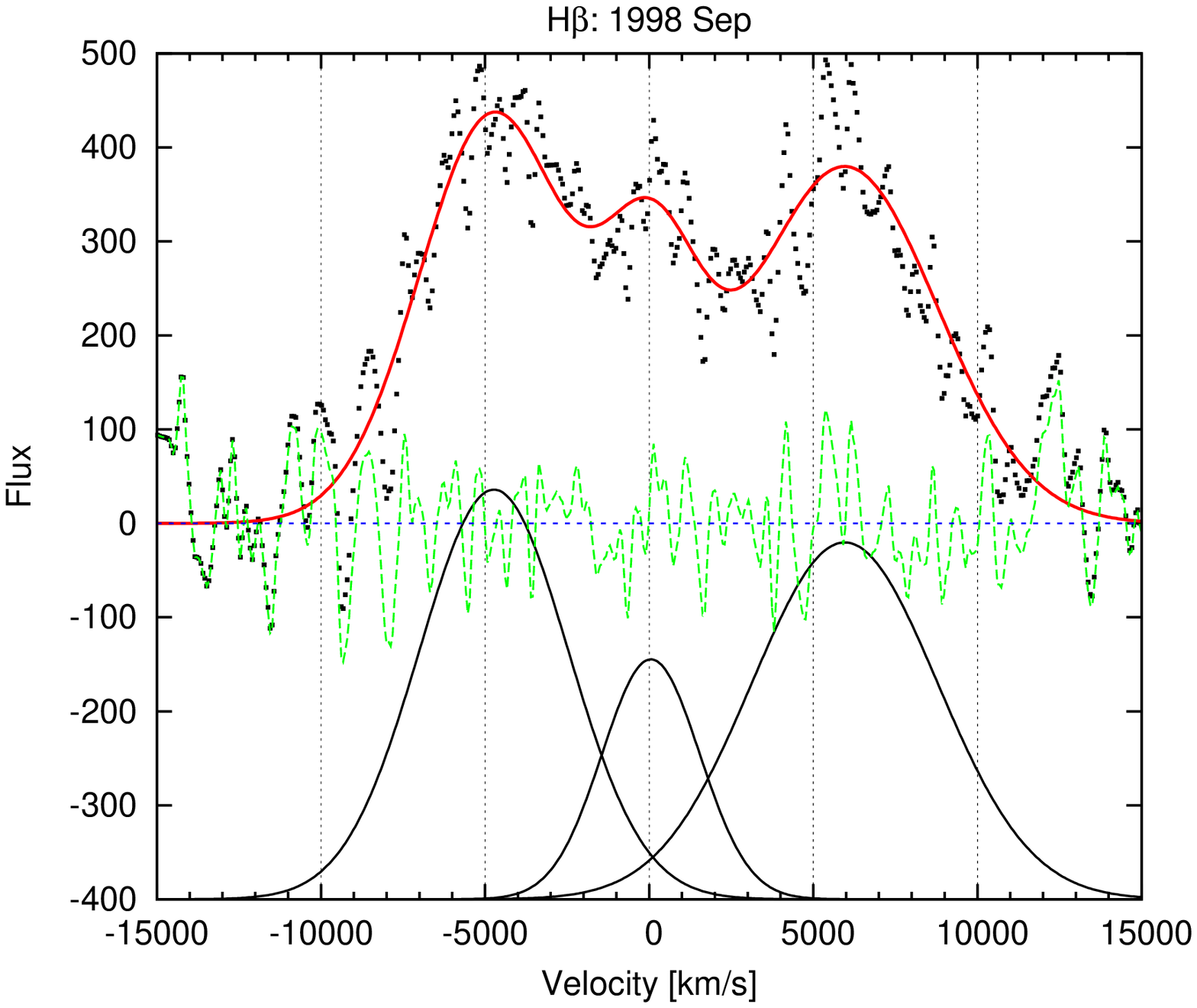}
\includegraphics[width=4cm]{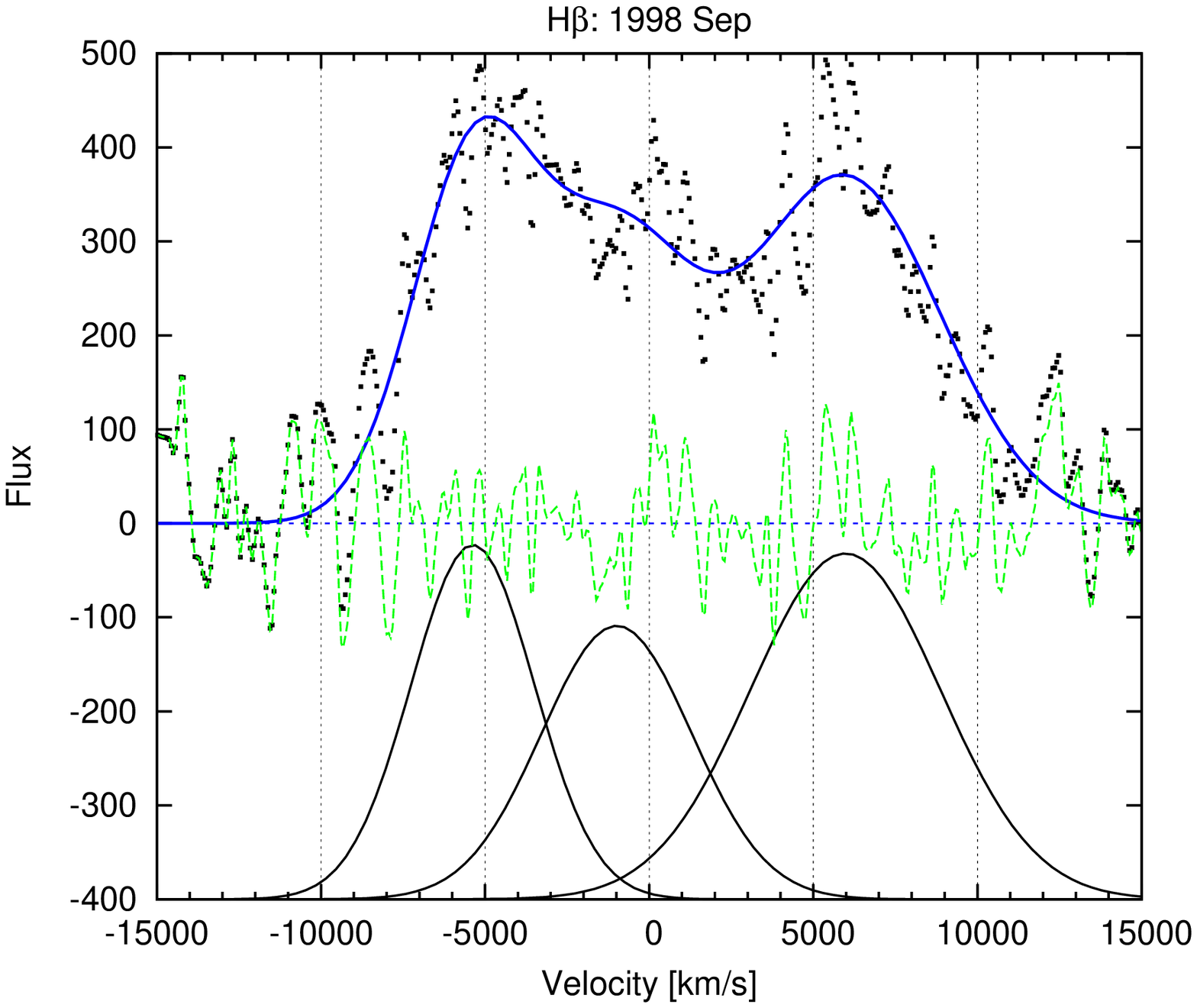}
\caption{The same as in Fig. \ref{fig_ha_broad_gauss} but for H$\beta$ line.}\label{fig_hb_broad_gauss}
\end{figure}
%%%%%%%%%%%%%%%%%%%%%%%%%%%%%%%%

\section{The broad line profile variability}

In Paper I we measured and analyzed variations in the continuum and line fluxes 
using total of 118 spectra covering the H$\beta$ wavelength region, and 90 
spectra covering the H$\alpha$ line. We showed  that the fluxes in the H$\alpha$ and H$\beta$ lines,
and in continuum were not strongly varying  (around 20\% in line and $\sim$30\% in continuum),
and that several flare-like events were observed during the monitored period (1987--2010). 
Considering spectra from new observations,
it seems that there is a minimum in the line emission in the 2012-2013 period (see Fig. \ref{lc}), and also it can be seen in
Fig. \ref{month2} that both lines are weak, moreover, H$\beta$ is very weak with two weak peaks comparable to noise.

During the monitored period, the broad  H$\alpha$ and H$\beta$ lines have double-peaked profiles 
(even in the minimum), but, as it can be seen in Figs. \ref{month1} and \ref{month2}, 
their line shapes and widths are different. 
Comparing the month-averaged spectra of the H$\alpha$ and H$\beta$ broad lines (see Figs.
\ref{month1} and \ref{month2}), one can note, that there is a difference between the peak ratio of 
 H$\alpha$ and H$\beta$ lines. The H$\alpha$ line mostly has stronger blue peak, while very
often the H$\beta$ line has almost the same intensities of the blue and red peak. On the other side, 
the H$\beta$  FWHM and FWQM (FWHM$_{\rm mean}$=16100$\pm$700 
km s$^{-1}$, FWQM$_{\rm mean}=$17900 $\pm$700 km s$^{-1}$) are significantly broader than the
H$\alpha$ ones (FWHM$_{\rm mean}=$14400$\pm$400 km s$^{-1}$, 
$FWQM_{\rm mean}=$16600 $\pm$800 km s$^{-1}$). This is in a contradiction with the disc model predictions, 
i.e. {if H$\beta$ is originating closer to the central black hole (as indicated by larger line widths)}, 
it is expected that it has a more pronounced blue peak than H$\alpha$.

%%%%%%%%%%%%%%%%%%%%%%%%%%%%%%%%
\begin{figure*}
\centering
\includegraphics[width=9cm]{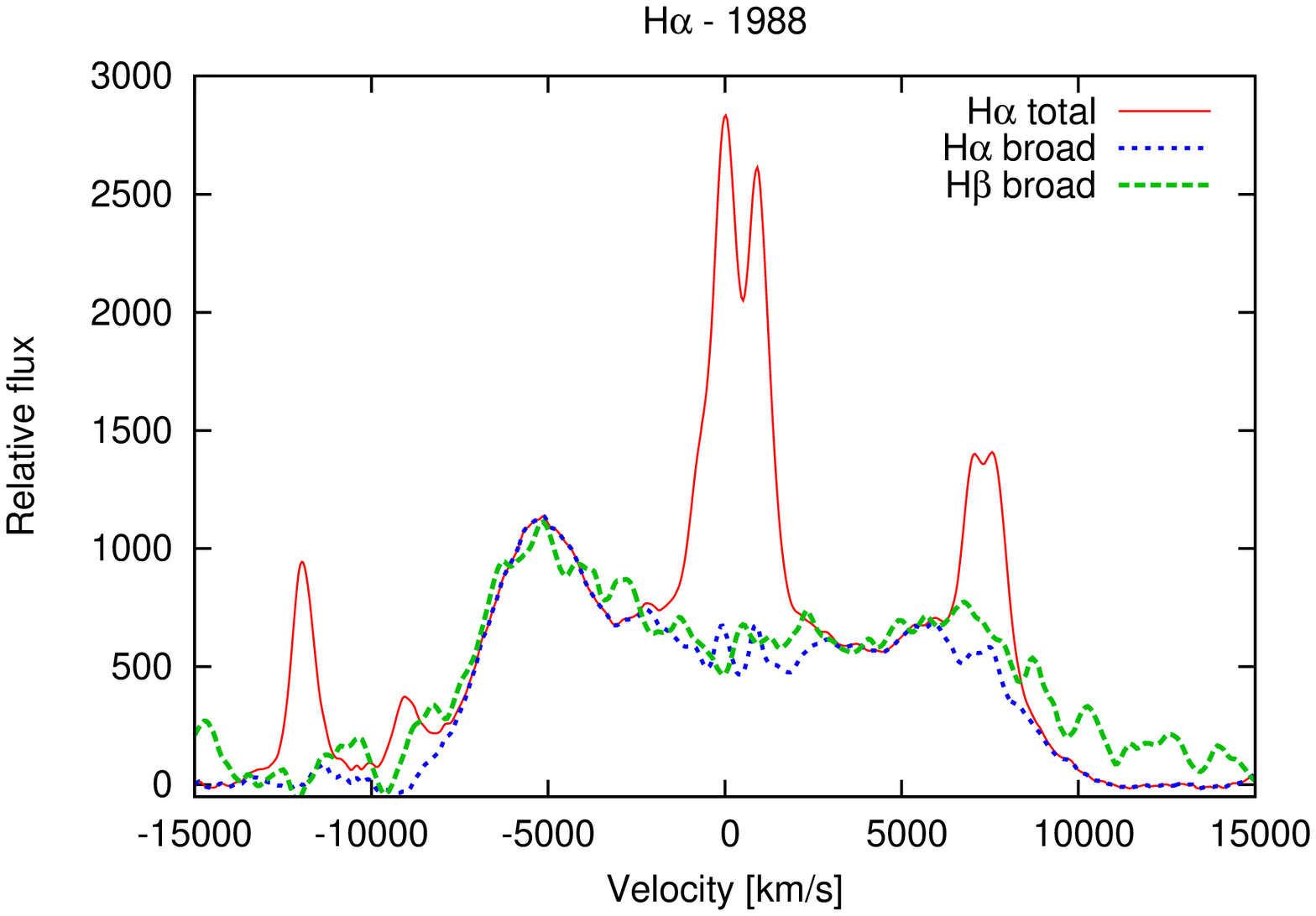}
\includegraphics[width=9cm]{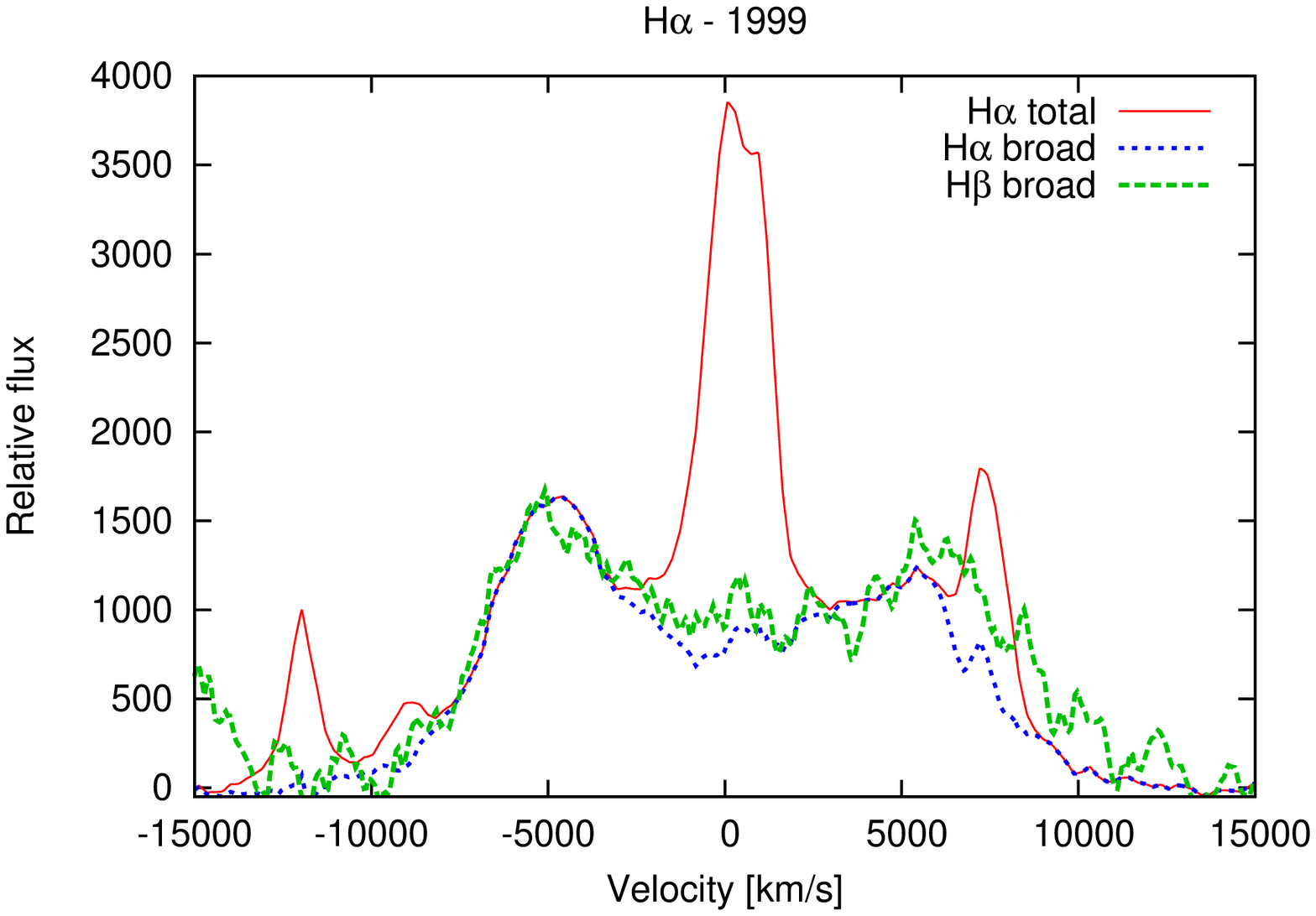}
\includegraphics[width=9cm]{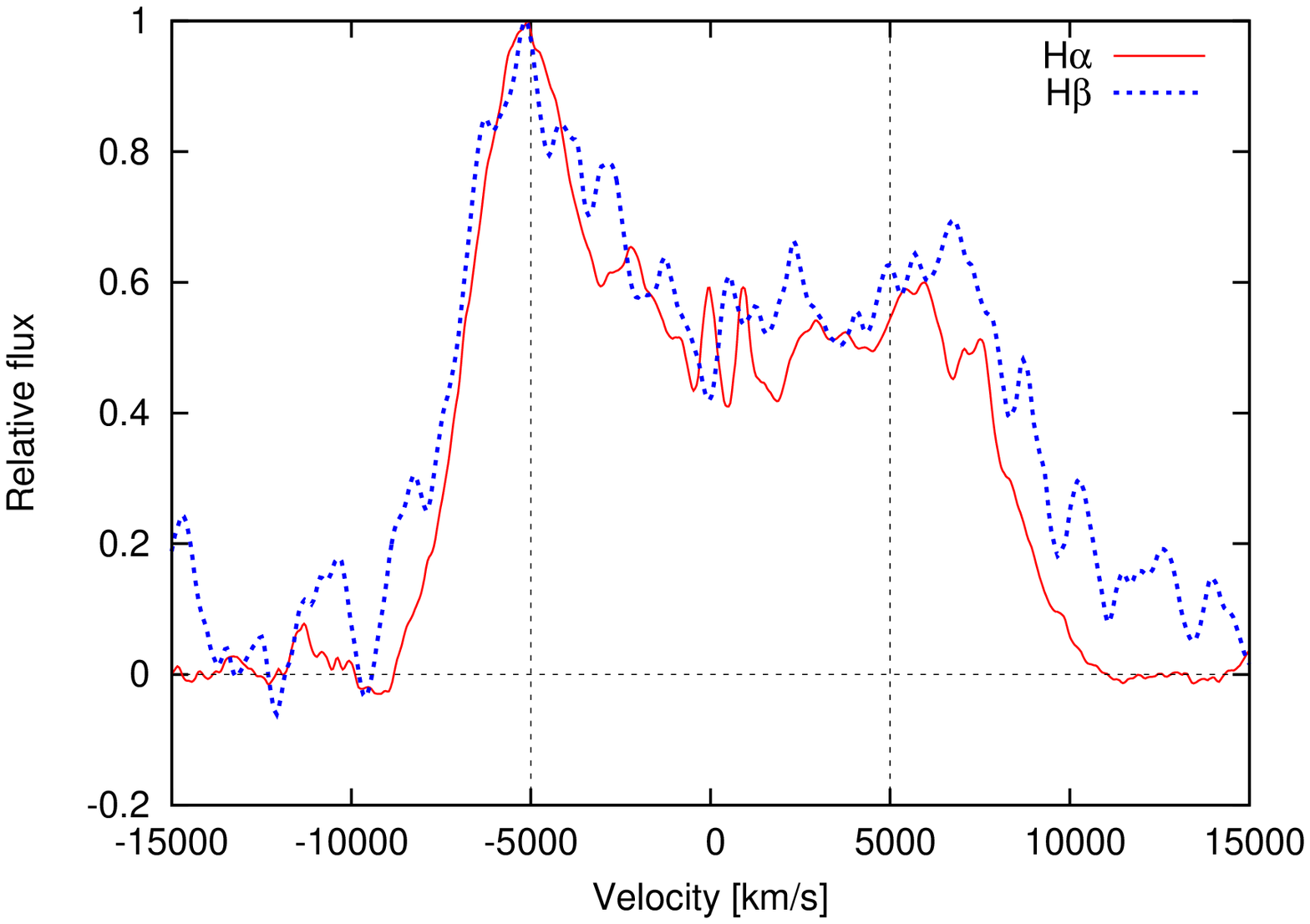}
\includegraphics[width=9cm]{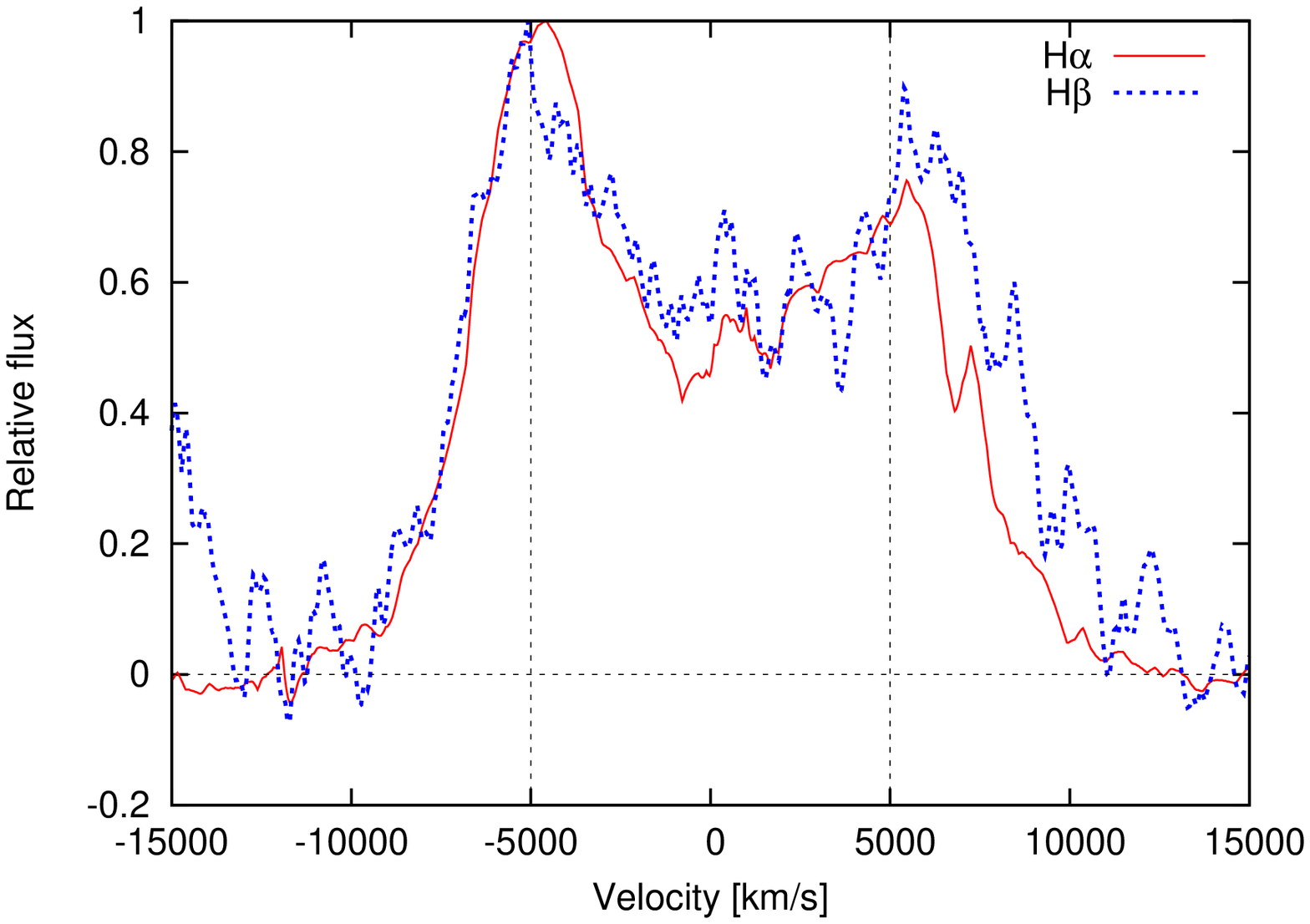}
\includegraphics[width=9cm]{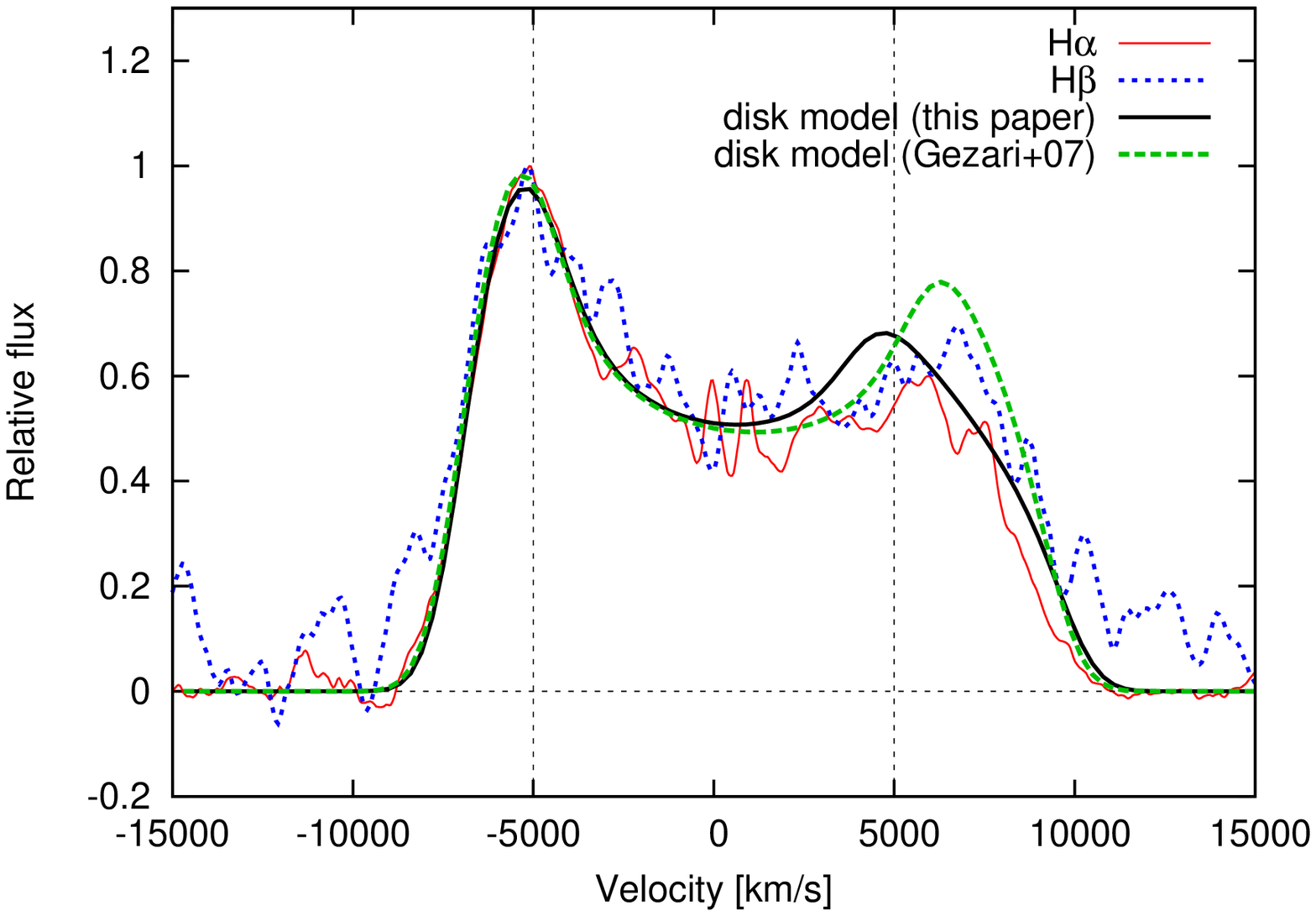}
\includegraphics[width=9cm]{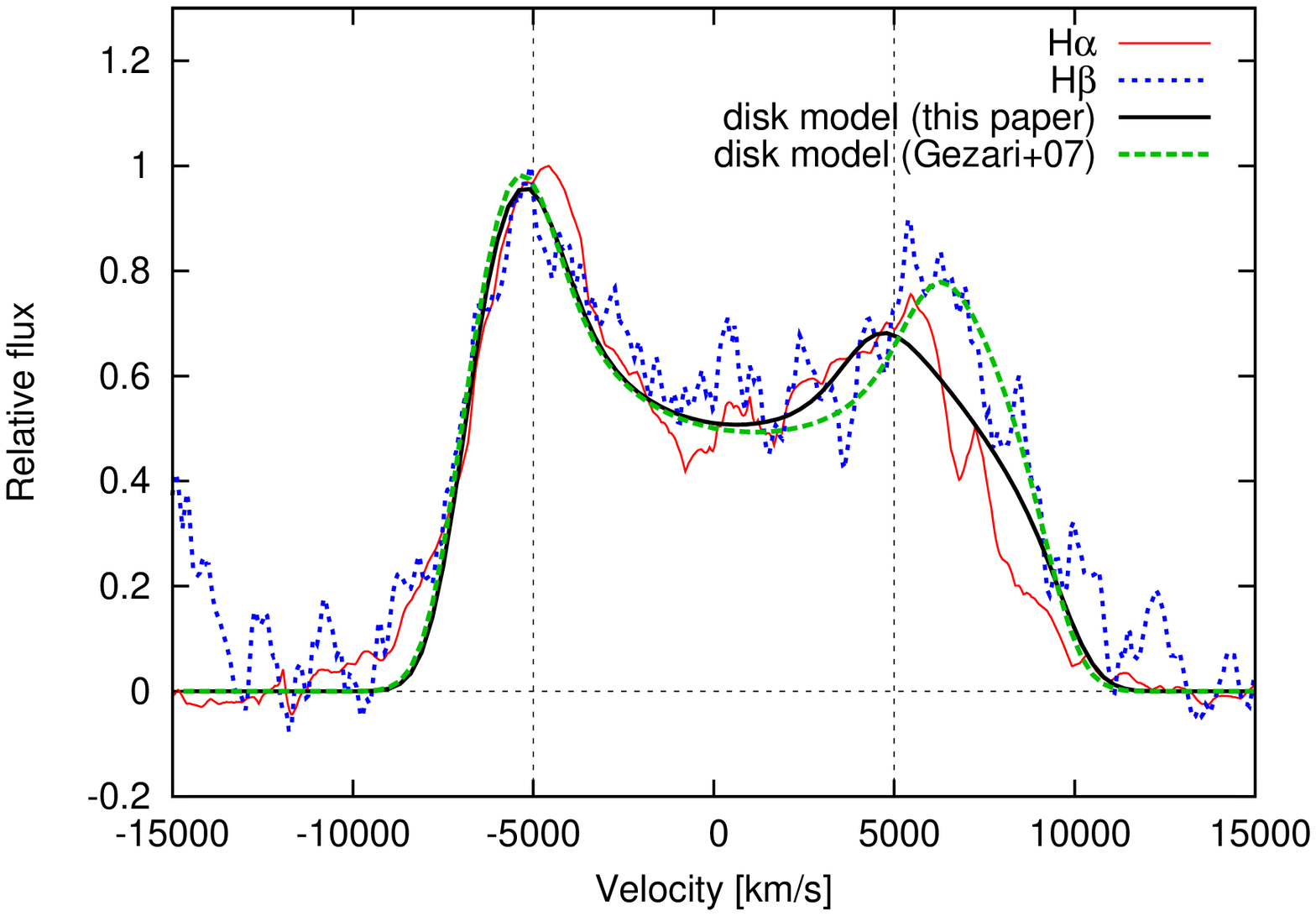}
\caption{The H$\alpha$ line profiles in 1988 (left panels) and 1999 (right panels).
Upper panels give total H$\alpha$ line (with narrow lines) compared with broad
H$\alpha$ and H$\beta$ component (multiplied with the constant to match the H$\alpha$ 
blue peak), middle panels give normalized broad H$\alpha$ and H$\beta$ component,
while bottom panels include disc models from this paper and \citet{g07}. }\label{comp}
\end{figure*}
%%%%%%%%%%%%%%%%%%%%%%%%%%%%%%%%

%%%%%%%%%%%%%%%%%%%%%%%%%%%%%%%%
\begin{figure}
\centering
\includegraphics[width=9cm]{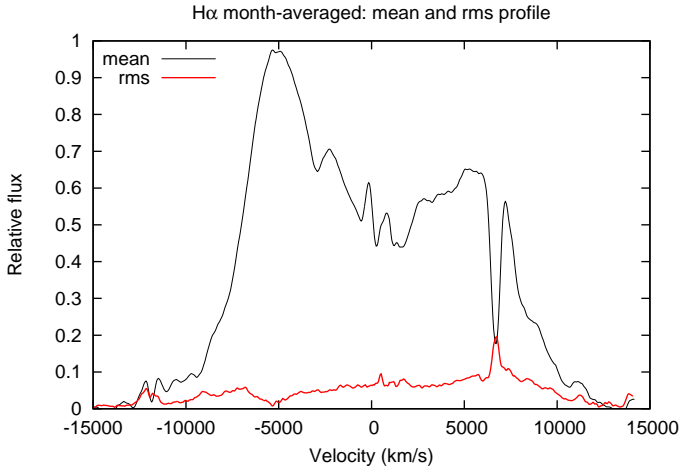}
\caption{The mean and rms of the  H$\alpha$ line profile obtained from the monthly-averaged profiles
normalized to the intensity of the blue peak.
}\label{mean}
\end{figure}
%%%%%%%%%%%%%%%%%%%%%%%%%%%%%%%%

On average, the line profiles of the H$\alpha$ and H$\beta$ lines are not changing significantly, beside some flare-like 
variations seen in short periods (as e.g. in the period 1987--1992, 1998, and 2012--2013). In order
to test the line profile changes we normalized the blue peak of the month-averaged H$\alpha$ profiles to unity 
and calculated the mean and rms profiles (Fig. \ref{mean}). The normalized H$\alpha$ rms profile
shows that practically there is no change in the line profile (below 10\% in the red wing, 
{  where some variations may be due to uncertainty of the narrow line subtraction}). Thus, considering the whole 26-year period, the averaged H$\alpha$ line profile variation is practically neglected.

\subsection{Gaussian analysis - Changes in the broad line profiles}

\begin{figure}
\centering
\includegraphics[width=9cm]{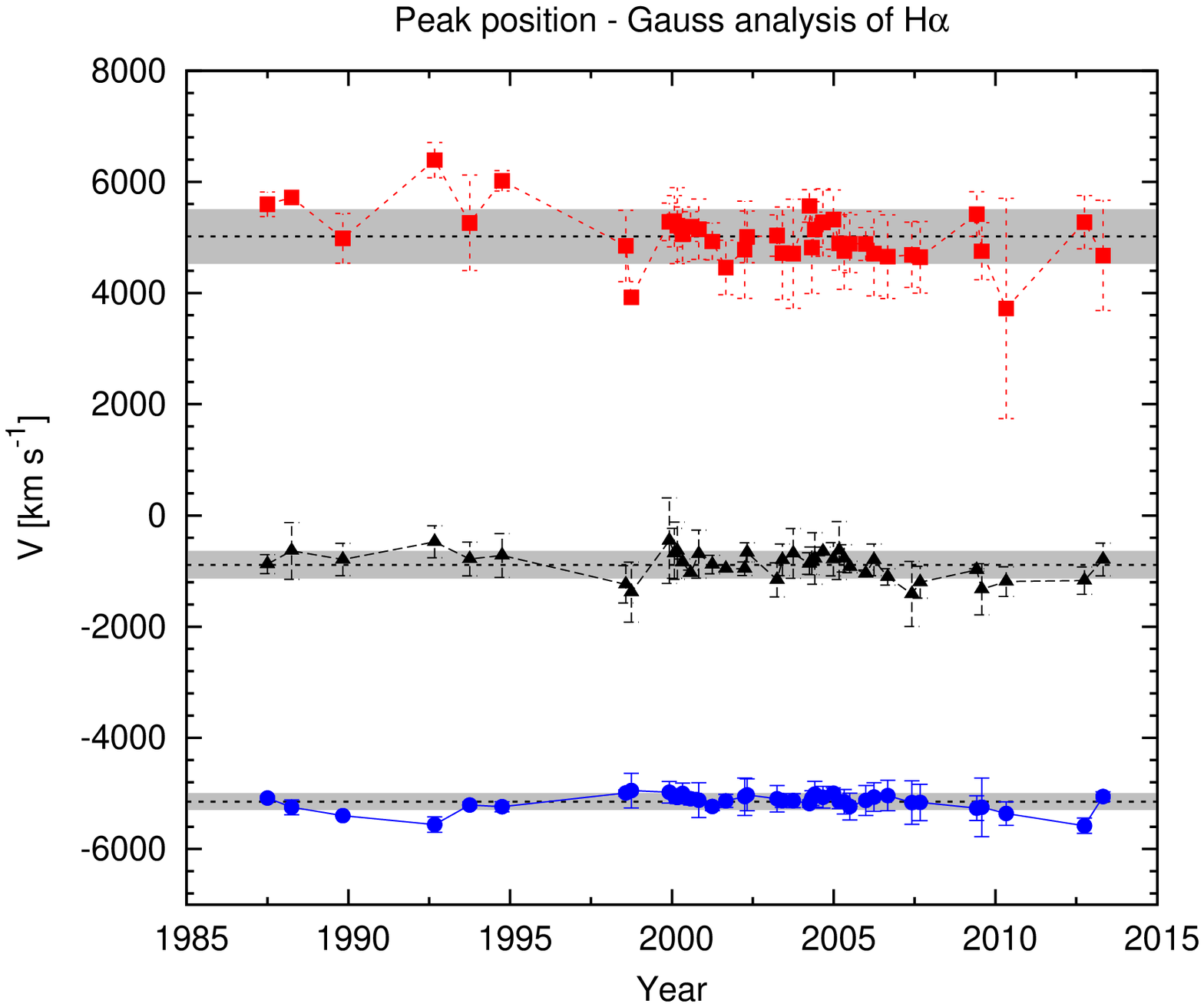}
\includegraphics[width=9cm]{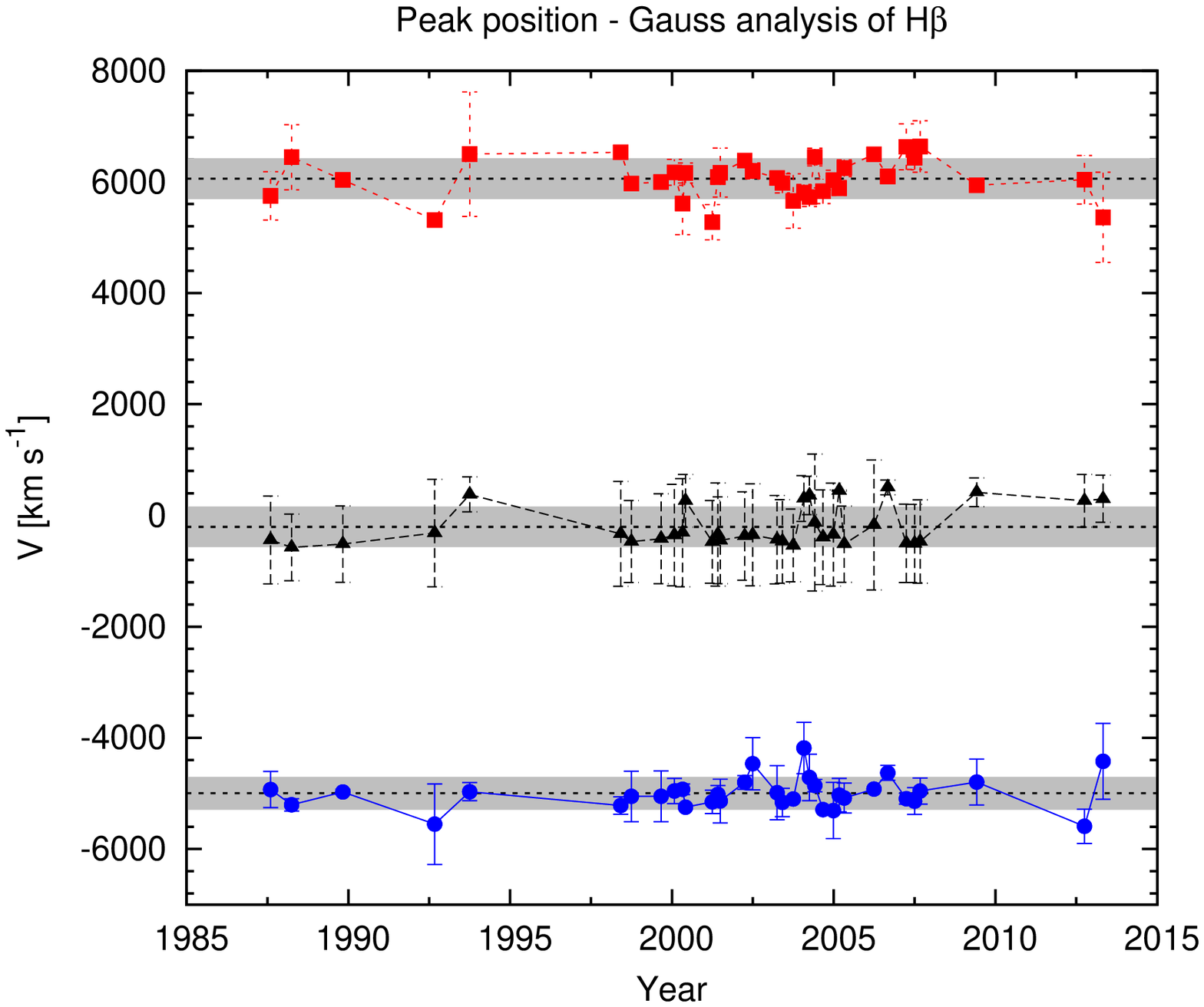}
\caption{The variability of the peak velocity
obtained from the Gaussian fitting (red, central, blue) of H$\alpha$ (upper)
and H$\beta$ line (bottom).} \label{pv-hab}
\end{figure}

The parameters from the Gaussian fittings are given in Tables \ref{tab02} and \ref{tab03}. 
As it can be seen from Tables, the {  H$\alpha$ and H$\beta$ blue} peak-positions  
are changing during the observed period for 
about $\pm$500 km s$^{-1}$ around the averaged values (see Fig. \ref{pv-hab} and Table \ref{tab02}),
{  while the red-peak positions are changing for  $\pm\sim$1000 km s$^{-1}$ around averaged ones, that may also 
 be due to the measurement uncertainties}.

\begin{figure}
\centering
\includegraphics[width=8cm]{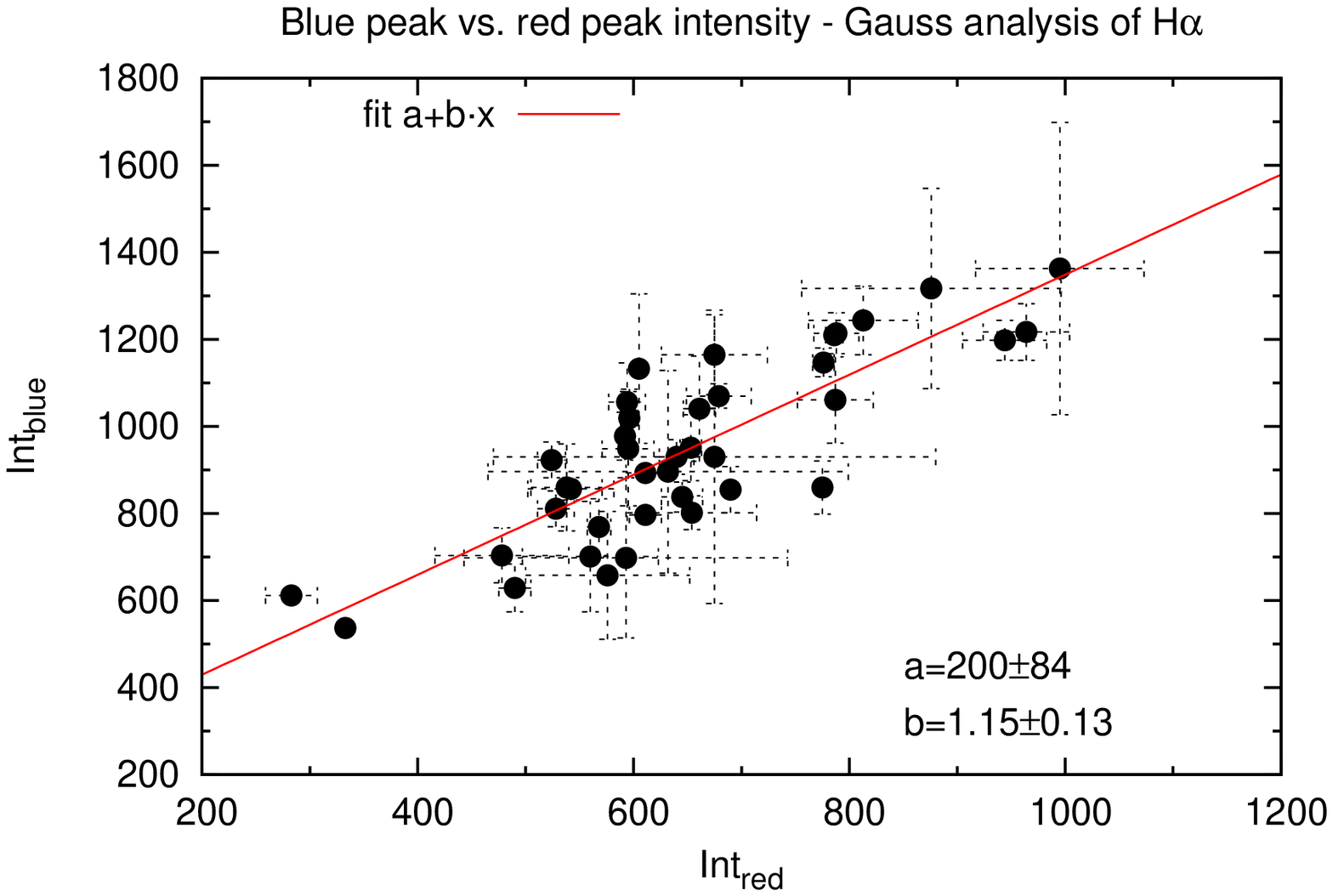}
\includegraphics[width=8cm]{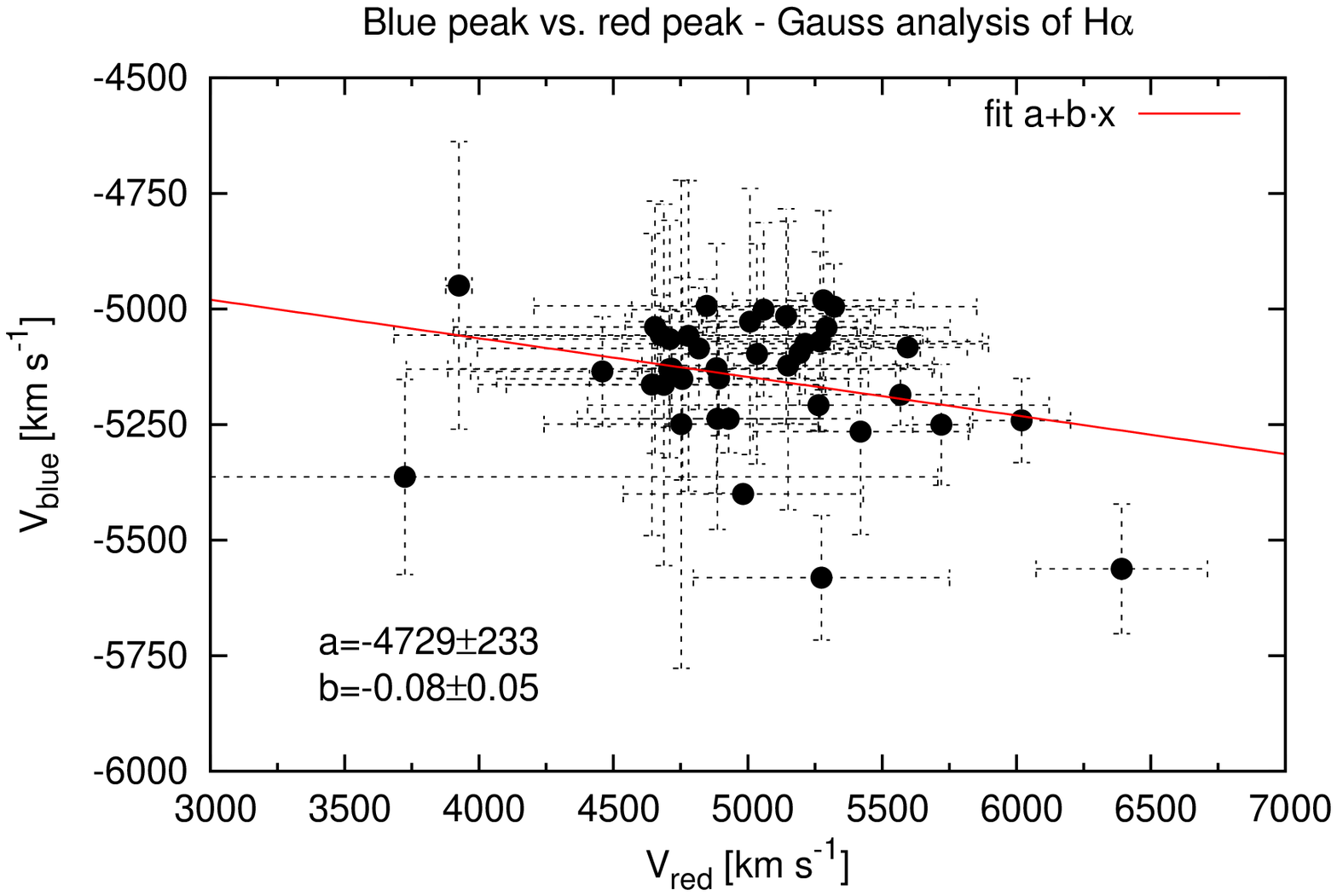}
\caption{The blue- to red-peak intensity (upper) 
and velocity (bottom) of H$\alpha$. 
The {  correlation between the intensities is significant (r=0.83 and P$_0$=0.5E-10),
while the anti-correlation between velocities is weak (r=-0.28)  and statistically is not 
significant (P$_0$=0.08)}.}\label{Ha}
\end{figure}

In general, the H$\alpha$ profiles are of much better quality than H$\beta$,
thus we plotted in Fig. \ref{Ha} the results of the peak-measurements of
the H$\alpha$ line: the blue- vs. red-peak intensity (upper panel), and the blue- vs. red-peak
velocity (down). It is interesting that there is no correlation between 
the velocities of the red and blue peaks (see Fig. \ref{Ha}, bottom panel), 
i.e. {  there may be a trend of an anticorrelation (but statistically not significant,
r=-0.28, P$_0$=0.08), and it seems that} they vary independently of each other.  But there is a good correlation 
between two peak intensities (see Fig. \ref{Ha}, upper panel) and the blue- 
and red-peak intensities are responding to each other. 

\begin{figure}
\centering
\includegraphics[width=8.5cm]{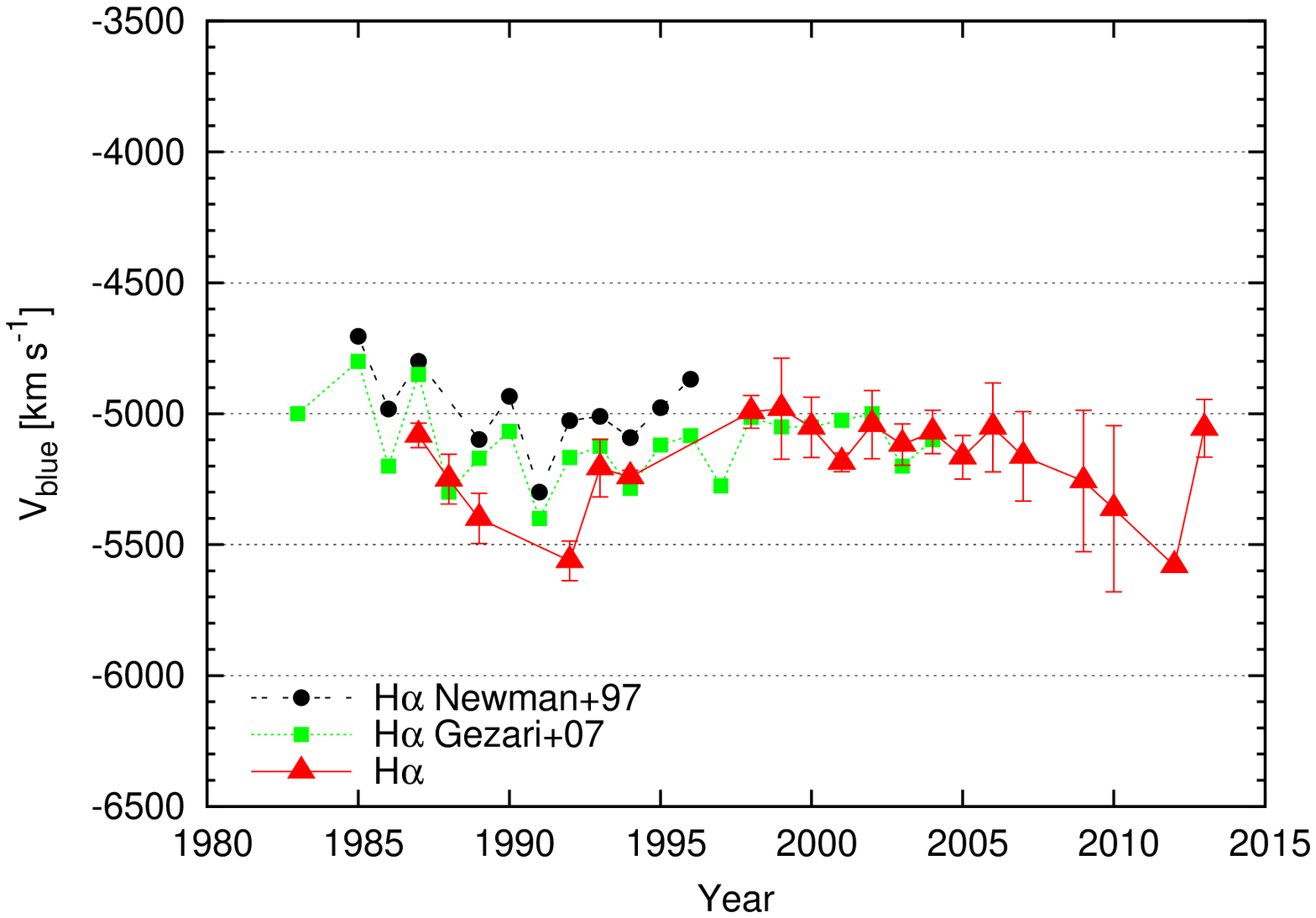}
\includegraphics[width=8.5cm]{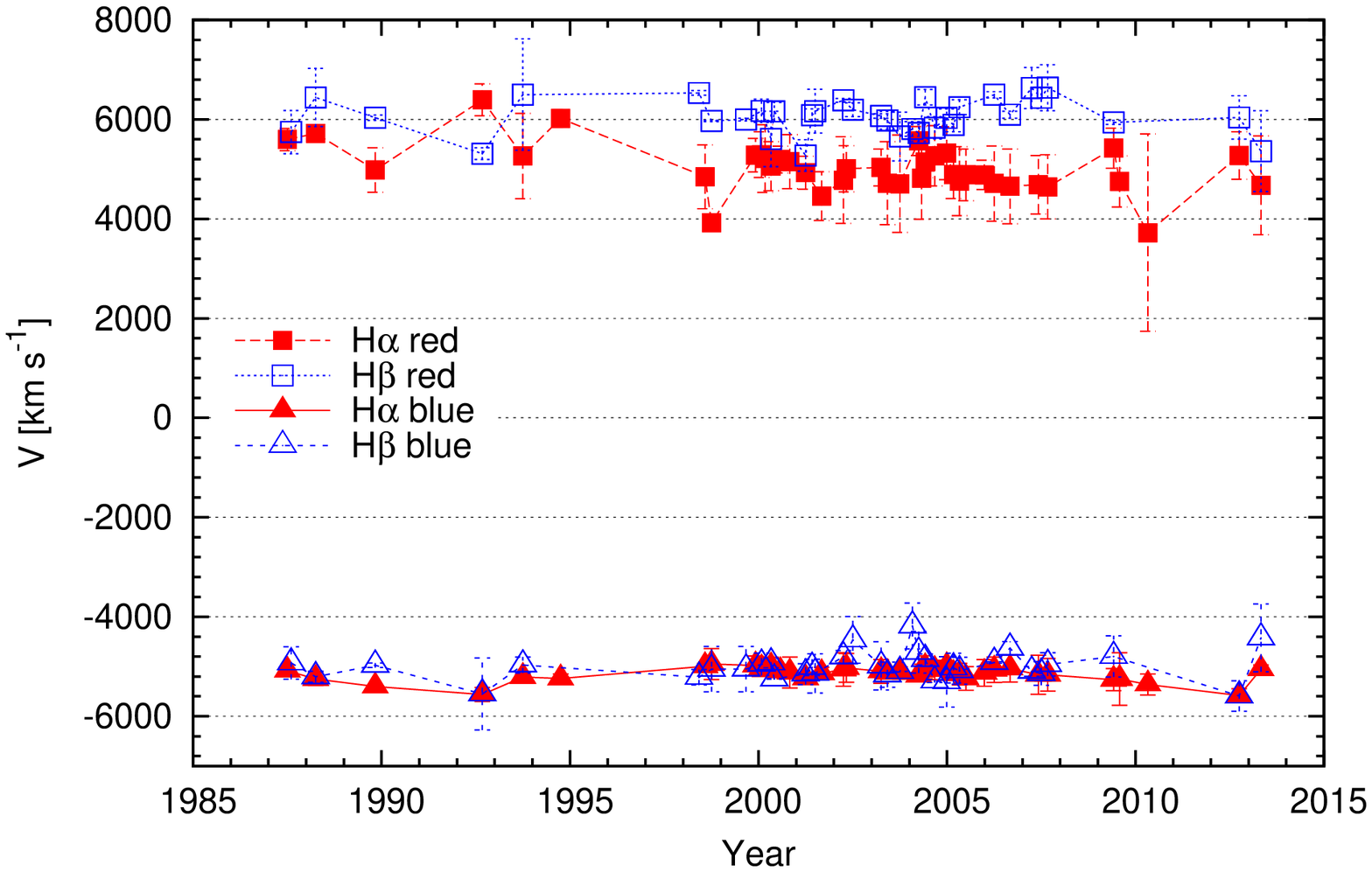}
\caption{Upper: Year-averaged radial velocities of the blue peak. The
different symbols represent: full squares - the H$\alpha$ data from
\citet{g07} year-averaged to match our data, full circles - the H$\alpha$ given in
\citet{n97} year-averaged to match our data, full triangles - our data
for H$\alpha$. Bottom: Month-averaged radial velocities of the blue 
and red peak positions for H$\alpha$ and H$\beta$. 
}\label{peak-y}
\end{figure}

In order to compare the position of the blue peak from our monitoring campaign
with previous campaigns, we plot in  Fig. \ref{peak-y} (upper panel)
our measurements of the blue peak velocity
for the H$\alpha$ line together with the measurements
of \citet{n97} and \citet{g07}. As it can seen from Fig. \ref{peak-y}, the agreement
between different measurements is good, and differences are within the error-bars 
($\pm$100-200 km s$^{-1}$). In Fig. \ref{peak-y} (bottom panel) we compared the blue- and red-peak positions
between H$\alpha$ and H$\beta$ during monitored period. As it can be seen from Fig. \ref{peak-y} (bottom panel), 
there 
is a discrepancy between the positions of the red peaks of H$\alpha$ and H$\beta$ (in some periods for about 
2000 km s$^{-1}$), while the blue-peak position in both lines has no big change. There are small differences between 
the H$\alpha$ and H$\beta$ blue-peak positions, which are below $\sim$500 km s$^{-1}$
(see also Tables \ref{tab02} and \ref{tab03}).

%%%%%%%%%%%%%%%%%%%%%%%%%%%%%%%%
%%%%%%%%%%%%%%%%%%%%%%%%%%%%%%%%

%%%%%%%%%%%%%%%%%%%%%%%%%%%%%%%%
%%%%%%%%%%%%%%%%%%%%%%%%%%%%%%%%

%%%%%%%%%%%%%%%%%%%%%%%%%%%%%%%%
%%%%%%%%%%%%%%%%%%%%%%%%%%%%%%%%

\begin{figure}
\centering
\includegraphics[width=8cm]{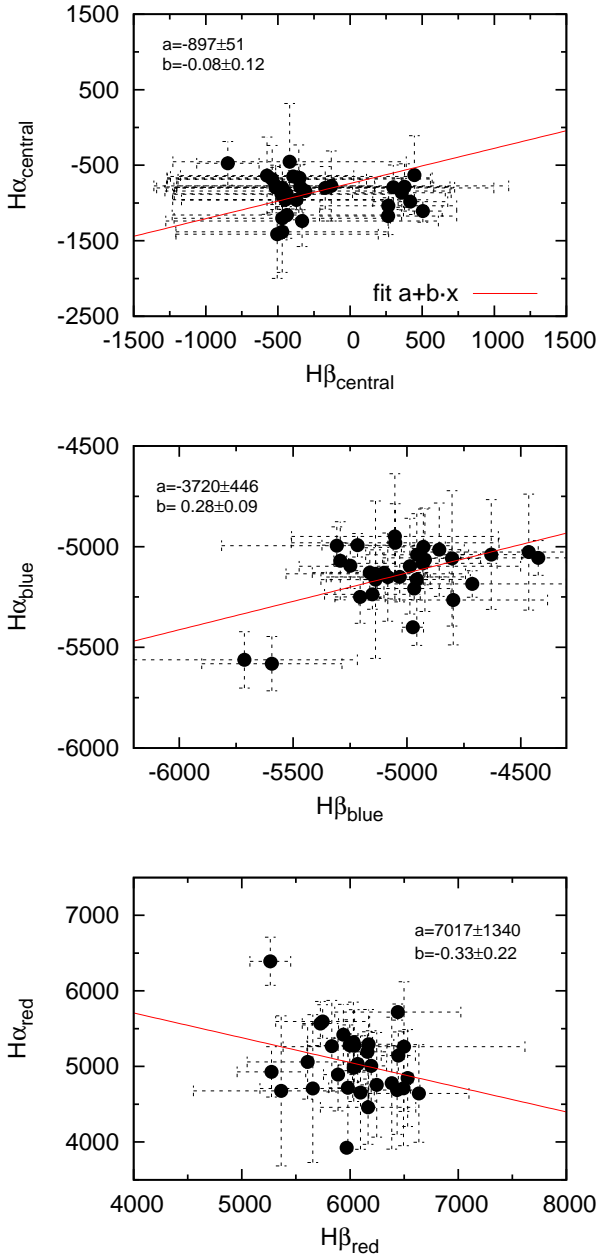}
\caption{H$\alpha$ vs. H$\beta$ peak velocity for the central (upper),
blue (middle) and red (bottom) peak from the month-averaged profiles. 
The dashed line represents the expected function for an equal peak position 
in both lines, while solid line represents the best fit of the measured data.
{  The correlation coefficients are: r=-0.12 (P$_0$ = 0.5) for the central, r =0.51  
(P$_0$=0.3E-02) for the blue peak, and  r = -0.26 (P$_0$ = 0.15) for the red peak.
The parameters of the best fit are given on the plots.}
}\label{f04}
\end{figure}

In Fig. \ref{f04} we plot the H$\alpha$ vs. H$\beta$ peak velocity for the central (upper) ,
blue (middle) and red (bottom)  peak of Gaussians from the month-averaged profile fits.
As it can be seen from Fig. \ref{f04} there are no expected 
correlations. 
However, it seems that a trend of anti-correlation is present in the red H$\alpha$ vs. 
H$\beta$ velocities (solid line on the plot), {  while a correlation trend 
may be present between the blue-peak velocities of H$\alpha$ and H$\beta$.
We should note that a trend of anti-correlation between the red-peak position may be caused by the
measurement uncertainties, but that is not the case for the blue-peak positions. 
A theoretical expectation is that the velocity of the blue peak of the 
H$\alpha$ line follows the one in the H$\beta$ line, and, as can be seen in Fig.\ref{f04} (middle) there may be 
a slight dependence between the blue-peak velocities of H$\alpha$ and H$\beta$, but it seems to be very
weak (even taking into account  the uncertainties in the  measurements which are shown on the plots).
This confirms that there are no expected correlations between the peak velocities of two broad lines.}

%%%%%%%%%%%%%%%%%%%%%%%%%%%%%%%%
%%%%%%%%%%%%%%%%%%%%%%%%%%%%%%%%

%%%%%%%%%%%%%%%%%%%%%%%%%%%%%%%%
\begin{figure}
\centering
\includegraphics[width=9cm]{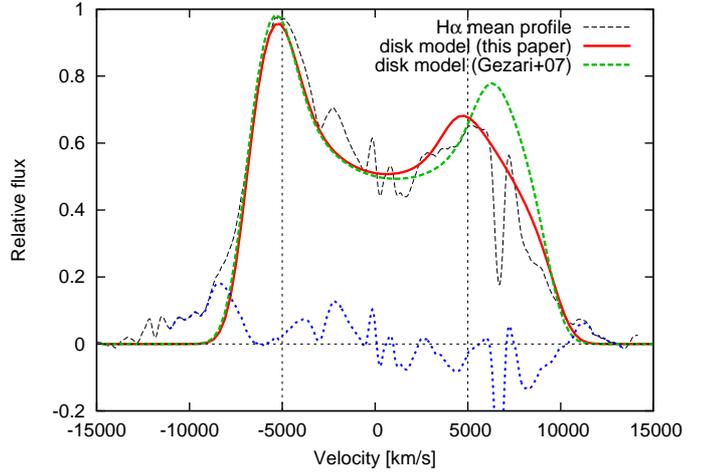}
\caption{The mean  normalized H$\alpha$ line profile fitted with
 the disc-model:  solid line -- our assumption (see text) and thick dashed line -- parameters taken from \cite{g07}.
The residual at the bottom corresponds to the
difference between the disc model from this paper and mean profile. }\label{mean1}
\end{figure}
%%%%%%%%%%%%%%%%%%%%%%%%%%%%%%%%

\subsection{Disc-model of the broad emission lines}

As we noted above, the relativistic disc model \citep{c89,ch89,g07} can explain the broad double-peaked profiles of 
Arp 102B. We performed several fits, and it was interesting that the month-averaged profiles can be 
successfully fitted with the simple \cite{c89} disc model.  The problem is that in some periods we could fit
the line profiles only if we change the inclination of the disc for about 10 degrees, that is not expected.
{  Note here that in the case when the inclination was fixed we 
obtain the satisfied fit by varying the other parameters (e.g. the emissivity and inner radius).}
Here we will not present all fits, only we present examples of the fits  with parameters given by \citet{g07}:
the inner radius $R_{\rm inn} \sim 325\ R_{\rm g}$, the outer radius 
$R_{\rm out} \sim 825\ R_{\rm g}$, the disc inclination $i \sim 31 \deg$, the random velocity in the disc 
$\sigma = 1050\ {\rm km\ s^{-1}}$, with the emissivity $r^{-q}$, where $q=3.0$ is assumed
(see Figs. \ref{comp} and \ref{mean1}).
Additionally, starting from the parameters given above, we find the best fit of  the mean H$\alpha$ profile, 
only changing
the inclination  taking  $i=23 \deg$, and that the line is blue-shifted at -900 km s$^{-1}$. As an example, 
two fits are compared 
with observations from two periods for the H$\alpha$ and H$\beta$ line in Fig. \ref{comp} (bottom panel). 
As it can be seen in 
Fig. \ref{comp}, the agreement with the blue peak of observations and both models are good, but differences 
are in the red wing. {  But we emphasize again that the narrow line subtraction may 
contribute to the uncertainty in the red-peak properties that can thus affect the obtained disc parameters.} 

If we consider the mean profile during the whole monitored period, one can conclude that 
the disc inclination is around $i=23 \deg$, while the disc dimensions are around  500 $\ R_{\rm g}$. 
There is agreement between the disc parameters from our fit of the mean H$\alpha$ profile and one given in 
\citet{g07}, {apart from} the disc inclination. {  However, by varying other parameters we were
able to obtain $i\sim 30 \deg$, but in all fittings the disc dimensions seem to be very compact
(about several 100s gravitational radii)}. In Fig. \ref{mean1} we plotted both models together with the normalized mean H$\alpha$ profile.

As a conclusion, the line shape (double-peaked profiles) can be mainly fitted with a simple 
relativistic disc model, but
the problem is, that, to fit the same line from different periods, one should change the disc {  parameters,
as e.g. inclination or emissivity}.  This, in principle is not expected, {  but we should point out 
that the obtained parameters from the fit are strongly depending on the positions and intensities of the blue and red
peaks, and as it was noted above there are relatively big uncertainties in the red-peak properties.}

\section{Discussion}

Double-peaked broad emission lines signature expected from the disc-like BLR, are observed in the spectrum of 
Arp 102B \citep{ch89,eh94,s00,g07,sh13}. The double-peaked lines are presented in a small fraction of AGNs \citep[see][]{st03}.
The double-peaked line profiles have been explained with a number of different models:
a) binary black holes \citep[see e.g.][]{ga83}, b) bipolar outflows \citep[see e.g.][]{zh91},
c) anisotropically illuminated spherical distribution of optically thick clouds \citep{gw96}, d) circular or 
elliptical accretion disc \citep[see e.g.][]{c89, ch89, er95,er96}, and e) other more sophisticated models
like precessing elliptical disc or a circular disc with a long-lived, single-armed spiral or warp
\citep[see e.g][]{g07,jov10}.

Here we can exclude the binary black hole \citep[see e.g.][]{ga83} model, since there is no significant changes in the position of the line peaks. Additionally, the model of anisotropically illuminated spherical 
distribution of optically thick clouds \citep{gw96} has a problem to explain a big distance between line peaks (more than 10000 
${\rm km\ s^{-1}}$), therefore here we will consider the disc-like geometry of the BLR and possibility of outflows in the BLR.

\subsubsection{The BLR disc emission: pro and contra}

As we noted above, the disc emission has been usually assumed to
model double-peaked profiles of Arp 102B, however some observational facts from
monitoring of the broad lines are in contradictions with the disc model.
Recall here some of them \citep[see][]{mp90,g04,g07}:
\begin{itemize}
 \item Flares in the broad-line flux, that can be seen also in our observations (see Paper I), but this can
 be expected if there are some transient processes in the disc \citep[see e.g.][]{jov10}.
  \item Systematic variations in full width at quarter maximum reported in  \citet{g04}, that also may be explained in the 
 disc structure variation.
\item Periodic oscillation of the red-to-blue wing flux ratio observed by \citet{n97} and \citet{sh13}, also can be caused 
by rotating structure in the disc.
 \item Intensity of the red peak is sometimes higher than the blue one (see Figs. \ref{month1} and \ref{month2}), also can be explained
 by perturbations in the disc structure \citep[see e.g.][]{jov10,pop11}, as well as with a model of precessing elliptical disc 
 \citep[][]{g07}.
\end{itemize}

As one can see, that some observed facts, mentioned above, can be explained by 
 a more complex and non stable 
configuration of the accretion disc, i.e. that  emissivity enhancements such as transient shocks induced by tidal perturbations could
be present, and that the change in line profile can be reproduced by changing the inner and outer radii 
of the line-emitting portion of the disc.
 
\begin{figure}
\centering
\includegraphics[width=9cm]{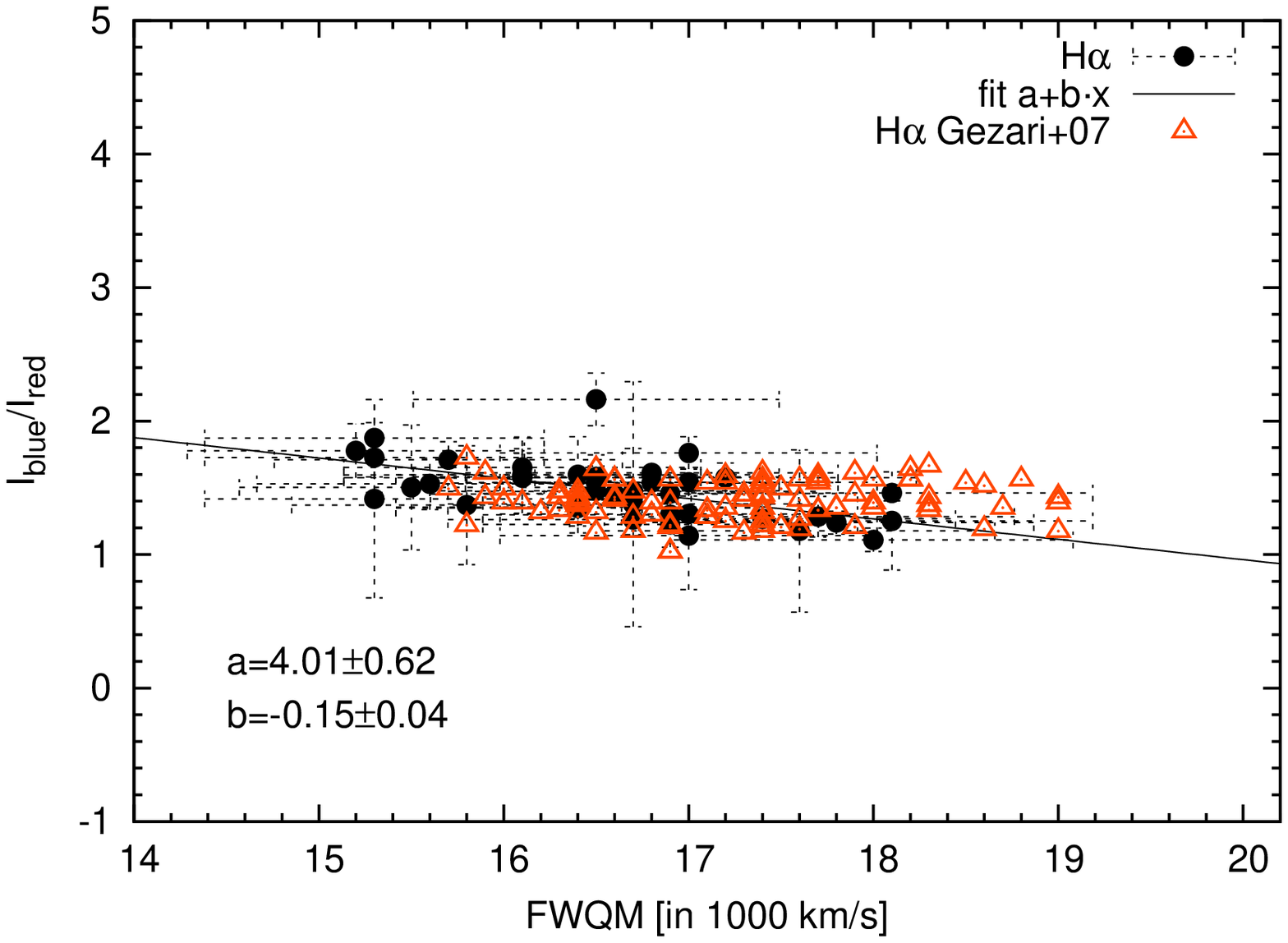}
\includegraphics[width=9cm]{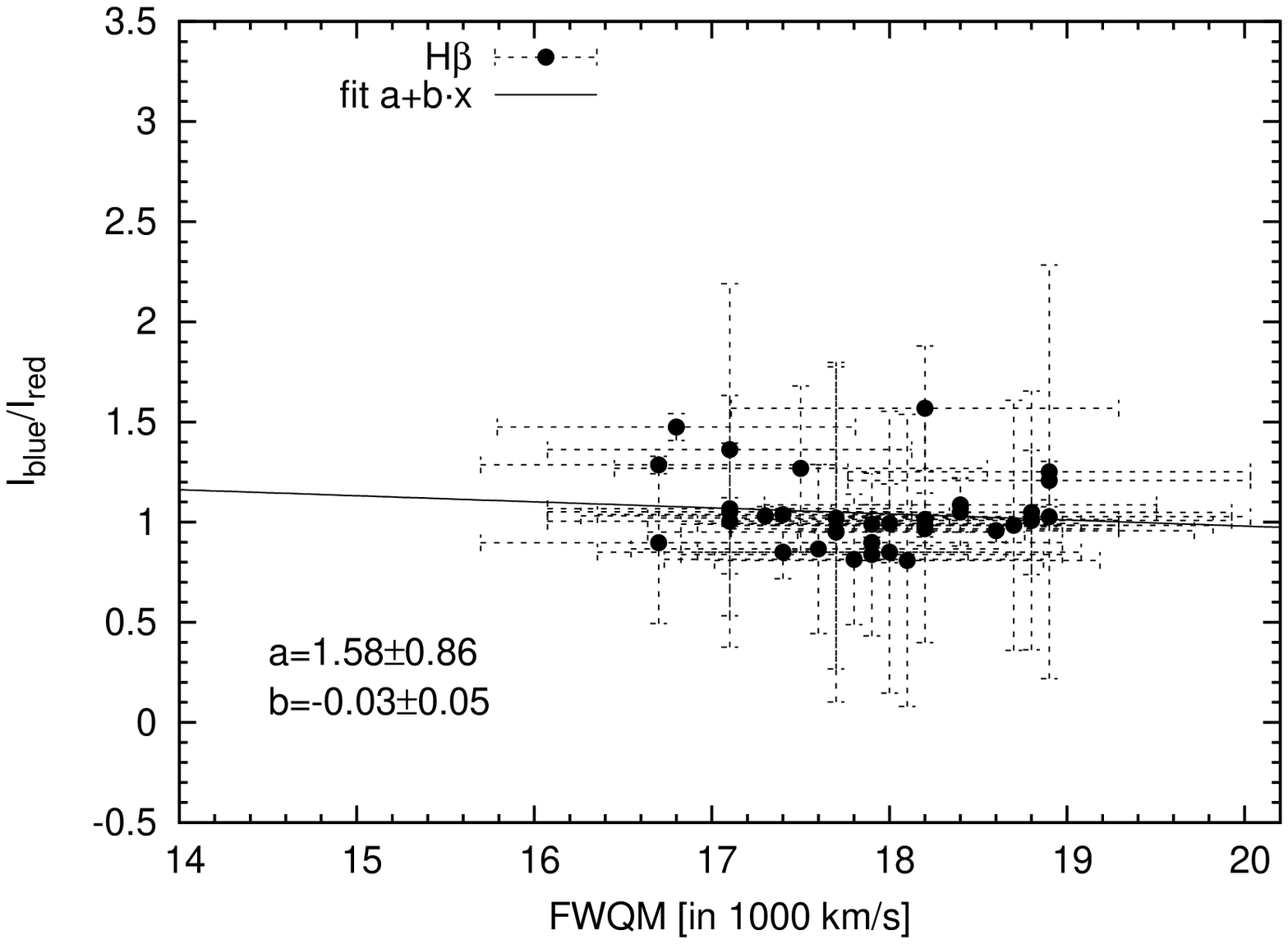}
\caption{The blue-to-red peak intensity ratio as a function of Full Width at Quarter
of Maximum intensity (FWQM) for H$\alpha$ (upper) and H$\beta$ (bottom).
{  The correlation coefficients are r =-0.55 (P$_0$=0.2E-03) and
r=-0.14  (P$_0$=0.4) for H$\alpha$ and H$\beta$ respectively. The best fit parameters are given 
on the plots. The measurements for the H$\alpha$ of \cite{g07} are given on upper plot (open triangles). 
}}\label{fig-w}
\end{figure}

However, there are some other observational facts, which are in disagreement with the disc model hypothesis:

a) A {  lack of (or slightly anti-)} correlations of the blue-to-red intensity ratio with FWQM,
as it is shown in Fig. \ref{fig-w}. In principle, one can expect in the relativistic disc
that the ratio between the blue and red peak should increase with increasing of the FWQM,
since, there should be no change in the inclination, the intensive blue peak indicates that 
the inner radius is closer to the central black hole. This should result in more extensive (and less intensive)
red part, that should contribute to the broader line measured at FWQM. I.e. the
observed tendency in $I_{blue}/I_{red}$ vs. FWQM {  should correlate, but it seems this has an opposite
trend (see Fig.  \ref{fig-w}). However, \cite{le10} demonstrated that in the case of a model of
the non-axisymmetric disc, in some cases the red-to-blue peak ratio increases as the width of the profile 
increases. Also, note here that some indication of such anti-correlation in other
double peaked AGNs can be seen in
\cite{le10}. They showed plots of 
red-to-blue peak ratio as a function of time and from a quick inspection one can see
a few cases where the changes do not follow the expectation that the FWQM 
is decreasing as the red-to-blue peak ratio increases (or the blue-to-red peak ratio decreases). 
}

b) The distance between the position of the red and blue peak is different for the H$\alpha$ ($\sim10500\ \rm km\ s^{-1}$)
and H$\beta$ ($\sim12000\ \rm km\ s^{-1}$), i.e. the  H$\beta$ shows larger distance for about 1500 $\rm km\ s^{-1}$, that,
in the frame of the disc hypothesis, {should give} that the line is closer to the central black hole, and consequently that 
relativistic effects are more observable in the H$\beta$ line profile. But, as it can be seen in Figs. \ref{month1} and
\ref{month2}, the blue boosted peak is more intensive in H$\alpha$, almost during whole monitored period.
On the other side, such huge distances between the
peaks indicate a fast rotating disc, that is probably close to the
black hole. From the rms profile (see Paper I, Fig. 11) one can see that the  change in the line 
profile is also two-peaked, i.e. there is a change in 
the broad line profile in red and blue peak in both lines, where the variability 
in the blue wing is significantly bigger than in the red one.
Note here, that there is one central peak (more intensive in the H$\beta$ line) in the rms profile, that may
be caused by a central component (also see Fig. 11 from Paper I).

%%%%%%%%%%%%%%%%%%%%%%%%%%%%%%%%
%%%%%%%%%%%%%%%%%%%%%%%%%%%%%%%%

c) Additionally, it is confusing that there is no big change in the H$\alpha$ normalized line profile during this 
long monitored period (see Figs. \ref{month1} and \ref{mean}), as well as in the displacement of the peaks.
If the emission is originating in the disc, {the disc emission is very stable during time, i.e. there is no
big change in the disc which affect the emission. {  On the contrary,  the different peak intensity ratio (e.g. 
in some periods red peak of H$\beta$ is more intensive than the blue one) indicates that 
long living changes in the disc structure should be present \citep[see e.g.][]{jov10,pop11}. These changes
should affect a compact disc ($\sim$500 gravitational radii, that is several light days), and consequently 
a stronger variability in the broad line profiles is expected.}

d) Finally, the spectropolarimetric observations are also in contradictions with the 
BLR disc model.  Spectropolarimetric
observations  \citep[see][]{a96,co98,co00} did not confirm the disc-like structure of
the BLR in Arp 102B. Especially it is interesting that the angle of the
jet direction ($\sim$ 105 degrees) corresponds to the polarization angle in
the H$\alpha$ line \citep[$\sim$103 degrees, see][]{co98}. The H$\alpha$ polarized shape of 
Arp 102B indicates a  line-emitting biconical outflow surrounded by a cylindrical scattering region \citep[][]{co98,co00}.
However, note here two facts which do not exclude disc model. First, that \citet{c97} obtained a
qualitative agreement between the  disc model, but other possible explanations of the
broad line polarization properties of Arp 102B are possible, as e.g. that to the disc emission there is additional outflow 
emission (see Fig. \ref{segm}). Second, as it was mentioned in \citet{af13}, the role of the inter-stellar polarization in the 
polarization properties of the broad lines is very important and after taking it into account, the polarization angle may be
changed.

%%%%%%%%%%%%%%%%%%%%%%%%%%%%%%%%
\begin{figure}
\centering
\includegraphics[width=9cm]{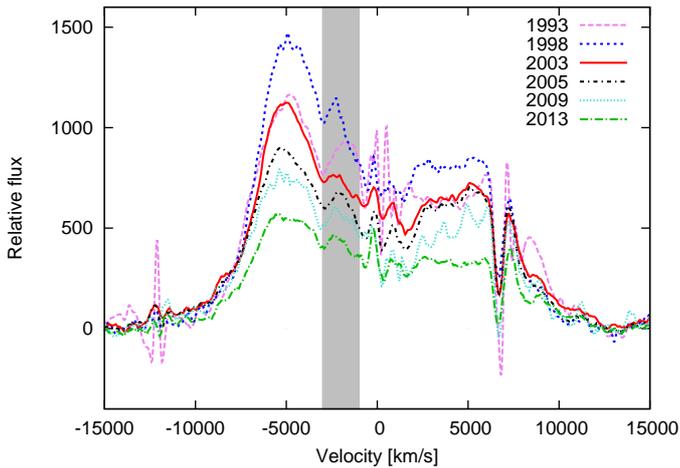}
\caption{Bump in the H$\alpha$ line profile at $\sim 2000$km s$^{-1}$
in years when it is most intensive.}\label{segm}
\end{figure}
%%%%%%%%%%%%%%%%%%%%%%%%%%%%%%%%

\subsubsection{Outflow(s) in the Arp 102B BLR}

Recently, \cite{co13} found two-armed nuclear spiral \citep[see also][]{fat11} in an extensive field around
the nucleus of
Arp 102B (2.5$\times$1.7  kpc$^2$) and they
estimated that the mass outflow rate along one (the east)  arm is between 0.26--0.32  M$\odot$ yr$^{-1}$,
that is  higher than the mass accretion rate. Therefore, one can expect that an outflow is present
in the BLR. As it can be seen in Fig. \ref{segm}, during some periods, there is a bump in the blue part of 
the H$\alpha$ line (around -2500 kms$^{-1}$), that may indicate that additionally to the disc emission, there is 
an outflow in the BLR.

In principle, a biconical outflow model can explain double-peaked line profiles
\citep[as e.g. Balmer line profiles of 3C 390.3, see][]{zh91}. 
The blue-shifted and red-shifted  peaks
are produced by the approaching and receding parts, respectively.
Additionally,  a model where an 
accelerating outflow together with an inflow of the emitting gas is dominating in the BLR can 
explain very complex line profiles \citep[also double-peaked][]{il10,le14}.

Also, spectropolarimetric observations seem to be in favor of the 
some type of a bi-conical model \citep[see][and corresponding scheme]{co98,co00}.

As a summary,  one can note that the outflow is probably present in the BLR, seen as a blue bump between the 
center of the line and blue peak
(around -2500 kms$^{-1}$, see Fig. \ref{segm}), but it is not clear if the emission 
from the outflow is dominant in the broad line profiles. Also,
one possible problem are relatively small changes in the broad line profiles, that, probably in the case of outflows should be higher.

\section{Conclusion}

We analyzed the variability in the broad line  properties in  a long period of 26 years of Arp
102B, an AGN with prominent double-peaked broad line profiles. 
We investigated the broad  line profile variations during this period
and from our investigations we can outline the following conclusions:

a) The broad line profiles have not been significantly changed in a period of 26 years in sense that during
monitored time, the shift of peaks stays almost unchanged and there are flare-like changes in the line intensity, 
and some time in the ratio of the blue-to-red peak intensity. This
is unusual if there is an emission of relatively compact accretion disc (dimension smaller than
1000 Rg), since, changes in the intensity ratio of peaks should result in the line widths as well as in the displacement of the 
peak velocities. Moreover, the intensity ratio of peaks shows {  a trend of}
an anticorrelation with the line width, 
that is opposite what one can expect from the disc emission.  Sometimes, the position of the red peak does not match 
the AD-model predictions (Fig. 6).

b) As an additional conclusion to our earlier one that the variation of fluxes of  H$\beta$ vs. H$\alpha$ has a 
small correlation \citep[see][]{sh13}, there is practically no correlation between {  velocities} of the blue
and red peaks between the  H$\beta$ and H$\alpha$ line. Moreover, the line profiles of  H$\beta$ and H$\alpha$ are different,
H$\alpha$ has more intensive blue peak, and H$\beta$ has almost equivalent intensity of peaks in the most of observations. Also, it is unexpected that the H$\beta$ is significantly broader ($\sim$1500 $\rm km\ s^{-1}$) than H$\alpha$, and that the relativistic effect of the blue-boosting is more prominent in the case of H$\alpha$

c)  An outflow in the BLR seems to be present (from time to time it can be noticed as the blue bump around -2500 kms$^{-1}$).
Although it is not clear how much the outflow can contribute to the broad line emission, it should be taken into account.

At the end, let us conclude that profiles of the broad double-peaked lines in Arp 102B can be fitted very well with 
the disc model, but there are several issues in the variation of the broad spectral line properties which 
are not in agreement with ones expected in the variability of the accretion disc structure. The
 better spectro-polarimetric observations/monitoring of Arp 102B
are needed to study the polarization properties of the ordinary
broad-line component in order to clarify of the BLR nature.

\section*{Acknowledgments}

This work was supported by INTAS (grant N96-0328), RFBR (grants
N97-02-17625 N00-02-16272, N03-02-17123, 06-02-16843, N09-02-01136,
12-02-00857a, 12-02-01237a),  CONACYT research grants 39560, 54480, and 151494 (Mexico),
PROMEP/103.5/08/4722 grant, the Ministry of
Education and Science of Republic of Serbia through the project
Astrophysical Spectroscopy of Extragalactic Objects (176001), and
DFG grant Ko 857/32-1. L. \v C. P., W. K. and D. I. are grateful 
to the Alexander von Humboldt foundation for support in the frame of program "Research Group
Linkage". {  We would like to thank the anonymous referee for very useful comments.}

%We would like to thank the anonymous referee for very useful
%comments and suggestions.

\clearpage

\appendix 

\section{The line fitting procedure and  narrow line subtraction}

{  One of the problems in the measurements of the broad line parameters is
the uncertainty in the narrow line subtraction, especially in the
red peak of H$\alpha$ and H$\beta$, since the narrow lines are right on top of 
the red peak in both broad lines. Therefore, to find uncertainties of the narrow 
line subtraction we did several tests. Additionally, we consider the B-band
absorption observed near the [SII] doublet in the H$\alpha$ wavelength range.

\subsection{Correction of the B-band absorption near [SII]}

The B-band absorption is present in the red wing of the H$\alpha$ line, near the narrow [SII] doublet
(see Fig. \ref{b-band}). To correct this absorption we used the template spectrum of  NGC 4339, which was 
observed at the same night as Arp 102B with 2.1m GHO telescope with resolution $\sim$8-9 \AA\ on Mar 26, 2003.
Then we corrected the B-band absorption near the [SII] lines in the H$\alpha$ spectral region.

In Fig. \ref{b-band} we illustrate the correction of the B-band absorption. As one can see in Fig. \ref{b-band}
the absorption is well corrected, but in some spectra the residuals are still present. However, the residuals are weak 
and can not affect the narrow line estimates, as well as the estimates of the red-peak position 
in H$\alpha$ (see following section and Figs. \ref{1}). 
Therefore, we accepted the parameters from the fits where the B-band absorption iwas not corrected.

\begin{figure}
	\centering
	\includegraphics[width=8.cm]{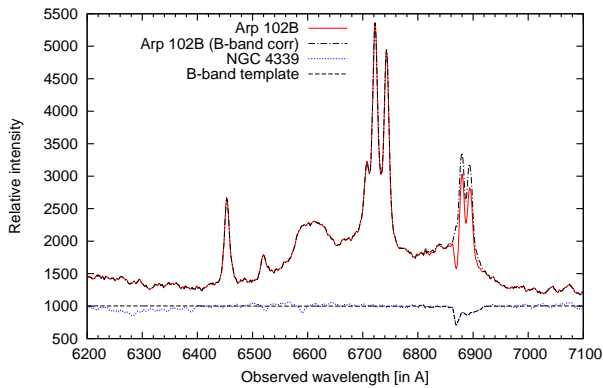}
	\caption{  The correction of the H$\alpha$ line on B-band absorption.
	The spectrum of NGC 4399 (shown below) is observed in the same epoch (Mar 26, 2003) as the 
	spectra of Arp 102B.}
	\label{b-band}
\end{figure}

\subsection{Estimation of the narrow line emission}

As we mentioned above, the uncertainties in the estimations of the broad line 
parameters come mainly from the subtraction of the narrow lines in the H$\beta$ and H$\alpha$ spectral range. It is 
hard to handle the removal of the narrow emission lines in Arp 102B, because the narrow lines 
are right on top of the red peak in the H$\alpha$ and H$\beta$ broad lines. To 
explore the uncertainty in the narrow line subtractions we performed two  approaches: 
i) fitting the narrow lines with Gaussian functions before and after the B-band absorption
has been corrected (as described in \S2.2), and ii) estimation of the 
broad profile using spline fitting (using DIPSO). 
In Figs. \ref{1} and \ref{2} we illustrated the subtraction of the narrow lines using these two methods.
In Fig. \ref{1} (fourth panel) we compared the broad H$\alpha$ line profile obtained after the subtraction of the
narrow lines using these two procedures (note that we fitted the spectra before and after the
correction for the B-band absorption). As it can be seen in Fig. \ref{2} (bottom panel), the fitted position of 
the blue and red peaks is practically the same also for H$\beta$ for both procedures, therefore we used the Gaussian decomposition in our estimates of the narrow line contribution.

\subsection{Correction of the fits using the ratio of the narrow lines}

As we mentioned above, some parameters of the narrow lines are fixed in the spectral fit; first of all
the Gaussian velocity dispersions and shift for all narrow lines in the line wavelength range, as well as
the ratio of the [OIII]4959,5007, that is fixed to 1:3 \citep{dim07}. Also, one cannot expect that the ratio 
of narrow lines is changing during the period of several years. Therefore one
very clear way to test the robustness of the narrow line fittings is to compute the
emission line ratios for each combined fit. 

The timescale of the observations is long enough that there could be some variation in the narrow emission line ratios, 
but certainly over timescales of $\sim$5-10 years there
should not be major changes in the narrow line ratios (except of the random scatter). 
We calculated the narrow line
ratios of [NII]/H$\alpha$, [OI]6300/H$\alpha$, [SII]6715/H$\alpha$, 
and  [OIII]5007/H$\beta$. The line ratios in the H$\alpha$ and H$\beta$ wavelength range are presented in Fig. \ref{nel},
where dashed lines represent the averaged ratio and shaded regions the $\pm$ 10\% of the averaged value.
As it can be seen in the upper panels of Fig. \ref{nel}, there is not any trend in the line ratio variability 
during the monitored period. 
Consequently, we  corrected those fits for which the narrow line ratios were significantly
different from the mean value, i.e. the spectra with a big narrow line ratio difference have been re-fitted 
taking the constraint in the fit that the line ratio has to be in the frame of $\pm$ 15\% 
of the corresponding  mean value. In Fig. \ref{nel}, bottom panels, the line ratios are shown after 
the correction of a number of spectra.

\subsection{The broad line fit and estimates of the parameter uncertainties}

To find the broad line parameters of 
the H$\alpha$ and H$\beta$ broad lines, we performed  the Gaussian
analysis, fitting the month-averaged broad line profiles. 
The broad profile was fitted with three Gaussian functions, 
corresponding to the blue, central, and red component (Fig. \ref{fig_ha_broad_gauss}). Each Gaussian was
described with three free parameters, i.e. in the fit procedure we have nine free parameters: three intensities,
 widths and shifts of Gaussian functions which describe the blue, central and red part of the broad line.
 
Inspection of the obtained parameters from the fit, as well as several additional tests
(changing slightly parameters), showed that the central component, is
often shifted to the blue (especially in H$\alpha$) and that, in some cases,
this component has a big (unexpected) change in the shift. Additionally,
 we found that the position of the central component can significantly affect parameters of the  peaks, especially 
the red one. Therefore, we repeated the fitting procedure with three broad Gaussian functions, but
putting the limits on the shift of the central component to be between -1000 kms$^{-1}$  and 600 kms$^{-1}$. 
Using this procedure we obtained an additional set of broad line parameters (parameters of the 
blue, central and red Gaussians). 
Inspection of differences between the parameters obtained from the fits with and without 
constraint of the central component shift showed that the peak positions are not significantly changed, but the 
intensities of the red peak, the shift of the central component, and the widths of the components 
have been significantly changed.

We compared the error-bars from the fits with differences between the parameters from these two fitting procedures,
and found that the error-bars of parameters are often significantly smaller than the
differences between the parameters from both fits. Therefore, we calculated the averaged parameters from the two fits, and
accepted uncertainties (error-bars) as discrepancy between the parameters from the two fits.
The averaged broad line parameters and corresponding estimated
uncertainties are given in Tables \ref{tab02} (for H$\alpha$) and  \ref{tab03} (for H$\beta$).

}

\begin{figure*}
	\centering
	\includegraphics[width=8.cm]{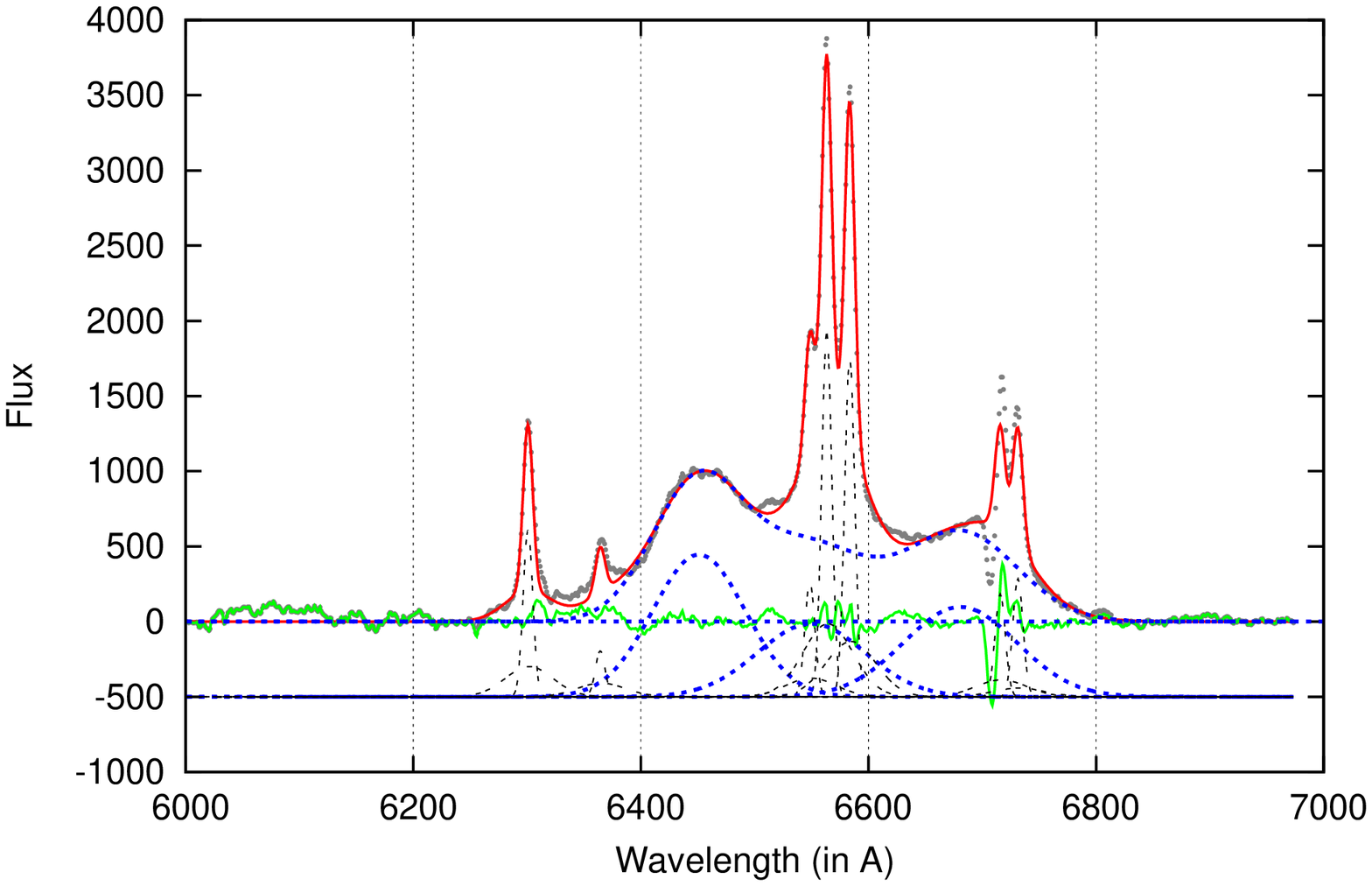}
	\includegraphics[width=8.cm]{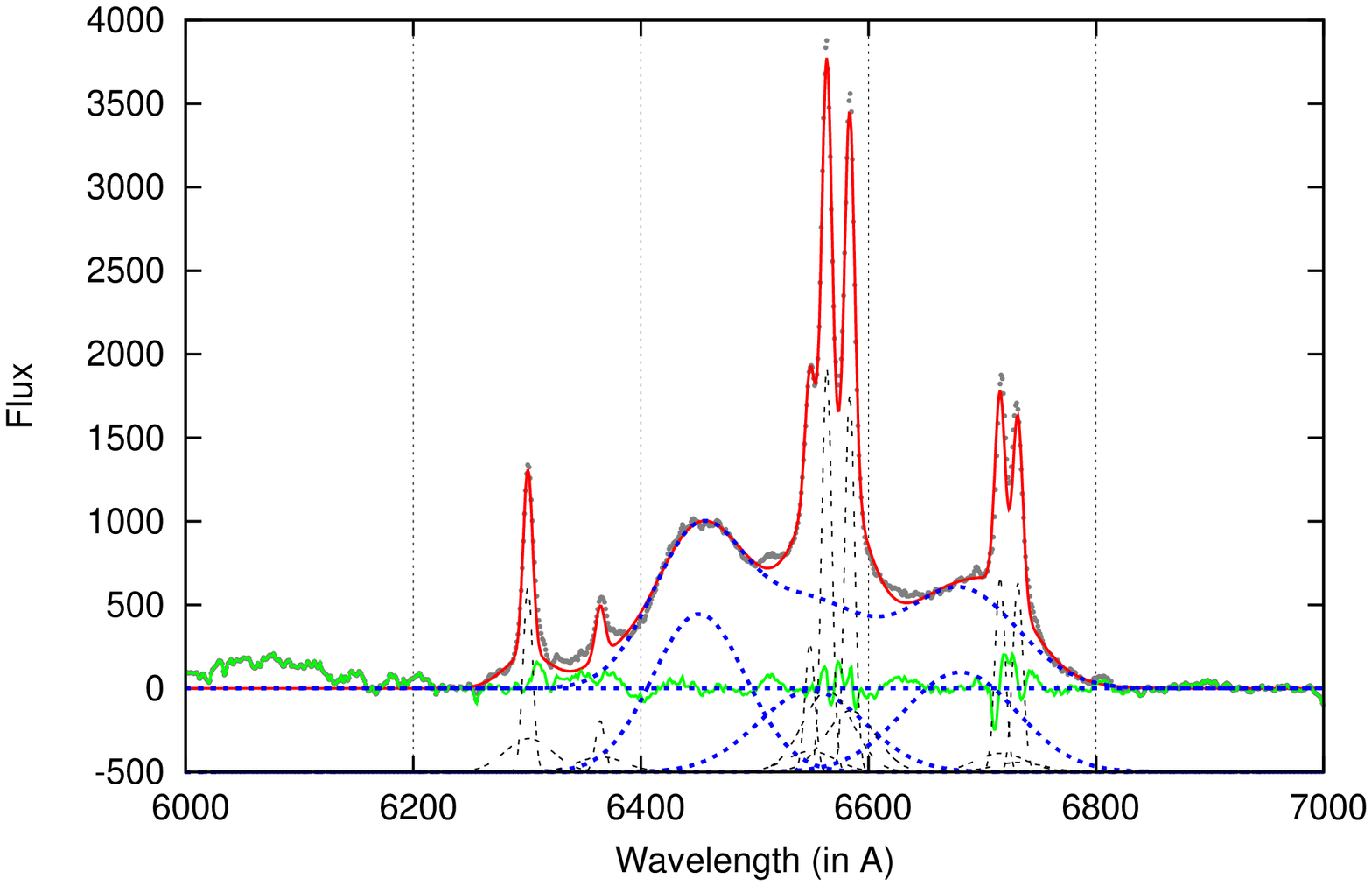}	
	\includegraphics[width=8.cm]{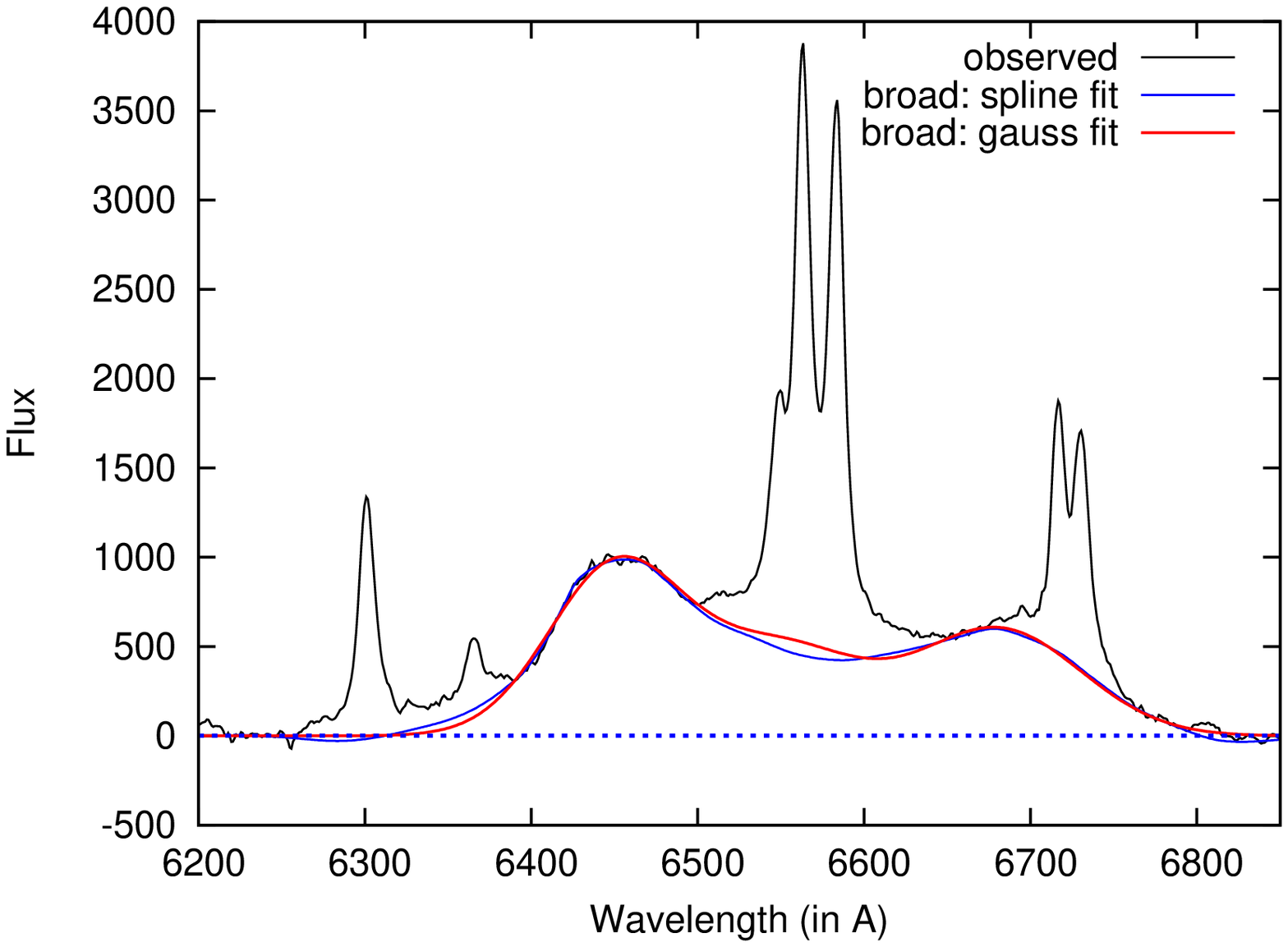}
	\includegraphics[width=8.cm]{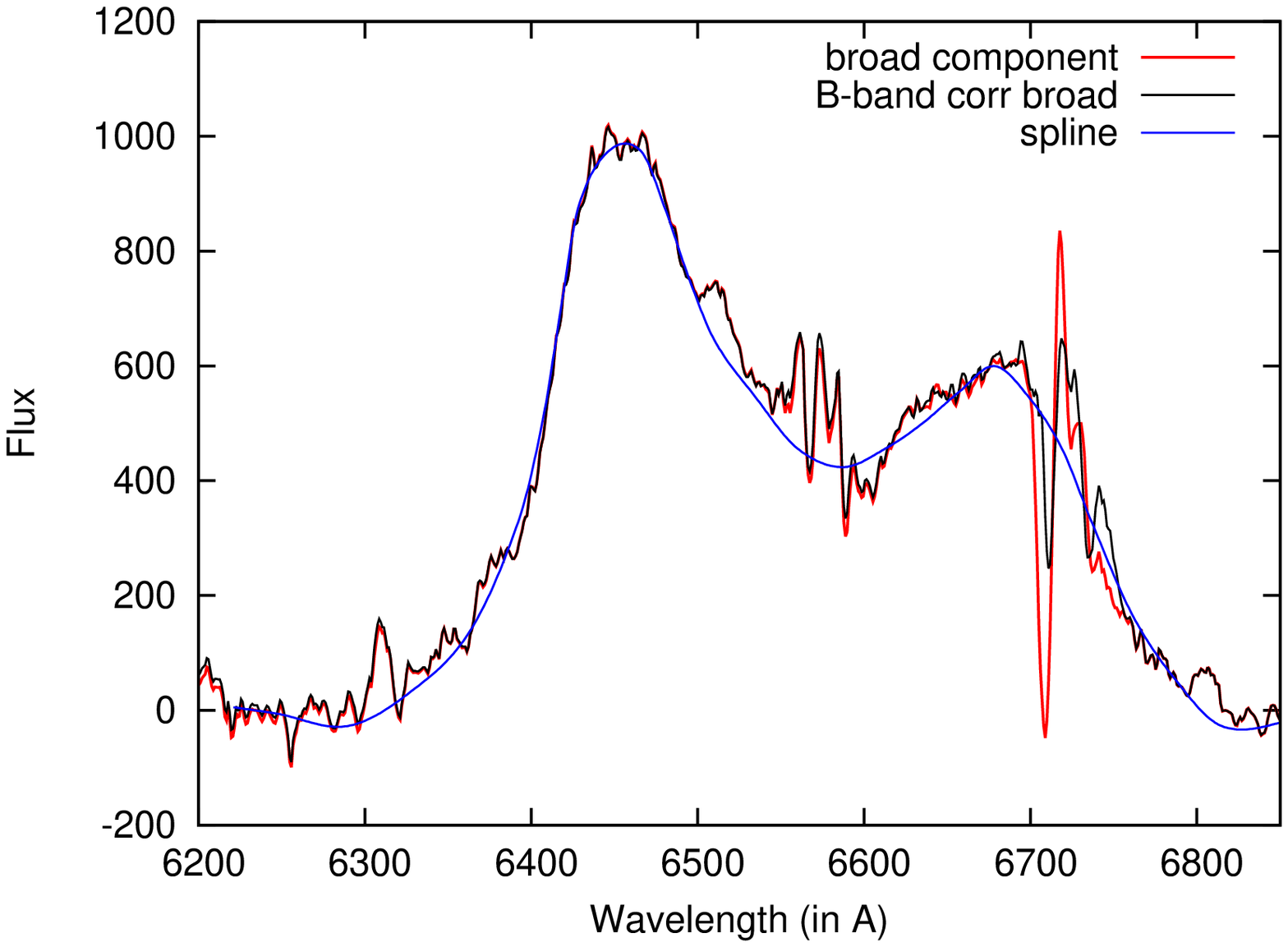}	
	\caption{  Multi-gaussian fitting of the H$\alpha$ wavelength range of the spectrum observed on Apr 02, 2003. 
	{  Upper left}: Fitting of the spectra where the B-band absorption is not corrected.
	{  Upper right}: The same but for the B-band corrected observed spectrum. 
	In these two plots same Gaussians parameters (apart for the [SII] lines which
	intensities are only increased) have been used.
	{  Bottom left:} Narrow lines removed using the DIPSO spline fitting of the broad component, compared with the 
	3-gaussian broad-component fitting.
	{  Bottom right:} Comparison of the H$\alpha$ broad component (after narrow lines subtraction) before
	and after the B-band absorption correction and the broad component obtained
	using the DIPSO spline fitting. The blue peak position is the same, a slight difference is seen in the
	red one.}
	\label{1}
\end{figure*}

\begin{figure}
	\centering
	\includegraphics[width=8.cm]{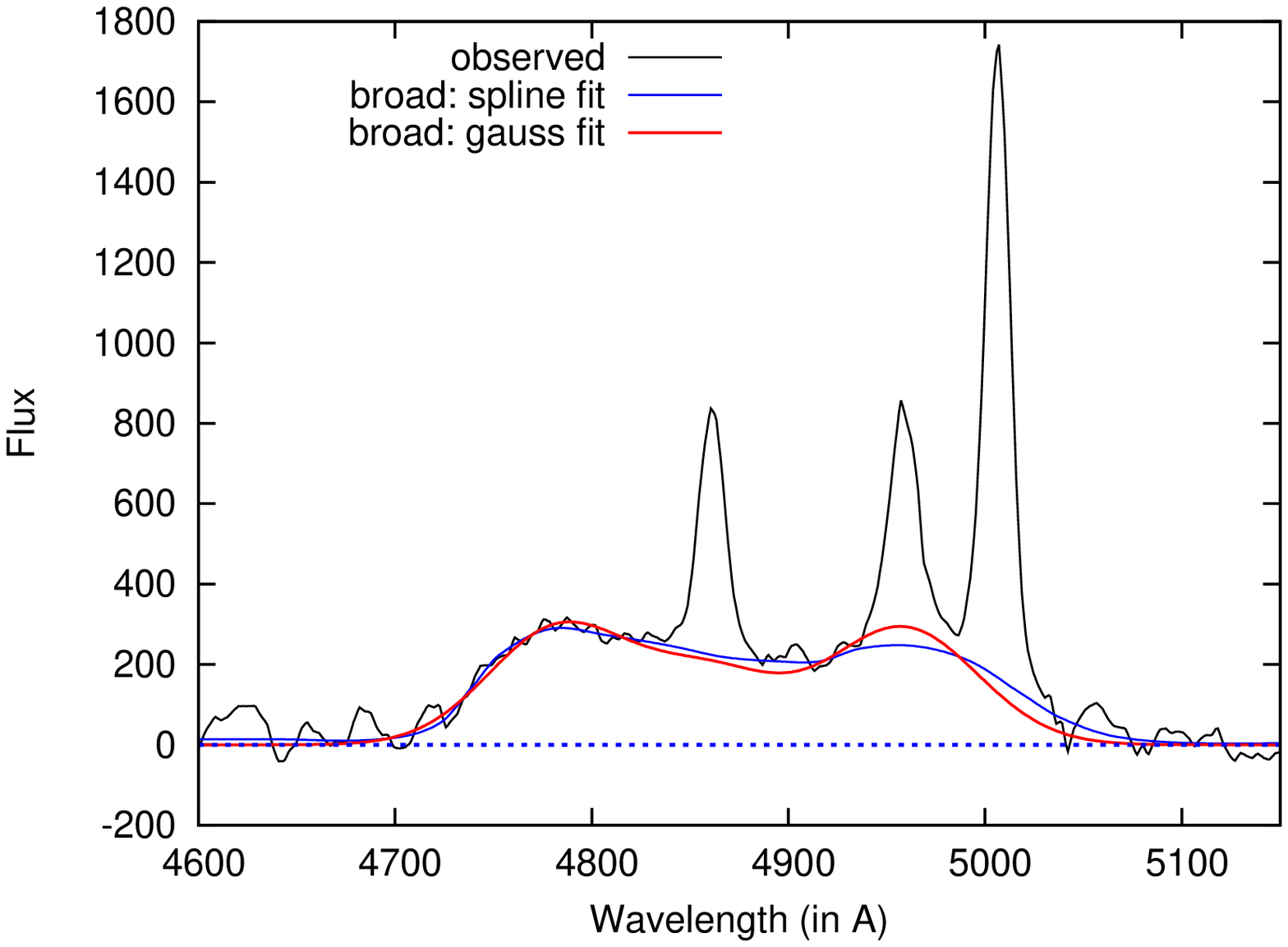}
	\includegraphics[width=8.cm]{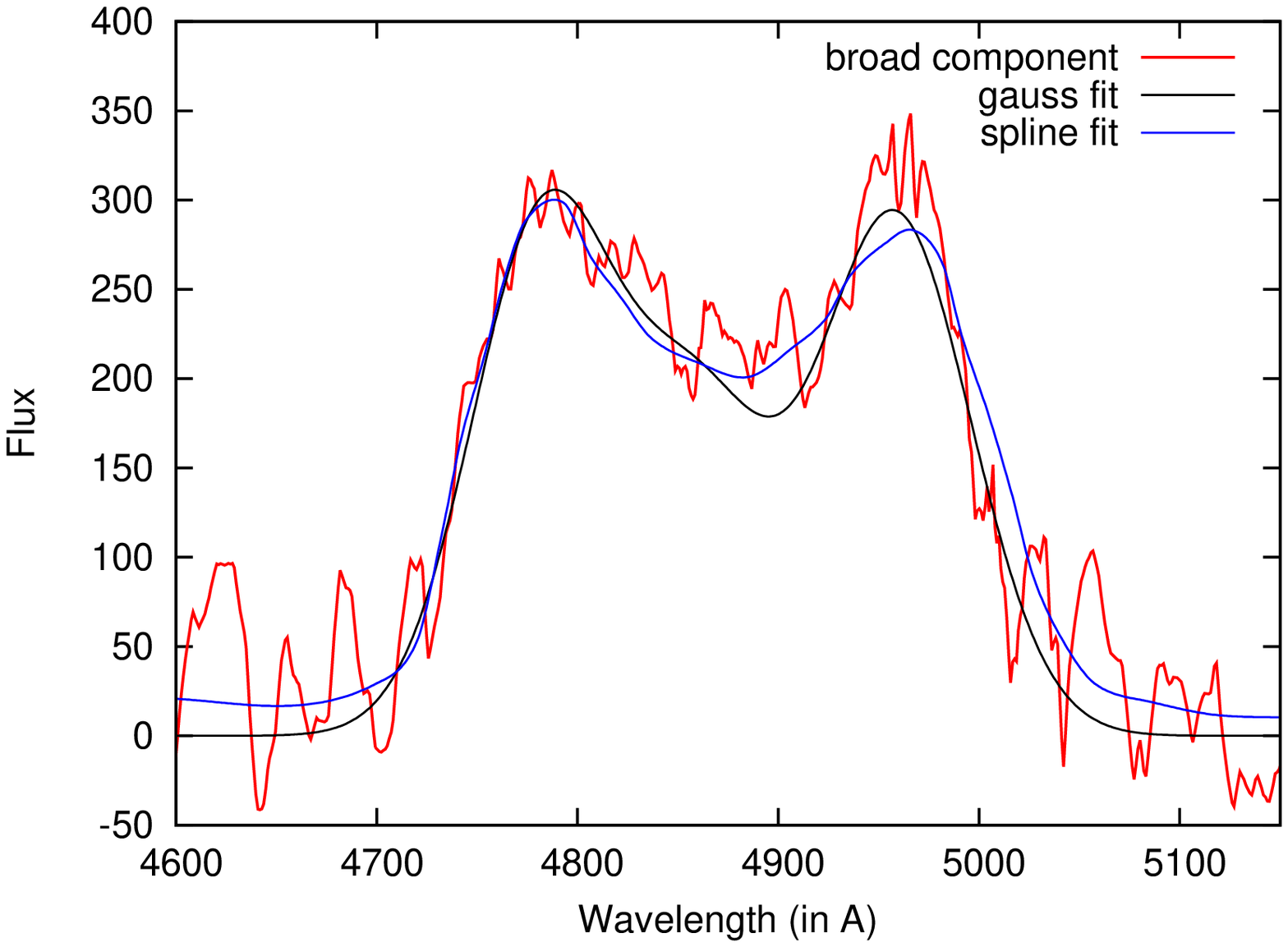}
	\caption{  Upper panel: Narrow lines removed in H$\beta$ using the DIPSO spline fitting of 
	the broad component, compared with the 3-gaussian broad-component fitting.
    Bottom panel: Comparison of the H$\beta$ broad component (after narrow lines subtraction) 
    using the 3-gaussian and DIPSO spline fitting. The blue peak position is the same, a slight difference is seen in the
	red one.	}
	\label{2}
\end{figure}

\begin{figure*}
	\centering
	\includegraphics[width=8.cm]{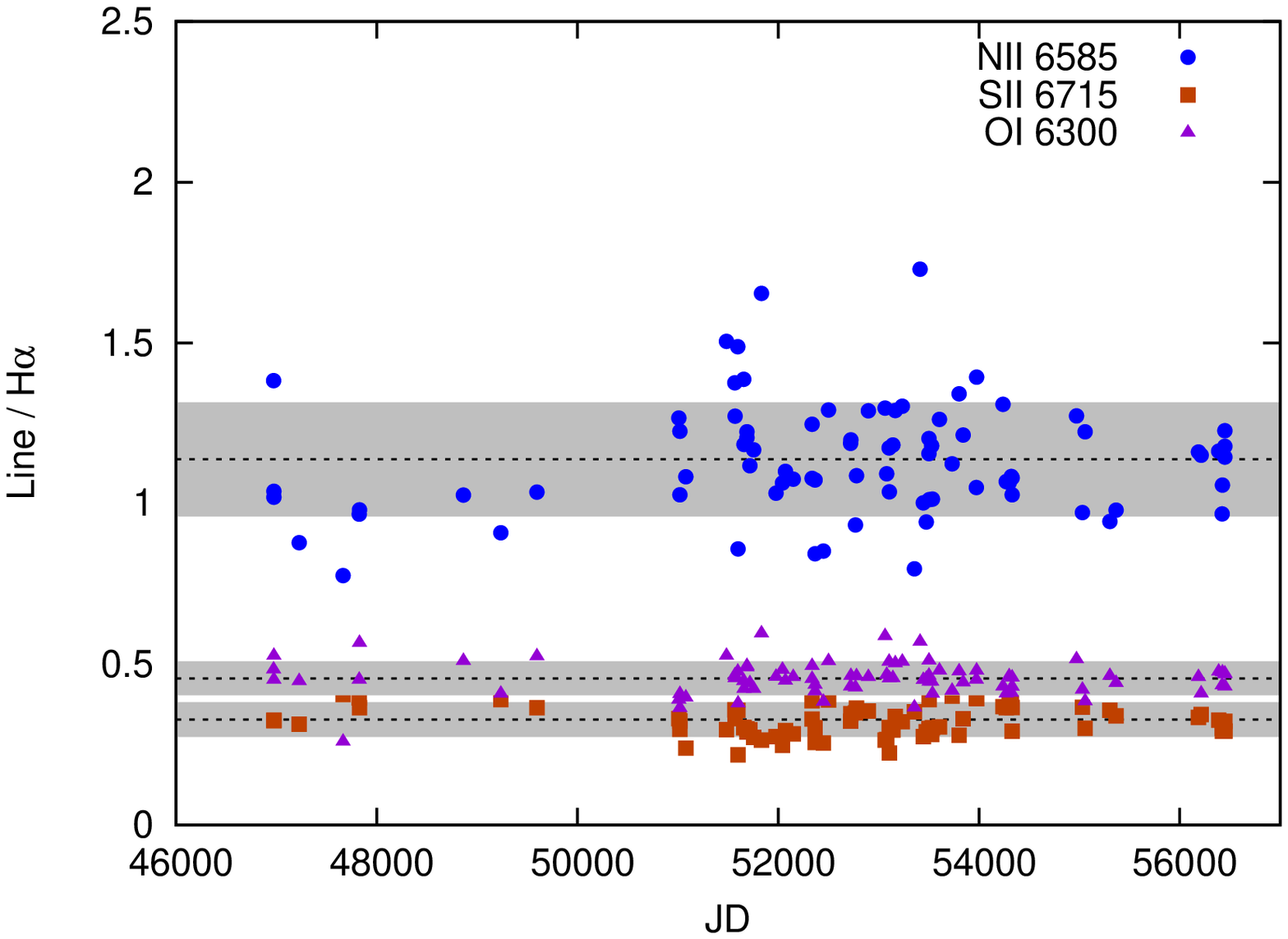}
	\includegraphics[width=8.cm]{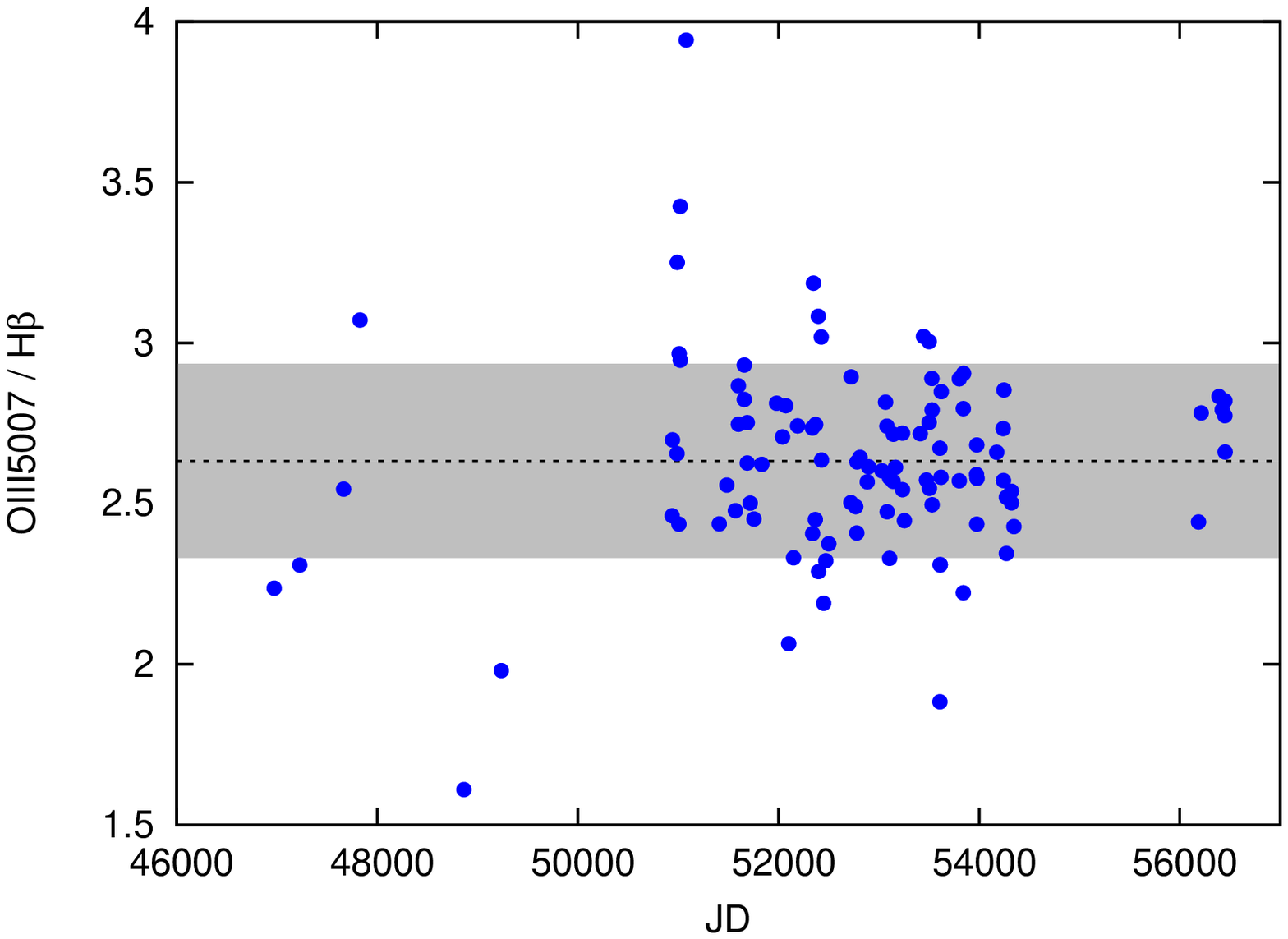}
	\includegraphics[width=8.cm]{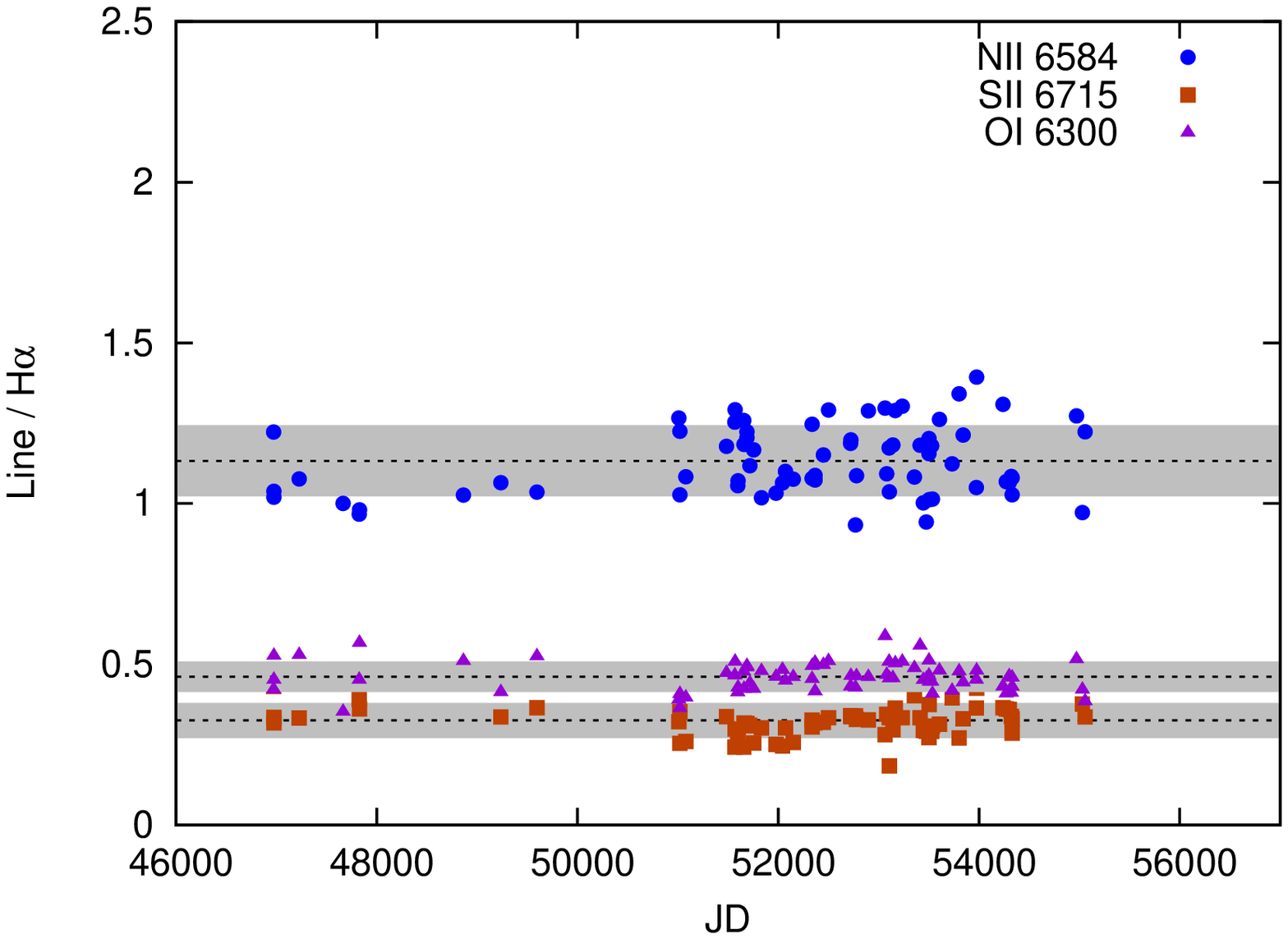}	
	\includegraphics[width=8.cm]{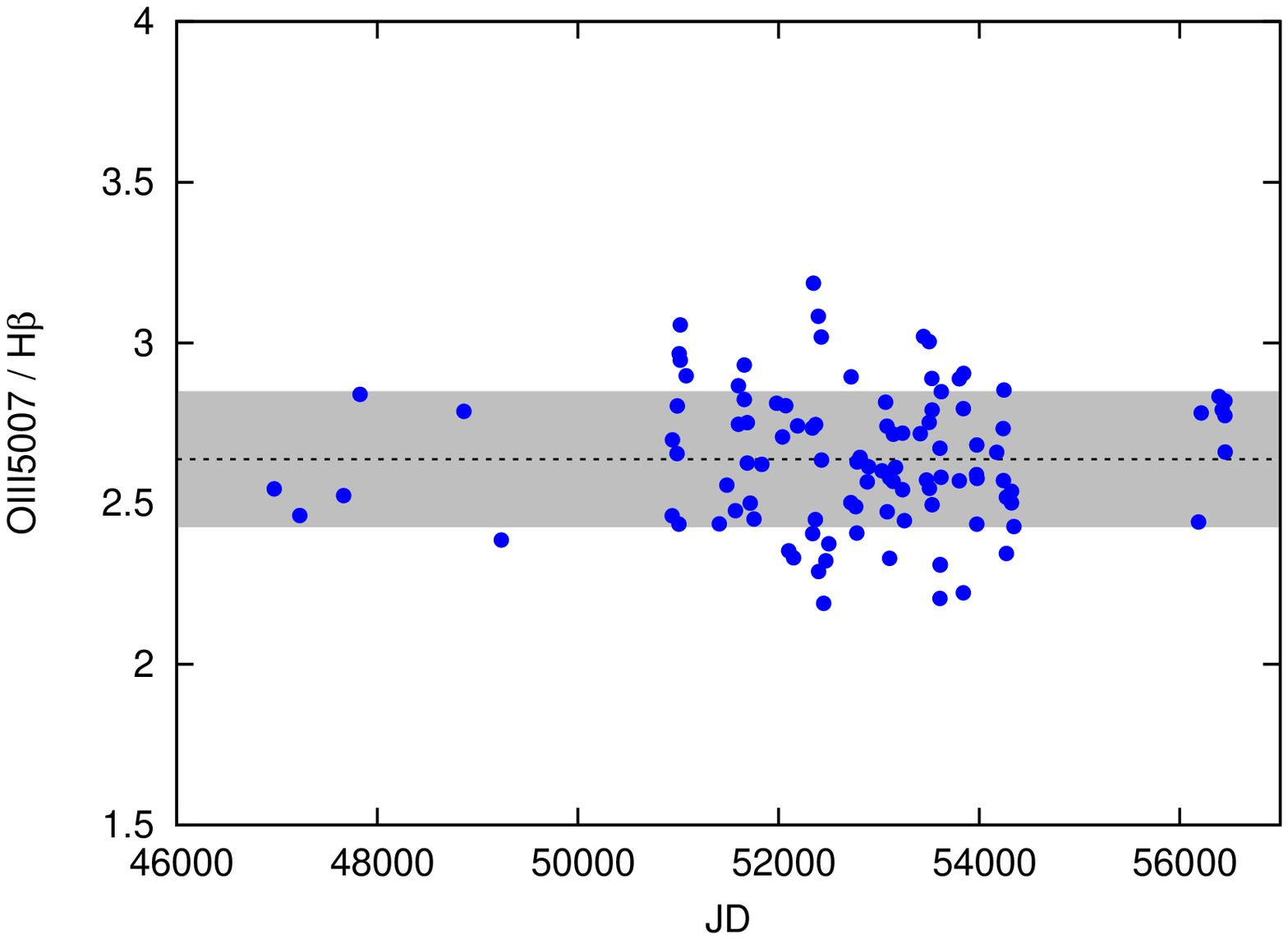}	
	\caption{  The ratio of the narrow emission line 
	(labeled on plots) fluxes (in the H$\alpha$ and H$\beta$ wavelength range) during the monitored period. The narrow lines
	fluxes are calculated from the Gaussian fitting parameters as a sum of two components (one narrower fitting
	the line core and one broader fitting the line wings, see \S2.2) The dashed lines with the
	shaded regions represent the mean value and the deviation of 10\% from this. On upper panels, the ratios
	obtained from the fit without any constraint, while in the bottom panels, the points with big scattering (from panel
	up) have been corrected.
	}
	\label{nel}
\end{figure*}


\begin{thebibliography}{}


\bibitem[\protect\citeauthoryear{Afanasiev et al.}{2014}]{af13}
Afanasiev, V. L., Popovi\'c, L. \v C., Shapovalova, A. I., Borisov, N. V., 
Ili\'c, D. 2014, MNRAS, 440, 519 

\bibitem[\protect\citeauthoryear{Antonucci et al.}{1996}]{a96} Antonucci, R., Hurt, T.,  Agol, E. 1996, ApJ, 456, L20


\bibitem[\protect\citeauthoryear{Chen \& Halpern}{1989}]{ch89} Chen, K. \& Halpern, J. 1989, ApJ, 344, 115

%\bibitem[\protect\citeauthoryear{Chen \& Halpern}{1990}]{ch90} Chen, K., \& Halpern, J.~P.\
%1990, \apjl, 354, L1

\bibitem[\protect\citeauthoryear{Chen et al.}{1989}]{c89} Chen, K. Halpern, J. P., Filippenko, A. V. 1989, ApJ, 339, 742

\bibitem[\protect\citeauthoryear{Chen et al.}{1997}]{c97} Chen, K, Halpern, J. P., Titarchuk, L. G. 1997, ApJ, 483, 194

\bibitem[\protect\citeauthoryear{Corbett et al.}{2000}]{co00}
Corbett, E. A., Robinson, A., Axon, D. J., Young, S. 2000, MNRAS, 319, 685

\bibitem[\protect\citeauthoryear{Corbett et al.}{1998}]{co98} 
Corbett, E. A., Robinson, A., Axon, D. J., Young, S., Hough, J. H. 1998, MNRAS, 296, 721

\bibitem[\protect\citeauthoryear{Couto et al.}{2013}]{co13}
Couto, G. S., Storchi-Bergmann, T., Axon, D. J., Robinson, A., Kharb, P.,; Riffel, R. A. 2013, MNRAS, 435, 2982.

\bibitem[\protect\citeauthoryear{Dimitrijevi\'c et al.}{2007}]{dim07} Dimitrijevi\'c, M. S., Popovi\'c, L. \v C., Kova\v cevi\'c, J., Da\v ci\'c, M.,
 Ili\'c, D. 2007, \mnras, 374, 1181

 \bibitem[\protect\citeauthoryear{Eracleous \& Halpern}{1994}]{eh94} Eracleous, M., \& Halpern, J. P. 1994, ApJS, 90, 1

\bibitem[\protect\citeauthoryear{Eracleous et al.}{1997}]{e97} Eracleous, M., Halpern, J., Gilbert, A., Newman, J. A.,  Filippenko, A.V. 1997, ApJ,
490, 216

\bibitem[\protect\citeauthoryear{Eracleous et al.}{1995}]{er95} Eracleous, M., Livio, M., Halpern, J.P., \& Storchi-Bergmann, T. 1995,
 \apj, 438, 610

\bibitem[\protect\citeauthoryear{Eracleous et al.}{1996}]{er96} Eracleous, M., Halpern, J.P., \& Livio, M. 1996, \apj, 459, 89

\bibitem[\protect\citeauthoryear{Fathi et al.}{2011}] {fat11}  Fathi, K., Axon, D. J., Storchi-Bergmann, T., Kharb, P.,
Robinson, A., Marconi, A., Maciejewski, W., Capetti, A. 2011, ApJ,
736, 77

\bibitem[\protect\citeauthoryear{Gaskell}{1983}]{ga83} Gaskell, C. M. 1983,  Liege
International Astrophysical Colloquia, 24, 473

\bibitem[\protect\citeauthoryear{Gezari et al.}{2007}]{g07} Gezari, S., Halpern, J.~P., \& Eracleous, M.\ 2007, \apjs, 169, 167

\bibitem[\protect\citeauthoryear{Gezari et al.}{2004}] {g04} Gezari, S., Halpern, J. P., Eracleous, M., Filippenko, A. V. 2004, IAUS, 222, 95

\bibitem[\protect\citeauthoryear{Goad \& Wanders}{1996}]{gw96} Goad, M., \& Wanders, I.\ 1996, \apj, 469, 113 

\bibitem[\protect\citeauthoryear{Halpern et al.}{1996}]{h96} Halpern, J. P., Eracleous, M., Filippenko, A. V., \& Chen, K. 1996, ApJ, 464, 704

\bibitem[\protect\citeauthoryear{Ili\'c et al.}{2010}]{il10} Ili\'c, D., Popovi\'c, L. \v C., Shapovalova, A. I.,
Kova\v cevi\'c, A., Le\'on-Tavares, J., Chavushyan, V. H. 2010, Mem. Soc. Ast. It. Supp., 15, 166

\bibitem[\protect\citeauthoryear{Jovanovi\'c et al.}{2010}]{jov10}
Jovanovi\'c, P., Popovi\'c, L. \'c., Stalevski, M., Shapovalova, A. I. 2010, ApJ, 718, 168

\bibitem[\protect\citeauthoryear{Le\'on-Tavares et al.}{2014}]{le14}
Le\'on-Tavares, J., Popovi\'c, L. \v C., Ili\'c, D.,Chavushyan, V. H. 2014, in preparation

\bibitem[\protect\citeauthoryear{Lewis et al.}{2010}]{le10}
Lewis, K. T., Eracleous, M., Storchi-Bergmann, T.  2010, ApJS, 187, 416

\bibitem[\protect\citeauthoryear{Miller \& Peterson}{1990}]{mp90} Miller, J. S., Peterson, B. M. 1990, ApJ, 361. 98

\bibitem[\protect\citeauthoryear{Newman et al.}{1997}]{n97} Newman, J. A., Eracleous, M., Filippenko, A. V.,  Halpern, J. 1997, ApJ, 485, 570

\bibitem[\protect\citeauthoryear{Popovi\'c et al.}{2004}]{pop04}
Popovi\'c, L. \v C., Mediavilla, E., Bon, E., \& Ili\'c, D. 2004
A\&A 423, 909

\bibitem[\protect\citeauthoryear{Popovi\'c et al.}{2001}]{pop01} Popovi\'c, L. \v C.. 
Mediavilla, E. G.,  Mu\~noz, J. A.2001, A\&A, 378, 295

\bibitem[\protect\citeauthoryear{Popovi\'c et al.}{2011}]{pop11}
Popovi\'c, L. \v C.,  Shapovalova, A. I., Ili\'c, D., Kova\v cevi\'c, A., Kollatschny, W., Burenkov,
A. N., Chavushyan, V. H., Bochkarev, N. G., Le\'on-Tavares, J.  2011, A\&A,  528A, 130

\bibitem[\protect\citeauthoryear{Riffel et al.}{2006}]{ri06} Riffel, R., Rodr{\'{\i}}guez-Ardila, A.,
\& Pastoriza, M.~G.\ 2006, \aap, 457, 61 

\bibitem[\protect\citeauthoryear{Shapovalova et al.}{2013}]{sh13} 
Shapovalova, A.~I., Popovi{\'c}, L.~{\v C}., Burenkov, A.~N., et al.\ 2013, \aap, 559, A10 

\bibitem[\protect\citeauthoryear{Sergeev et al.}{2000}]{s00} Sergeev, S. G., Pronik, V. I.,  Sergeeva, E. A. 2000, A\&A, 356, 41


\bibitem[\protect\citeauthoryear{Strateva et al.}{2003}]{st03} Strateva, I.~V., 
Strauss, M.~A., Hao, L., et al.\ 2003, \aj, 126, 1720 


\bibitem[\protect\citeauthoryear{Sulentic et al.}{1990}]{s90} Sulentic, J. W., Zheng, W., Calvani, M., Marziani, P. 1990, ApJ, 355, 15


\bibitem[\protect\citeauthoryear{Zheng et al.}{1991}]{zh91}	Zheng, W., Veilleux, S., Grandi, S. A. 1991, ApJ, 381, 418.


\end{thebibliography}
\end{document}